\documentclass[authoryear]{elsarticle}

\usepackage{lineno,hyperref}

\journal{Journal of Computational Physics}

\usepackage{geometry}
\biboptions{authoryear}

\usepackage{natbib}

\usepackage{url}
\usepackage{amsmath}
\usepackage{amssymb}

\usepackage[dvipsnames]{xcolor}

\usepackage{graphicx}

\newcommand{\RR}{\mathrm{Re}}

\def\doubleunderline#1{\underline{\underline{#1}}}
\usepackage{tensor}
\usepackage{dsfont}

\usepackage{amssymb}

\usepackage{bbm}

\usepackage{algorithmicx}
\usepackage{algorithm}
\usepackage[noend]{algpseudocode}
\makeatletter
\renewcommand{\ALG@beginalgorithmic}{\footnotesize}
\makeatother

\usepackage{booktabs}

\usepackage{amsmath}

\makeatletter
\newcommand{\vast}{\bBigg@{4}}
\newcommand{\Vast}{\bBigg@{7}}
\makeatother

\usepackage[]{float}

\begin{document}

\begin{frontmatter}

\title{A conservative and non-dissipative Eulerian formulation for the simulation of soft solids in fluids}

\author[1]{Suhas S. Jain}
\ead{sjsuresh@stanford.edu}

\author[2]{Ken Kamrin}
\ead{kkamrin@mit.edu}

\author[1]{Ali Mani\corref{mycorrespondingauthor}}
\cortext[mycorrespondingauthor]{Corresponding author}
\ead{alimani@stanford.edu}

\address[1]{Center for Turbulence Research, Stanford University, Stanford, CA, 94305, USA}
\address[2]{Deparment of Mechanical Engineering, Massachusetts Institute of Technology, Cambridge, MA, 02139 USA}

\begin{abstract}
Soft solids in fluids find wide range of applications in science and engineering, especially in the study of biological tissues and membranes. In this study, an Eulerian finite volume approach has been developed to simulate fully resolved incompressible hyperelastic solids immersed in a fluid. We have adopted the recently developed reference-map technique (RMT) by Valkov et. al (J. Appl. Mech., 82, 2015) and assessed multiple improvements for this approach. 
These modifications maintain the numerical robustness of the solver and allow the simulations without any artificial viscosity in the solid regions (to stabilize the solver). This has also resulted in eliminating the striations (``wrinkles") of the fluid-solid interface that was seen before and hence obviates the need for any additional routines to achieve a smooth interface. An approximate projection method has been used to project the velocity field onto a divergence free field. Cost and accuracy improvements of the modifications on the method have also been discussed. 
\end{abstract}

\begin{keyword}
fluid-structure interaction, conservation, central differences, Eulerian approach, level set.
\end{keyword}

\end{frontmatter}


\section{Introduction}

Soft solids in fluids are ubiquitous in nature. Study of these systems is of practical relevance in science and engineering, especially in the field of biomedicine \citep{turitto1972platelet,wootton1999fluid,andrews1999role,fogelson2004platelet}. Some of such applications involve the study of the interaction between micro-bubble collapse-induced shock waves with the tissue in an animal body \citep{Adami2016}, study of the electroporation phenomenon \citep{neumann1982gene}, study of hemodynamics and suspension of blood cells \citep{pozrikidis2003modeling,pozrikidis2010computational}.

\begin{figure}[h]
    \centering
    \includegraphics[width=\textwidth]{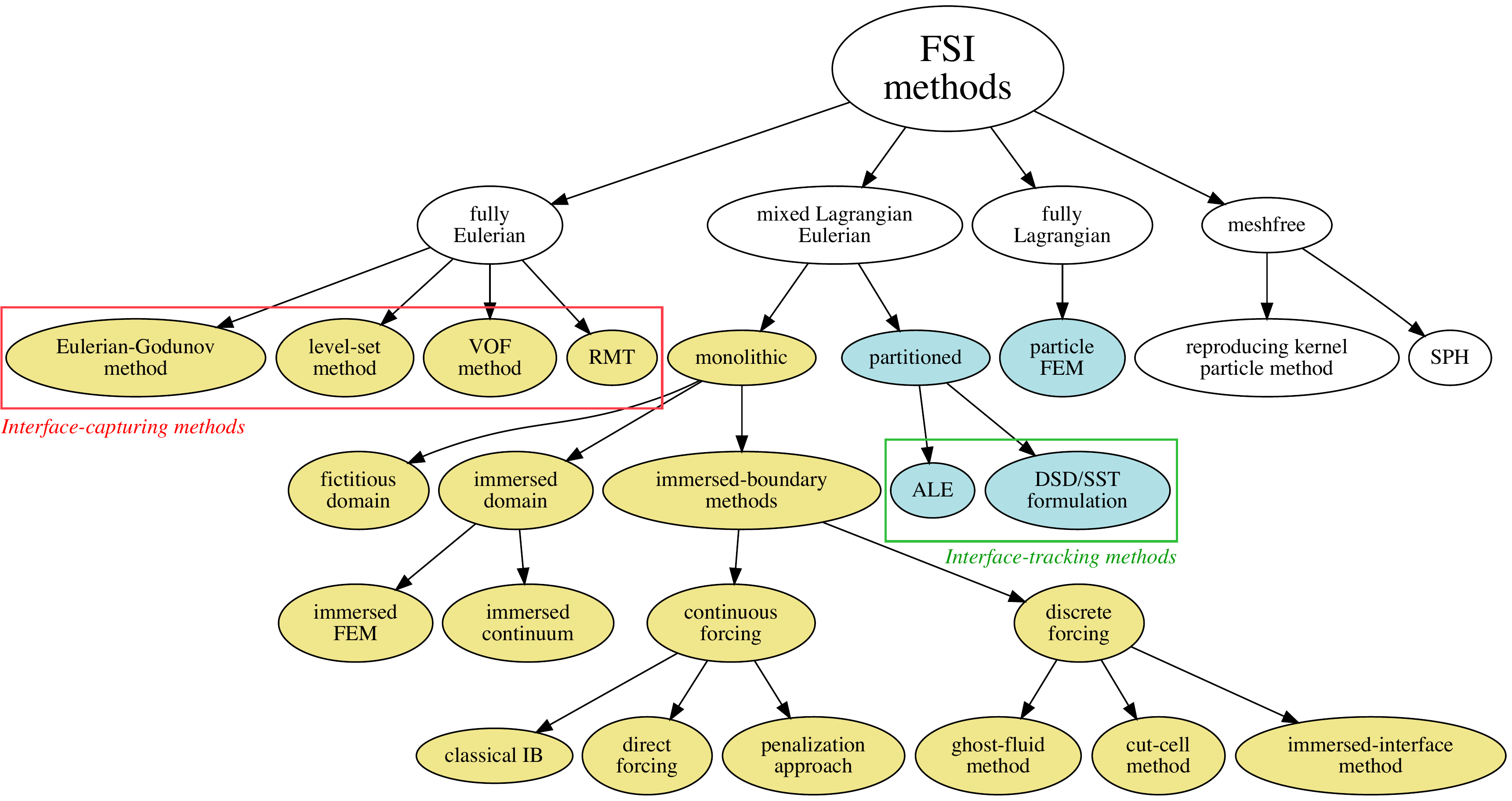}
    \caption{A broad classification of widely used FSI methods in the literature. Methods with a yellow background color represents ``\textit{non-conforming mesh methods}", where the mesh does not comply with the shape of the solid structures and the methods with a blue background color represents ``\textit{conforming mesh methods}", where the mesh complies with the shape of the structures. Fully Eulerian approaches are also referred to as ``\textit{interface-capturing methods}" and the partitioned-based mixed-Lagrangian-Eulerian methods are referred to as ``\textit{interface-tracking methods}" in the literature.}
    \label{fig:classification}
\end{figure}

Numerical methods to simulate a fluid-solid coupled system, also known as a fluid-structure interaction (FSI) problem can be broadly classified into mesh-based methods and meshfree methods. Further, the mesh-based methods can be subdivided into (a) fully-Eulerian approach, where both fluids and solids are solved on an Eulerian grid, (b) mixed Lagrangian Eulerian approach, where typically fluids are solved on an Eulerian grid and solids are represented using a Lagrangian grid, (c) fully Lagrangian approach, where both fluids and solids are solved using a Lagrangian grid. A detailed classification of various methods used to study FSI problems is shown in the Figure \ref{fig:classification}. 

FSI has historically been studied using a partitioned-based mixed-Lagrangian-Eulerian approaches, where fluid and solid regions are solved separately on different meshes using different methods (see the, arbitrary Lagrangian-Eulerian (ALE) approach of \cite{Hu2001,hirt1974arbitrary,nitikitpaiboon1993arbitrary,hughes1981lagrangian,belytschko1980fluid}, deforming-spatial-domain/stabilized-space-time approach (DSD/SST) of \cite{tezduyar1992new,hughes1996space}). These methods have been widely used to study problems such as flapping wings \citep{mittal1995parallel,takizawa2012space}, fluid-particle interaction (FPI) \citep{mittal1994massively,johnson1996simulation,johnson1997parallel,johnson19973d,johnson1999advanced,johnson2001methods}, patient-specific arterial modeling of cerebral aneurysms \citep{torii2004influence,torii2006computer,torii2006fluid,torii2007influence,torii2007numerical,torii2008fluid,torii2009fluid,torii2010influence,torii2010role,torii2011influencing}, parachute modeling \citep{kalro2000parallel,stein2000parachute,stein2001fluid} wind-turbine rotor aerodynamics \citep{takizawa2011stabilized,takizawa2011numerical}, moving hyperelastic particles \citep{gao2009deformation} and modeling flow in the heart \citep{watanabe2004multiphysics}. However, these methods found success mostly in the stiff limit of the solids \citep{hu1996direct,johnson19973d} and was found to be too cumbersome for highly deforming solids, since it requires generating a new grid at each time step. 

To overcome the cost of partitioned-based approaches for highly deforming solids, monolithic-based solvers were developed, where a single system of equations are solved simultaneously in a coupled manner \citep{hubner2004monolithic,michler2004monolithic,ryzhakov2010monolithic}. To account for the effect of presence of solids, a force term that is computed based on the structural configuration of the solid is added to the fluid equations. These methods have been applied to study problems such as modeling rigid particles \citep{yuki2007efficient}, modeling flexible bodies \citep{mori2008implicit,zhao2008fixed}, red blood cell \citep{mori2008implicit,eggleton1998large,gong2009deformation}. Further, monolithic-based solvers can be sub-divided into (i) fictitious domain (FD) method / distributed Lagrange multiplier method \citep{glowinski1999distributed,glowinski2001fictitious,patankar2001formulation}, (ii) immersed-boundary methods, where the solid region is represented as a boundary \citep{mittal2005immersed}, (iii) immersed-domain method, where the solid region is represented as a body with a finite volume, for example the immersed finite-element method \citep{liu2006immersed,liu2007mathematical,zhang2007immersed,wang2013modified} and the immersed continuum method of \citep{wang2006immersed,wang2007iterative,wang2010immersed}

Immersed-boundary methods are known to be the simplest of all the methods. For example, the classical immersed-boundary (IB) method \citep{peskin1972flow,Peskin1982,peskin2002immersed,kim2006immersed,huang2009immersed}, the direct-forcing method \citep{mohd1997simulations,luo2007modified,mark2008derivation,guy2010accuracy}, the penalization approach \citep{kim2007penalty}, the ghost-cell method, the cut-cell finite volume approach \citep{Clarke1986} and the immersed-interface method \citep{Leveque1994,li2001immersed,li2003overview,li2006immersed,layton2009using} all use an Eulerian grid for the fluid region and the boundary of the deforming solid is considered as a forcing term in the fluid equations either in the continuous form (continuous forcing methods) or in discrete form (discrete forcing methods). Owing to its simplicity, these methods have been used to study a wide variety of problems such as magnetohydrodynamics of liquid metals \citep{grigoriadis2009immersed}, complex flows in irregular domains \citep{fadlun2000combined,udaykumar1996elafint,le2008implicit,kim2001immersed,iaccarino2003numerical}, turbulent flows \citep{yang2006embedded}, modeling cochlea \citep{beyer1992computational}, modeling flexible fibers \citep{stockie1998simulating,wang2009numerical}, rigid bodies \citep{mittal2004flutter}, flow-induced vibration \citep{mittal2003cartesian}, biomimetic flight mechanism \citep{mittal2002computational}, flow past an airfoil \citep{ghias2004non},  flexible filaments \citep{zhu2003interaction}, modeling mechanics of heart \citep{peskin1982fluid,griffith2005simulating}, sperm motility near boundaries \citep{fauci1995sperm}, microswimmers \citep{dillon1995microscale} and ship hydrodynamics \citep{weymouth2006advancements,weymouth2008physics} where it's been coupled with two-fluid solvers. However, these methods are known to not capture the realistic structural response of the solid and often use a linear theory (infinitesimal strain theory) to approximate the stresses and the deformation of the solid. 

On the other hand, relatively less popular class of methods to solve FSI problems are the fully Lagrangian approach such as a particle-finite-element method (PFEM) and meshfree methods such as the reproducing kernel particle method (RKPM) and smoothed-particle hydrodynamics (SPH). These methods have been used for applications such as large number of floating bodies in fluid, bed erosion etc. \citep{onate2008advances}.

Finally, the relatively newer class of methods are the fully Eulerian approaches. These methods typically use an interface-capturing method that was initially developed to track material interfaces in two-fluid flows (see \citet{Mirjalili2017}). These approaches are inherently cost effective due to a fixed mesh and results in a easily parallelizable computer programs and is particularly very advantageous compared to other methods for highly deforming solids. Some of previous applications of these methods are modeling linear elastic materials \citep{xiao1999computation}, elasto-plastic materials \citep{udaykumar2003eulerian,okazawa2007eulerian} and neo-hookean materials \citep{liu2001eulerian,van1994eulerian,dunne2006eulerian,sugiyama2010full}, modeling flow in complex domains \citep{nagano2010full}. One of the earliest known Eulerian approach that can solve solid regions using ``\textit{true nonlinear solid constitutive laws}" coupled with fluid flow is the Eulerian Godunov method of \cite{Miller2001}). This has been applied to study mostly elastic-plastic solids \citep{barton2010eulerian,ghaisas2018unified}. However, the main disadvantage of this method is that the method is limited to unbounded domains. 

Other well-known fully Eulerian approaches are the level-set method \citep{cottet2008eulerian,cottet2016semi} and the volume-of-fluid (VOF) method \citep{sugiyama2011full,ii2011implicit,ii2012full}. These methods have gained popularity recently and has been used to study problems such as modeling phospholipidic vesicles and cardiomyocyte membrane \citep{maitre2009applications}, modeling large number of red blood cells (RBCs) and platelets in a capillary vessel \citep{sugiyama2017full} and turbulent channel flow over hyperelastic walls \citep{rosti2017numerical}. A recent work by \cite{Kamrin2012} introduced the ``reference map technique" (RMT), a fully Eulerian approach for the simulation of solids and an extension to coupled fluid-solid problems \citep{Valkov2015}. In this work, visco-elastic solids were successfully simulated on a staggered grid coupled with a Newtonian fluid in a compressible flow setting using hyperelastic constitutive laws. The main differences of the RMT and VOF methods are (i) a reference-map vector field is transported to track the deformation of the solid in RMT approach as opposed to a tensor field (left Cauchy-Green deformation tensor) in the VOF method. (ii) RMT method has been extended to account for solid-solid contact conditions. We therefore adopt the RMT approach and extend this formulation for incompressible settings \citep{Jain2017,rycroft2018reference} and assess multiple improvements to the original RMT method \citep{Valkov2015}. Other approaches in the literature similar to RMT that is worth mentioning are \citep{dunne2006eulerian,govindjee1996computational}. Further, we point the readers towards excellent review articles by \citet{hou2012numerical,takizawa2012space,takagi2012review,mittal2005immersed} for additional details on the methods and many more applications of FSI problems which could not be included here for the sake of brevity.


In the present paper, we describe a conservative and non-dissipative reference-map technique (RMT) for the simulation of incompressible soft solids in fluids. We discuss the improvements made for this model in terms of the accuracy, cost, ease of implementation, and robustness of the method and also discuss some of the best modeling practices. Some of the important features of our approach compared to the state-of-the-art RMT \citep{Valkov2015} are (a) discrete momentum conservation, (b) a least-squares extrapolation procedure that is accurate and cost-effective, (c) a modified advection equation for the reference map field that improves robustness of the method, (d) a non-dissipative central-difference scheme that eliminates any spurious dissipation of kinetic energy, and (d) projection method for incompressible flows. Rest of the paper is organized into sections as follows: Section \ref{sec:rmt} describes the basic formulation of the reference-map technique, governing equations that describe the motion of fluids and solids, and their respective constitutive laws. Section \ref{sec:numerics} describes the numerical method and introduces the conservative formulation, discretizations, projection method algorithm, a strategy to reconstruct level-set field, modifications to the reference-map advection equation, a new least-squares based extrapolation procedure and a closure model. Section \ref{sec:results} presents the verification of the solver against the results from a Lagrangian approach, presents the cost and accuracy improvements of the new extrapolation procedure, illustrates the importance of the use of a conservative formulation, and presents more complex test cases involving solid-solid and solid-wall contact situations. Finally, section \ref{sec:summary} presents the summary along with the concluding remarks.


\section{Eulerian formulation for solids and fluids}\label{sec:rmt}

\subsection{Reference map technique}
\begin{figure}
\vskip 0.1in
\centering
\includegraphics[width=0.6\textwidth]{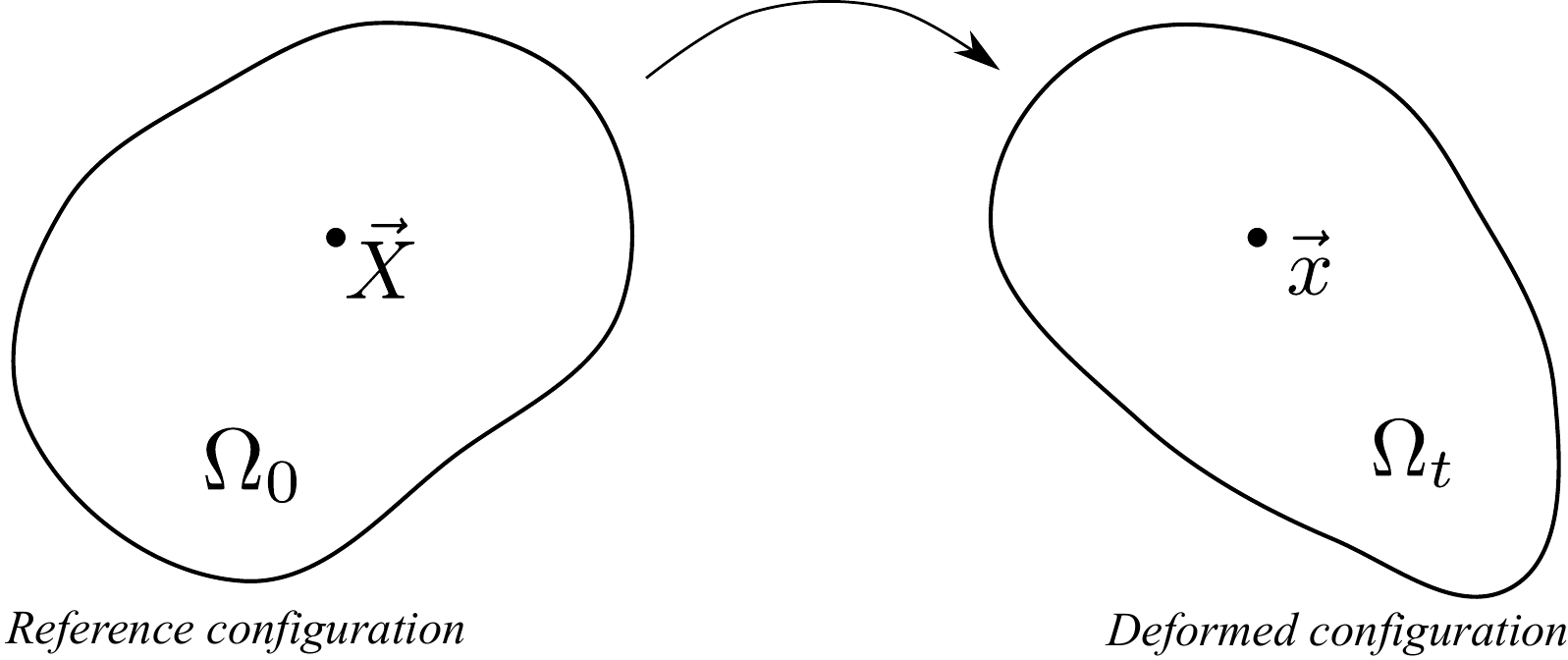}
\caption{Schematic of a deforming solid in convective coordinate system. $\Omega_0$ represents the solid in reference configuration and $\vec{X}$ the position vector of a material particle in $\Omega_0$. $\Omega_t$ represents the solid in deformed configuration at time $t$ and $\vec{x}$ the corresponding position vector of the same material particle in $\Omega_t$.}
\label{fig:convect}
\end{figure}

Consider a solid in convective coordinate system, as shown in Figure \ref{fig:convect}. At time $t=0$, the solid is in its initial configuration  (reference configuration), represented by $\Omega_0$, and at time $t>0$, the solid is in its deformed configuration, represented by $\Omega_t$ after being displaced and deformed by external forces. If, $\vec{X}$ represents a position vector in $\Omega_0$ that points to a  material particle, then this material particle in $\Omega_t$ has the same $\vec{X}$ associated with it, since $\vec{X}$ represents the initial coordinates of the point in $\Omega_0$. Hence $\vec{X}$ acts as a tag for all the material particles in the solid. If $\vec{x}$ represents the corresponding position vector in $\Omega_t$, then we can define a vector map $\vec{\xi}:\mathbb{R}^4\rightarrow\mathbb{R}^3$ (a reference map) as
\begin{equation}
\vec{\xi}(\vec{x},t)=\vec{X},
\label{equ:reference_map}
\end{equation}
such that $\vec{\xi}$ remains constant for a material particle in the solid (as long as the solid doesn't deform plastically) but varies from particle to particle. Hence the material derivative of $\vec{\xi}$ field yields
\begin{equation}
\frac{D\vec{\xi}(\vec{x},t)}{Dt} = 0.
\label{equ:advect}
\end{equation}
Expressing this in terms of the local derivatives, we obtain an advection equation for the $\vec{\xi}(\vec{x},t)$ field as
\begin{equation}
\frac{\partial\vec{\xi}(\vec{x},t)}{\partial t}+\vec{u}.\vec{\nabla}\vec{\xi}(\vec{x},t)=0.
\label{equ:advect_refer}
\end{equation}
This equation can be integrated in time given the initial condition $\vec{\xi}(\vec{x},t=0)=\vec{x}=\vec{X}$.  Thus, $\vec{\xi}(\vec{x},t)$ acts as a tag for all the points in the solid, and the kinematic condition in Eq. (\ref{equ:advect_refer}) can be used to track every point in the solid, given its initial coordinates. Stress and strain in solid constitutive laws are typically expressed in terms of the material deformation gradient $\mathbb{F}$. Hence, relating $\mathbb{F}$ to $\vec{\xi}(\vec{x},t)$ as 
\begin{equation}
\mathbb{F}(\vec{X},t)=\partial\vec{x}/\partial\vec{X}=[\overrightarrow{\nabla}\overrightarrow{\xi}(\vec{x},t)]^{-1},
\label{equ:def_grad}
\end{equation}
we can express the stress and strain tensors in terms of this new primitive variable $\vec{\xi}(\vec{x},t)$. Eqs. (\ref{equ:reference_map})-(\ref{equ:def_grad}) in combination give rise to a novel approach to track all the material points in a solid and close the system of equations to model a solid on an Eulerian grid.

\subsection{Governing equations for solids and fluids}\label{sec:govern_eq}


In an Eulerian formulation, momentum balance equation for both fluids and solids can be written as 
\begin{equation}
\frac{{\partial(\rho\vec{u})}}{\partial t}+\vec{\nabla}.({\rho\vec{u}\otimes\vec{u}})=\vec{\nabla}.\doubleunderline{\sigma},
\label{equ:mom_equ}
\end{equation}
where $\vec{u}$ is the global velocity field and $\doubleunderline{\sigma}$ is the Cauchy stress. Mass balance equation for fluids (continuity equation) can be written as 
\begin{equation}
\frac{\partial\rho}{\partial t}+\vec{\nabla}.(\vec{u}\rho)=0,
\label{equ:continuity}
\end{equation}
which in the incompressible limit simplifies to $\vec{\nabla}.\vec{u}=0$. Similarly, the mass balance for solids can be written as $\rho=\rho_{0}[det(\mathbb{F})]^{-1}$, and in the incompressible limit it simplifies to $det(\mathbb{F})=1$, implying that the density doesn't change ($\rho = \rho_0$). Here, $\rho$ and $\rho_{o}$ are the density of the deformed and reference configurations, respectively. It can be shown that the conditions $\vec{\nabla}.\vec{u}=0$ and $det(\mathbb{F})=1$ are equivalent (see, Appendix A).


For solids, the Cauchy stress can be expressed as a function of strain given by 
\begin{equation}
\doubleunderline{\sigma}^s=(det\mathbb{F})^{-1}\mathbb{F}\frac{\partial\bar{\psi}(\mathbb{E})}{\partial\mathbb{E}}\mathbb{F}^{T} - \lambda \mathds{1} =2(det\mathbb{F})^{-1}\mathbb{F}\frac{\partial\hat{\psi}(\mathbb{C})}{\partial\mathbb{C}}\mathbb{F}^{T} - \lambda \mathds{1},
\label{equ:cauchy_stress}
\end{equation}
where $\mathbb{E}=(1/2)(\mathbb{F}^{T}\mathbb{F}-\mathds{1})$ is the \textit{Green's (or Lagrangian) finite strain tensor}, $\mathbb{C}=\mathbb{F}^{T}\mathbb{F}$ is the \textit{right Cauchy-Green's deformation tensor} (or stretch tensor), $\psi(\mathbb{F})=\bar{\psi}(\mathbb{E})=\hat{\psi}(\mathbb{C})$ is the strain energy density (\textit{Helmhotz free-energy density}) function and $\lambda=P$ is the Lagrangian multiplier and is equal to pressure in the incompressible limit \citep{holzapfelnonlinear}. We use the incompressible neo-Hookean constitutive model for solids, given by 
\begin{equation}
\hat{\psi}(\mathbb{C})=\mu^s(tr\mathbb{C}-3),
\label{equ:incom_neo_hook}
\end{equation}
where $\mu^s={E}/{2(1+\nu)}$ is the shear modulus (\textit{Lame's first parameter}), $E$ is the \textit{Young's modulus} and $\nu$ is the \textit{Poisson's ratio}. Taking a partial derivative of this strain energy density function with respect to $\mathbb{C}$ yields
\begin{equation}
\frac{\partial\hat{\psi}(\mathbb{C})}{\partial\mathbb{C}}=\mu^s \mathds{1}.
\label{equ:incom_neo_hook}
\end{equation}
Using this and the incompressibility condition for solids ($det(\mathbb{F})=1$), the $\doubleunderline{\sigma}^s$ reduces to a simple form given by $\doubleunderline{\sigma}^s=2\mu^s\mathbbm{b} - P \mathds{1}$, where $\mathbbm{b} = \mathbb{F}\mathbb{F}^T$ is the \textit{left Cauchy-Green's deformation tensor} (or stretch tensor). Further more, expressing $\mathbb{F}$ in terms of $\vec{\xi}$, Cauchy stress can be expressed in terms of this new primitive variable $\vec{\xi}$ as
\begin{equation}
\doubleunderline{\sigma}^s=2 \mu^s [(\vec{\nabla}\vec{\xi})^{-1} (\vec{\nabla}\vec{\xi})^{-T}] - P\mathds{1} = 2\mu^s [(\vec{\nabla}\vec{\xi})^T (\vec{\nabla}\vec{\xi})]^{-1} - P \mathds{1}.
\label{equ:solid_cauchy1}
\end{equation}

Nonlinearity in the stress-strain relationship is more evident when $\vec{\xi}$ is expressed in terms of its components. For an incompressible solid in two-dimensions, the Cauchy stress reduces to the form (Appendix B)

\begin{equation}
\doubleunderline{\sigma}^s= 2\mu^s\begin{bmatrix}
   (\frac{\partial \alpha}{\partial y}) ^2 + (\frac{\partial \beta}{\partial y})^2       & -\Big[ (\frac{\partial \alpha}{\partial x}) (\frac{\partial \alpha}{\partial y}) + (\frac{\partial \beta}{\partial x}) (\frac{\partial \beta}{\partial y})\Big]  \\
    -\Big[ (\frac{\partial \alpha}{\partial x}) (\frac{\partial \alpha}{\partial y}) + (\frac{\partial \beta}{\partial x}) (\frac{\partial \beta}{\partial y})\Big]      &  (\frac{\partial \alpha}{\partial x}) ^2 + (\frac{\partial \beta}{\partial x})^2 
\end{bmatrix} - P \mathds{1},
\label{equ:solid_cauchy2}
\end{equation}
where $\alpha=\vec{\xi}.\hat{i}$ and $\beta=\vec{\xi}.\hat{j}$ are the components of $\vec{\xi}$. For fluids, the Cauchy stress can be expressed as a function of the rate of strain. We use the Newtonian constitutive model given by
\begin{equation}
\doubleunderline\sigma^f=\mu^f\left[\left(\vec{\nabla}\vec{u}\right)+\left(\vec{\nabla}\vec{u}\right)^{T}\right]-P\mathds{1} =\mu  \begin{bmatrix}
    2\frac{\partial u} {\partial x}       & \frac{\partial u} {\partial y} + \frac{\partial v} {\partial x}  \\ \\
    \frac{\partial v} {\partial x} + \frac{\partial u} {\partial y}      &  2\frac{\partial v} {\partial y}  
\end{bmatrix} - P\mathds{1},
\label{equ:fluid_cauchy}
\end{equation}
where $P$ is the pressure and the matrix form of the above system of equations is for a fluid in two-dimensions. Extension to three-dimensions is not included here, but is straightforward. We solve a conservative variable-density formulation of the above system of equations and to close the system of equations for fluid-solid coupled simulations, we use a mixture model derived based on the one-fluid formulation \citep{kataoka1986local} for two-phase flows (see, Section \ref{sec:closure_model}), or the so called ``\textit{one-continuum formulation}" \citep{sugiyama2011full}.

\section{Discretization, numerical method and conservative implementation}\label{sec:numerics}

\subsection{Basic methodology}\label{sec:basic}

\begin{figure}
\centering
\includegraphics[width=0.5\textwidth]{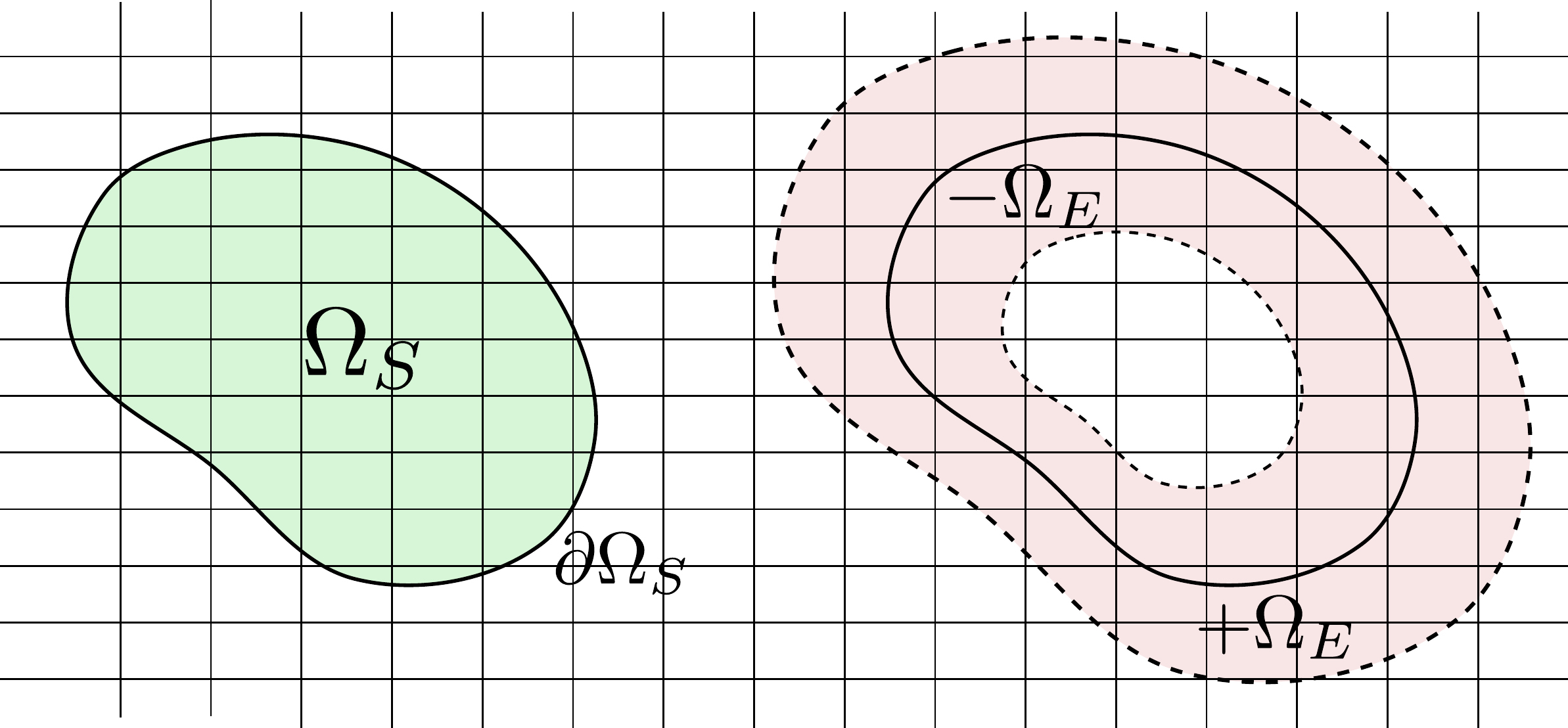}
\caption{Schematic of a solid on an Eulerian grid. $\Omega_S$ represents the solid region, $\partial \Omega_S$ the boundary of the solid. $\Omega_E\approx6\Delta x$ is the narrow band of extended solid region around $\partial \Omega_S$. }
\label{fig:solid}
\end{figure}
\begin{figure}
\centering
\includegraphics[width=0.3\textwidth]{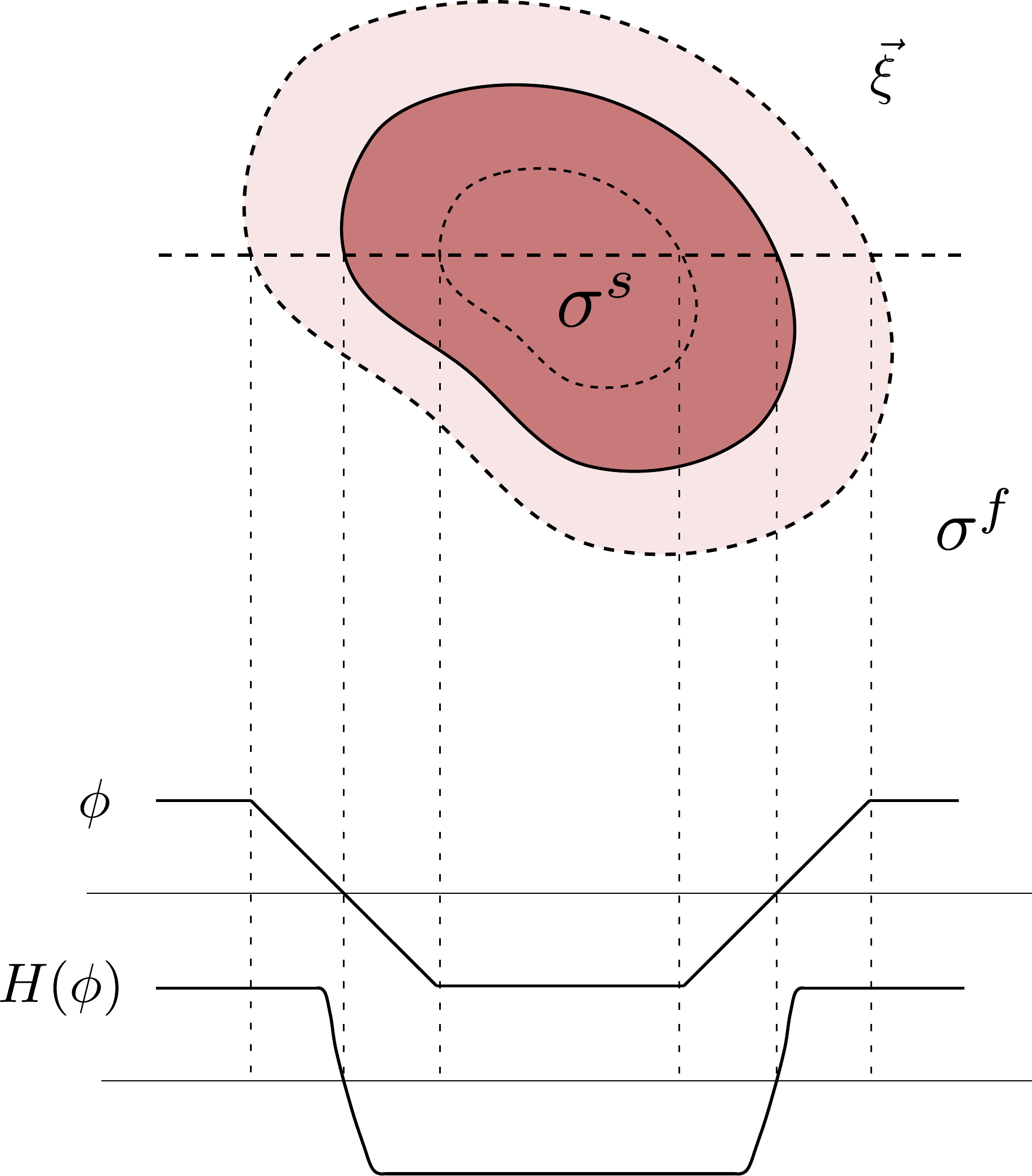}
\caption{Schematic showing the procedure to the compute global Cauchy stress $\doubleunderline\sigma$. Darker shaded region enclosed by a solid line is the solid region $\Omega_S$ defined by the reference map field $\vec{\xi}$. Solid Cauchy stress $\doubleunderline\sigma$ is evaluated in this region. Fluid Cauchy stress $\doubleunderline\sigma$ is evaluated outside the solid region. $\phi$ and $H(\phi)$ represents the constructed level-set and smoothed Heaviside field (Eq. \ref{eq:heavy}) using the $\xi$ field.}
\label{fig:procedure}
\end{figure}

Consider a solid on an Eulerian grid, as shown in Figure \ref{fig:solid}. Here, $\Omega_S$ represents the region inside the solid, $\partial \Omega_S$ represents the boundary of the solid, $\pm \Omega_E\approx4\Delta x$ represents a narrow band region (an extended solid region) around $\partial \Omega_S$, and $\pm \Omega_T\approx2\Delta x$ represents another narrow band region (a transition zone) around $\partial \Omega_S$ (not shown in the Figure \ref{fig:solid}). Since, both solid and fluid regions are solved together in a coupled fashion, they share the same grid and a global velocity field. In the regions of solid $\Omega_S$, solid Cauchy stress $\doubleunderline{\sigma}^s$ is computed using the solid constitutive law (Eq. \ref{equ:solid_cauchy2}), and outside this region, fluid Cauchy stress $\doubleunderline\sigma^f$ is computed using the fluid constitutive law (Eq. \ref{equ:fluid_cauchy}). Once the stresses for solid and fluid regions are evaluated, a level-set field $\phi$ and a smoothed Heaviside function $H(x)$ is constructed using the reference map field $\vec{\xi}$ as illustrated in Figure \ref{fig:procedure}. This Heaviside function is used to appropriately blend the solid and fluid stresses around the solid-fluid interface to compute the global Cauchy stress $\doubleunderline\sigma$. Finally the velocity field is updated by solving the discretized version of momentum equation and by projecting the velocity field onto a divergence-free field. Form of the equations used, discretization techniques and algorithms used in this approach are explained in detail in the subsequent sections. 

An important thing to note is that $\vec{\xi}$ is a variable that contains the information of the origin of the material. This field quickly becomes invalid in the region containing fluids due to the highly nonlinear deformation behavior of the fluids. Hence, $\vec{\xi}$ is defined only within the solid region, and to evaluate the solid stress in $\Omega_T$, the $\vec{\xi}$  field is appropriately extrapolated into the regions outside the solid (into $\Omega_E$).

\cite{Valkov2015}, used the hyperbolic partial-differential equation (PDE) approach of \cite{Aslam2004} to extrapolate the $\vec{\xi}$ field. This approach assumes that a level-set field is known in the region of extrapolation. By contrast, we use a least-squares-based extrapolation procedure (see, Section \ref{sec:extrapolation}) that does not require a known level-set field in the region of extrapolation. However, a local level-set field $\phi$ has to be defined in the $\Omega_E$ region, which is used in defining the Heaviside function required for the mixture model (see, Section \ref{sec:closure_model}) and also in enforcing the solid-solid and solid-wall contact boundaries. Hereafter, we refer to the approach by \cite{Valkov2015} as the original RMT. 

\subsection{Conservative formulation and discretization}\label{sec:conservative}

In the numerical solution of partial differential equations, divergence form of the equations is usually preferred over the primitive form (non-conservative form), since it results in discrete conservation of the quantities being solved. We solve the momentum equation in a conservative form as written in Eq. \ref{equ:mom_equ}, where both the inertial term and the stress term are in divergence form (see, Section \ref{sec:conserve} for the illustration of importance of the use of divergence form for the stress term). We also use the conservative form of the equation for the advection of $\vec{\xi}$ field. Though $\vec{\xi}$ is not a physically conservative field, volume enclosed in the solid region $\Omega_S$ bounded by the fluid-solid interface $\partial \Omega_S$ that is extracted using the $\vec{\xi}$ field should be conserved (see, Section \ref{sec:vol-error} for a description on the volumetric error of the solid). Notice that the Eq. \ref{equ:advect_refer} can be rewritten in a conservative form as 

\begin{equation}
\frac{\partial\vec{\xi}(\vec{x},t)}{\partial t}+\vec{\nabla}\cdot[\vec{u}\ \vec{\xi}(\vec{x},t)]=0,
\label{equ:conserv_advect_refer}
\end{equation}
in the incompressible limit, using the divergence-free condition ($\vec{\nabla}\cdot \vec{u}=0$). 

We use a finite-volume approach on a collocated uniform grid to discretize our system of equations. Hence, all our primary variables ($\vec{\xi}$,$\vec{u}$,$p$,$\rho$) are stored on the cell center. We modify the approximate projection method of \cite{Almgren2000} to incorporate the coupled solution of solid and fluid regions. The steps involved in our projection method are shown in detail in Algorithm \ref{alg:projection}.

\begin{algorithm}[t!]
\footnotesize
\caption{One full time-step iteration with the modified projection method}
\label{alg:projection}
\begin{algorithmic}[1]
\State Advect reference map $\vec{\xi}$ and extrapolate using least-squares method (see Sections \ref{sec:refer_advection},\ref{sec:extrapolation}).
\State Reconstruct level-set field $\phi$ and reinitialize using fast-marching method (see Section \ref{sec:level-set}).
\State Compute solid stress $\doubleunderline\sigma^s$ using Eq. (\ref{equ:solid_cauchy2}), fluid stress $\doubleunderline\sigma^f$ using Eq. (\ref{equ:fluid_cauchy}) and update $\rho$ and $\doubleunderline{\sigma}$ (see Section \ref{sec:closure_model}).

%

\State Solve advection and diffusion to obtain intermediate velocity $\vec{u}_P^{**}$
\begin{equation}
\frac{\rho^{n+1}\vec{u}_{P}^{*}-\rho^{n}\vec{u}_{P}^{n}}{\Delta t}=-\vec{\nabla}.\left(\rho\vec{u}_P\vec{u}_P\right)^{n},
\label{equ:advect}
\end{equation}
\begin{equation}
\frac{\rho^{n+1}\vec{u}_P^{**}-\rho^{n+1}\vec{u}_P^{*}}{\Delta t}=\vec{\nabla}.\doubleunderline{\sigma}(\mu^s,\mu^f,\vec{u}_P^*,\vec{\xi}).
\label{equ:diffuse}
\end{equation}
where subscript $P$ represents cell-centered values, $n$ and $n+1$ represents two consecutive time steps. Here, an Euler time-stepping scheme for advection step is shown for representation only, however an RK4 time-stepping is used in the implementation to achieve numerical stability. 

\State Interpolate to obtain face values (Rhie-Chow-like interpolation)
\begin{equation}
\vec{u}_{f}^{**}=\left\langle \vec{u}_{P}^{**}\right\rangle _{P\rightarrow f}-\left\lbrace\frac{\Delta t}{\rho_{f}^{n+1}}\left[ \left(\vec{\nabla}P\right)_{f}^{n}-F_{f}^{n+1}\right] \right\rbrace,
\label{equ:rhie_chow}
\end{equation}
where $\left\langle \right\rangle _{P\rightarrow f}$ is an interpolation from the cell center to the cell face, subscript $f$ represents face-centered values and $F$ is the body force computed using the balanced-force approach of \cite{Francois2006}).

\State Solve pressure Poisson equation
\begin{equation}
\vec{\nabla}.\left[\frac{\left(\vec{\nabla\delta}P\right)_{f}^{n+1}}{\rho_{f}^{n+1}}\right] = -\vec{\nabla}.\left(\frac{\vec{u}_{f}^{**}}{\Delta t}\right),
\label{equ:poisson}
\end{equation}
where $\delta P$ is the correction for pressure. 
\State Update the pressure
\begin{equation}
P^{n+1}=P^{n}+\delta P^{n+1}.
\end{equation}
\State Update the face velocity field \textemdash exactly divergence free (to be used in
calculating convective fluxes in the next time step)
\begin{equation}
\vec{u}_{f}^{n+1}=\vec{u}_{f}^{**}-\left[\frac{\Delta t}{\rho_{f}^{n+1}}\left(\vec{\nabla\delta}P\right)_{f}^{n+1}\right].
\end{equation}
\State Update the cell center velocity field \textemdash approximately divergence free
\begin{equation}
\vec{u}_{P}^{n+1}=\vec{u}_{P}^{**}-\Delta t\left\langle \frac{(\vec{\nabla}P)_{f}-F_{f}}{\rho_{f}}\right\rangle _{f\rightarrow P}^{n+1},
\end{equation}
where $\left\langle \right\rangle _{f\rightarrow P}$ is an interpolation from the cell face to the cell center.
\end{algorithmic}
\end{algorithm}

We split the momentum equation into an advection and diffusion part, since it allows us to use different time-stepping schemes. A second-order central differencing scheme is used to compute convective fluxes in the advection part of the momentum equation and the advection equation for $\vec{\xi}$, and they are solved using an RK4 time integration scheme. The use of central-difference scheme for the advection of both $\vec{\xi}$ and $\vec{u}$ fields not only yields a conservative and a non-dissipative approach but also results in solving momentum equation and reference map advection equations consistently, which is crucial for the simulation of high solid-to-fluid density-ratio flows (see Section 3.10 in \citet{tryggvason2011direct}). A forward-Euler time integration scheme is used to solve the diffusion part of the momentum equation (Eq. \ref{equ:diffuse}). We use the second-order central-difference approximation to evaluate the gradient tensor ($\vec{\nabla}\vec{\xi}$) in Eq. (\ref{equ:solid_cauchy1}), unlike the one-sided differences used in the original RMT. For example, in a Cartesian two-dimensional case, $\partial \alpha / \partial x$ in Eq. (\ref{equ:solid_cauchy2}) is approximated as 
\begin{equation}
\frac{\partial \alpha}{\partial x} = \frac{\alpha_{i+1,j} - \alpha_{i-1,j}}{2 \Delta x}, 
\label{}
\end{equation}
where $\alpha=\vec{\xi}.\hat{i}$. The divergence of Cauchy stress in Eq. (\ref{equ:diffuse}) is also computed using the second-order central-difference scheme. The exact form of discretization of the stress terms is crucial in obtaining a consistent and conservative formulation, hence the discretization used in the current work is presented in detail in Appendix C.

Finally, the use of collocated grid arrangement results in checkerboard pressure fields. To eliminate this, we use a ``Rhie-Chow like interpolation" for the intermediate velocity fields ($\vec{u}^{**}$) after the update of advection and diffusion

\begin{equation}
\vec{u}_{f}^{**}=\left\langle \vec{u}_{P}^{**}\right\rangle _{P\rightarrow f}-\left\lbrace\frac{\Delta t}{\rho_{f}^{n+1}}\left[ \left(\vec{\nabla}P\right)_{f}^{n}-F_{f}^{n+1}\right] \right\rbrace,
\label{equ:rhie_chow}
\end{equation}
where subscript $P$ represents cell-centered values, $f$ represents face-centered values, $\left\langle \right\rangle _{P\rightarrow f}$ is an interpolation from cell center to the cell face and $F$ is the body force computed using the balanced-force approach of \cite{Francois2006}). Interpolation from cell center to the cell face ($\left\langle \right\rangle _{P\rightarrow f}$) for a uniform Cartesian grid can be written as
\begin{equation}
  u_{i+1/2,j} = \frac{u_{i,j} + u_{i+1,j}}{2}  
\end{equation}
\begin{equation}
  v_{i,j+1/2} = \frac{v_{i,j} + v_{i,j+1}}{2}  
\end{equation}
where $u$ and $v$ are the $x$ and $y$ components of the velocity field. This Rhie-Chow like interpolation does not affect the discrete conservation of momentum. However, it does add a small amount of conservation error in the transport of kinetic energy which is of the order $O(\Delta t \Delta x^2)$. This has been previously shown to be of dissipative in nature, hence it does not affect the stability of the method \citep[see, Section 6.1 of ][]{morinishi1998fully}.

\subsection{Level-set reconstruction}\label{sec:level-set}

As explained in the section \ref{sec:basic}, a level-set field $\phi$ is required to be defined at every time step in the $\Omega_E$ region. One way to define $\phi(\vec{x},t)$ is to advect $\phi$ using the standard level-set advection equations given $\phi(\vec{X},t=0)$. This approach could lead to a mismatch between the $\phi(\vec{x},t)=0$ and the boundary of the solid defined by $\vec{\xi}(\vec{x},t)$ field, which in turn could result in a wrinkled solid-fluid interface, affecting the overall quality of $\vec{\xi}$ field in the extrapolated region (see Figures 9,10 in \citet{Valkov2015}). In the original RMT approach, this issue was resolved by performing additional smoothing routines to eliminate the striations in the extrapolated regions, which could potentially lead to additional mass conservation issues. To avoid this problem, we propose a simpler, exact, conservative and also cost-effective way to define the level-set field $\phi(\vec{x},t)$ at any time $t$. $\phi(\vec{x},t)$ can be reconstructed from the given $\phi(\vec{X},0)$ field at $t=0$, utilizing the known $\vec{\xi}(\vec{x},t)$ field at time $t$ using a simple condition given by 
\begin{equation}
\phi (\vec{x},t)=\phi[\vec{\xi}(\vec{x},t),t=0].
\label{equ:compatibility}
\end{equation}
Since an analytical expression can be defined for $\phi(\vec{X},t=0)$ for simple-shaped solids, the above equation yields an exact field for $\phi(\vec{x},t)$ for a given $\vec{\xi}(\vec{x},t)$, thus maintaining a perfect match between the $\phi(\vec{x},t)=0$ surface and the boundary of the solid defined by $\vec{\xi}(\vec{x},t)$, which is crucial in developing a robust solver. If an analytical expression for  $\phi(\vec{X},t=0)$ is not available, then a bilinear interpolation (in two-dimensions) can be used to calculate $\phi(\vec{\xi}(\vec{x},t),t=0)$. 

\subsection{Modified reference map advection}\label{sec:refer_advection}

\begin{figure}
\centering
\vskip 0.1in
\includegraphics[width=\textwidth]{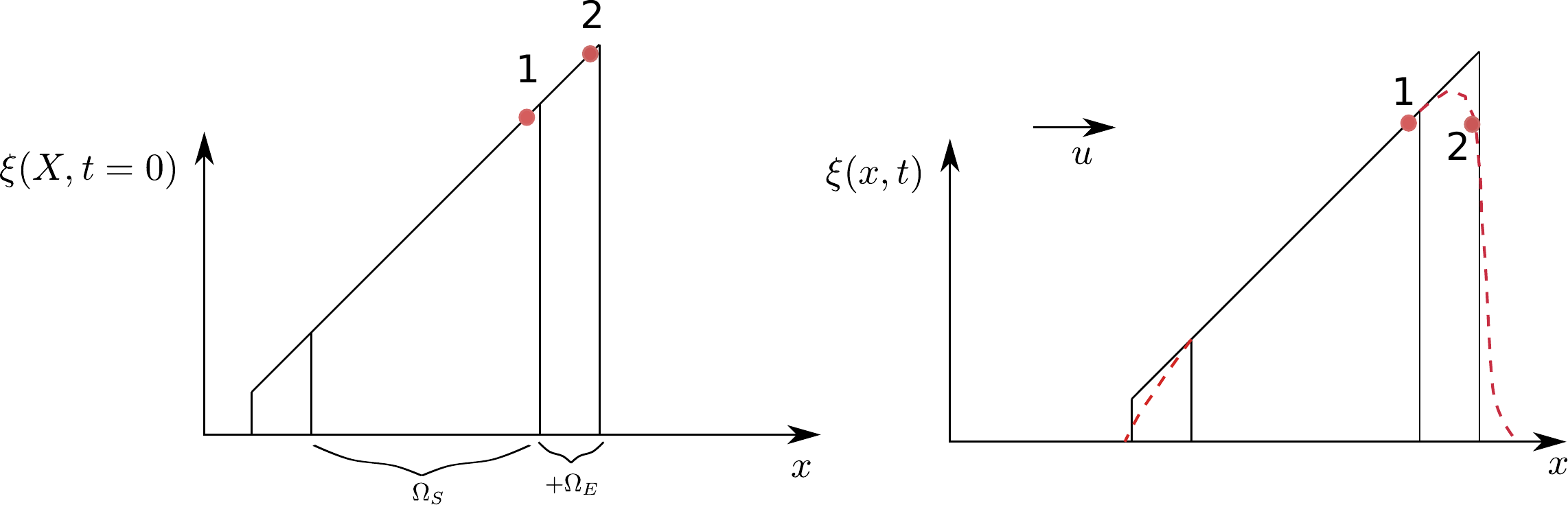}
\caption{\label{fig:mod_refer} $\vec{\xi}$ field of a one-dimensional solid at an initial time $t=0$ and at later time $t$ after being advected with a velocity $u$, illustrating the failure of the reference map technique solved without the use of modified advection equation. Circle 1 represents a location inside the solid and 2 outside the solid.}
\end{figure}

The reference map field $\vec{\xi}$ is advected using Eq. (\ref{equ:conserv_advect_refer}). As explained in Section \ref{sec:basic}, $\vec{\xi}$ is defined and advected only within the $\Omega_S$. This can be conveniently achieved by modifying Eq. (\ref{equ:conserv_advect_refer}) into 
\begin{equation}
\frac{\partial\vec{\xi}(\vec{x},t)}{\partial t}+H(\vec{x})\vec{\nabla}\cdot[\vec{u}\ \vec{\xi}(\vec{x},t)]=0,
\label{equ:mod_refer}
\end{equation}
where $H(\vec{x})$ is a Heaviside function defined as
\begin{equation}
H(\vec{x})= \Bigg\{ 
    \begin{aligned}
        1 & \hspace{10mm} \Omega_S\\
        0 & \hspace{10mm} else.
    \end{aligned}
 \label{}   
\end{equation}
This modification to the advection equation of $\vec{\xi}$ has multiple advantages; the very obvious one is that this approach effectively eliminates the high-frequency content in the $\vec{\xi}$ field, resulting in an ability to use simple schemes such as central-differences to compute the fluxes, without losing the accuracy of the solution to dispersion errors. To realize the second advantage, which is more subtle, consider the $\vec{\xi}$ field of a one-dimensional solid, as shown in Figure \ref{fig:mod_refer}.

At time $t=0$, $\vec{\xi}$ is a simple straight line ($\xi=x$) and is given as an input to the solver, as shown on the left. Let $u$ denote the velocity field; then after time $t$, an ideal solid would have advected to a new location, shown on the right, maintaining the shape (solid line). If the modified equation shown in Eq. (\ref{equ:mod_refer}) is not used to advect, then a Total Variation Diminishing (TVD) type scheme can be used to compute the fluxes, which artificially add diffusivity to the equation to stabilize the solver. This results in a non-monotonic $\vec{\xi}$ profile shown on the right with dotted lines. If this profile is obtained as a result of advection, then the reconstruction step of the level-set field using Eq. (\ref{equ:compatibility}) breaks down. To understand this, consider two solid circles 1 and 2 in Figure \ref{fig:mod_refer}. Circle 1 is inside the solid in $\Omega_S$ at $t=0$, but circle 2 is outside the solid in $\Omega_E$. After the advection step, circle 1 still represents a value of $\xi$ inside the solid region, whereas circle 2 now represents a value of $\xi$ inside solid region $\Omega_S$ due to numerical diffusion. To clip the values of $\xi$ outside the solid region before extrapolating $\xi$, the boundaries of the solid needs to be identified. This can be done using the level-set field constructed using Eq. (\ref{equ:compatibility}) (which takes $\xi$  as the input). This procedure (without the modification to the advection equation Eq. \ref{equ:mod_refer}) typically creates two boundaries for the solid, resulting in the failure of the method. Therefore using the modified advection equation for $\vec{\xi}$ effectively eliminates this issue by clipping the values of $\vec{\xi}$ outside the solid right in the advection step. As a result, $\vec{\xi}$ can be extrapolated without any need for explicit clipping. 

\subsection{Least-squares-based extrapolation}\label{sec:extrapolation}
\begin{figure}
\centering
\includegraphics[width=0.8\textwidth]{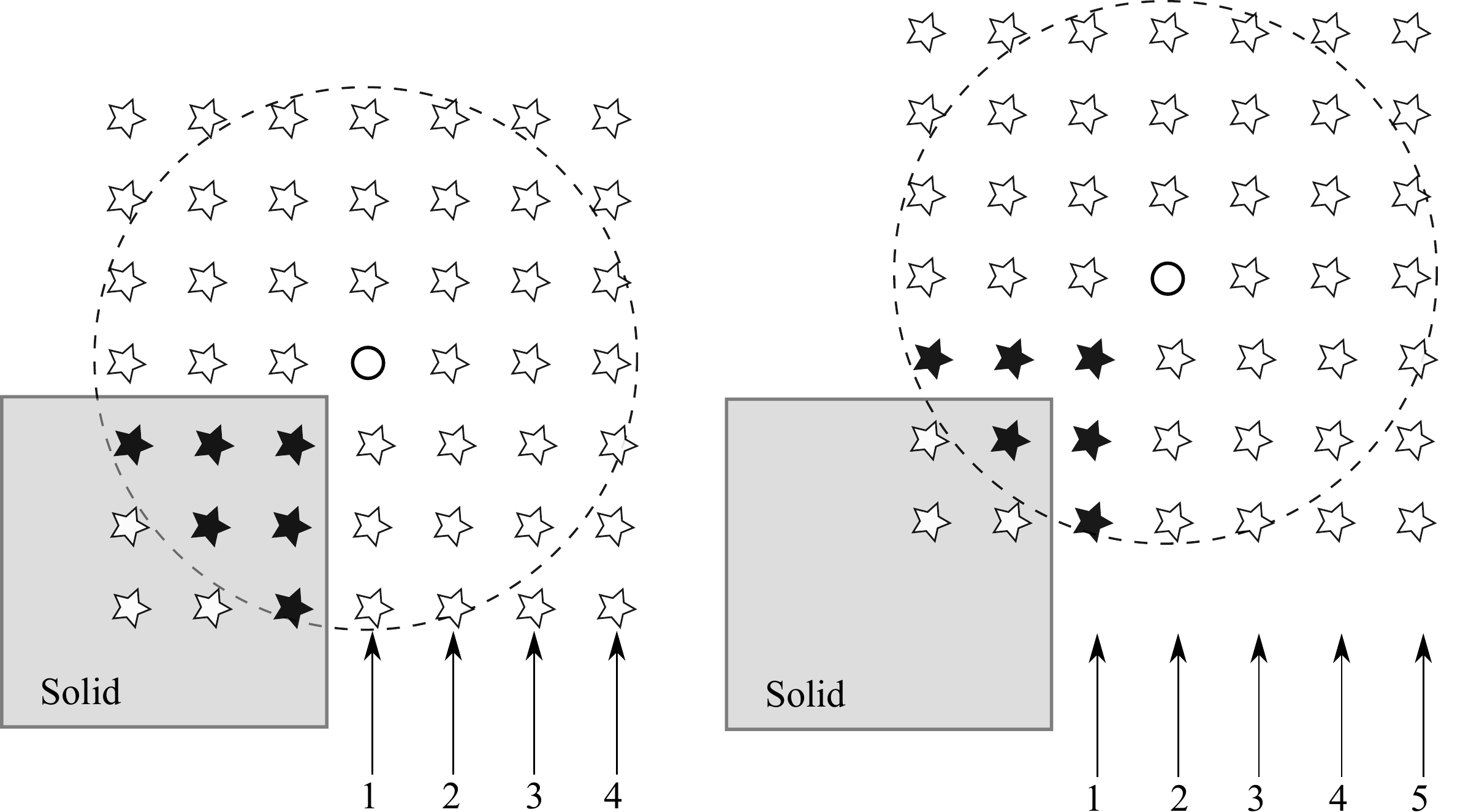}
\caption{Schematic of the cell-traversal procedure for least-squares extrapolation. Numbers represent the passes. The left figure shows the state of the system before the first pass, and the right figure is the state after the first pass. The stars represent cell-center locations, filled stars represent the cells where the values are already known and the solid circle represents the cell where the extrapolated value is being computed.}
\label{fig:least_square}
\end{figure}

The original RMT used a hyperbolic PDE approach to extrapolate the $\vec{\xi}$ field outside the solid regions into $\Omega_E$. We propose a simpler, more cost-effective approach to extrapolate the $\vec{\xi}$ field based on the assumption that the $\vec{\xi}$ field is locally linear, even in the deformed state of the solid since the $\vec{\xi}$ field represents a mapping from the current coordinates $\vec{x}$ to the original coordinates $\vec{X}$ of the solid, which is linear in $x$, $y$ and $z$. This assumption relies on the fact that the solids do not continuously deform under an applied stress. Unlike solids, liquids do not resist stress and continuously deform which would violate this locally linear assumption of $\vec{\xi}$ field. Therefore the $\vec{\xi}$ field is only defined within the solid region. This assumption of locally linear $\vec{\xi}$ field was also used in the hyperbolic PDE based extrapolation of the original RMT approach \citep{Valkov2015}.

Consider the solid represented by a square (in two-dimensions) in Figure \ref{fig:least_square}. Consider the dashed circle of radius 4$r$ as the stencil, where $r=\sqrt[]{\Delta x ^2 + \Delta y ^2}$. Hence a plane of the form $\xi=ax+by+c$ can be fit for the known cell values, where $x$, $y$ and $\xi$ represent the coordinate location and the reference map value of the cells and $a$, $b$ and $c$ are the coefficients to be determined, thus forming an over-determined system that can be solved using the least-squares approach. The stencil's radius was chosen to make the system over-determined for all the possible configurations. Once the coefficients are calculated, the value of $\xi$ at the solid circle can be computed. The procedure begins by repeatedly solving least-squares systems for all the cells adjacent to the cells for which the value of $\xi$ is already known. This is considered as the first pass. The values computed in the first pass are considered as good as the values inside the solid for the second pass. This procedure is repeated until the required width of the extrapolated region is obtained. This cell traversal procedure is summarized in Algorithm  \ref{traverse}.

\begin{algorithm}[H]
\footnotesize
\caption{Traversal algorithm}\label{traverse}
\begin{algorithmic}[1]

\State For all the cells in the domain set $flag1$ such that
\[
    flag1 = \Bigg\{
    \begin{aligned}
        & 1 \hspace{10mm} \Omega_S \\
        & 0 \hspace{10mm} else    
    \end{aligned}
    \label{flag1}
\]
\State For all the cells in the domain set $flag2$ such that
\[
    flag2 = \Bigg\{
    \begin{aligned} 
    & 1 \hspace{10mm} \Omega_S + \Omega_E\\
    & 0 \hspace{10mm} else
    \end{aligned}
    \label{flag2}
\]

\State Let $temp\_flag$ = $flag1$
\For{all cells with $temp_flag$ = $0$}
\If {adjacent neighbour or corner neighbour has $flag1$ == $1$}
	\State Solve least-squares system.
	\State Update $temp\_flag$ to $1$.
\EndIf
\EndFor
\State Set $flag1$ = $temp\_flag$.
\State Repeat steps $5$ to $9$ until $temp\_flag$ $\rightarrow$ $flag2$.

\end{algorithmic}
\end{algorithm}

\subsection{Closure model}\label{sec:closure_model}

The level-set field $\phi$ reconstructed using Eq. (\ref{equ:compatibility}) should be reinitialized to restore its signed-distance property. We solve the Eikonal equation by adopting the fast marching method (FMM) of \cite{Chopp2001} to reinitialize the $\phi$ field. The coupled fluid-solid system of equations is closed by defining the mixture model inspired by the ``one-fluid formulation" as
\[
\doubleunderline{\sigma} = \hat{H}[\hat{\phi}(\vec{x},t)]  \doubleunderline{\sigma}^f + \left\lbrace1 - \hat{H}[\hat{\phi}(\vec{x},t)]\right\rbrace\doubleunderline{\sigma}^s,
\]
\[
\rho = \hat{H}[\hat{\phi}(\vec{x},t)]  \rho^f + \left\lbrace1 - \hat{H}(\hat{\phi}(\vec{x},t))\right\rbrace \rho^s,
\]
where $\hat{\phi}$ is the reinitialized level-set field and $\hat{H}(x)$ represents a smoothed Heaviside function defined as
\begin{equation}
\hat{H}(x) = \vast\{
\begin{aligned}
0 \hspace{1in} x \le -w_T\\
\frac{1}{2}\bigg[1 + \frac{x}{w_T} + \frac{1}{\pi}sin(\frac{\pi x}{w_T})\bigg] \hspace{1in} |x|<w_T\\
1 \hspace{1in} x\ge w_T,
\end{aligned}
\label{eq:heavy}
\end{equation}
where $w_T$ represents the width of the transition region $\Omega_T$. For $n$ number of solids, this model can be extended accordingly
\begin{equation}
\doubleunderline{\sigma} = \Big\{ \sum_{i=1}^n H_i[\hat{\phi}(\vec{x},t)] - n + 1 \Big\} \doubleunderline{\sigma}^f + \sum_{i=1}^n \Big\{1 - H_i[\hat{\phi}(\vec{x},t)]\Big\}\doubleunderline{\sigma}^s.
\label{}
\end{equation}

When two solids collide in a fluid, a body force needs to be added to the momentum equation to keep them separated and to avoid the inter-penetration of solids. We use a similar procedure as described in \citet{Valkov2015} to calculate the body force $\vec{f}_{i,j}$ for solid-solid contact and solid-wall contact conditions. A level-set field $\phi_{12}$ is defined as 
\begin{equation}
\phi_{12}=\frac{\phi^{(1)} - \phi^{(2)}}{2}
\label{fig:midsurface}
\end{equation}
where $\phi_1$ and $\phi_2$ are the level-set fields associated with two colliding solids, hence $\phi_{12}=0$ represents a mid-surface between the two solids. The body force $\vec{f}_{i,j}$ can then be defined as 
\begin{equation}
\vec{f_{i,j}} = \Bigg\{
\begin{aligned}
\gamma_{i,j}\hat{n}_{12i,j}\hspace{1in} \phi^{(1)} < 0\ or\ \phi^{(2)}<0\\
0 \hspace{1.5in} otherwise
\end{aligned}
\end{equation}
\begin{equation}
\gamma_{i,j} = k_{rep} \delta_s(\phi_{12i,j})
\end{equation}
where $\hat{n}_{12i,j}$ is the unit vector normal to the level-sets of $\phi_{12}$ and pointing away from the mid-surface, $k_{rep}$ is a prefactor and $\delta_s(x)$ is a compactly supported \textit{influence function} given by
\begin{equation}
\delta_s(x) = \vast\{
\begin{aligned}
\frac{1+cos\frac{\pi x}{w_T}}{2w_T} \hspace{1in} |x| < -w_T\\
0 \hspace{1.5in} |x|\ge w_T.
\end{aligned}
\end{equation}
Note that, we need to define a separate $\vec{\xi}$ field and transport it for each object that undergoes collision in the simulation, which is required to evaluate Eq. (\ref{fig:midsurface}). However, if there are many objects in the domain and if we know that they do not collide with each other beforehand (when they are sufficiently far away), we can use the same $\vec{\xi}$ field for them to reduce the cost and memory requirements. 

Finally, the pressure Poisson equation (Eq. (\ref{equ:poisson})) which results in a linear system of equations is solved using a conjugate-gradient (CG) approach. Major differences and improvements to the original RMT method by \citet{Valkov2015} are listed in Table \ref{tab:differences}. We thus extended the original reference map technique (RMT) to solve for incompressible fluid-structure interaction problems on an Eulerian collocated grid. Modifications proposed in the extrapolation procedure of the reference map, reconstruction of the level-set field and consistent numerical discretization results in improved robustness, cost effectiveness and conservation properties of the approach. 

\begin{table}
\centering
\begin{tabular}{lll}
\hline
 & The original RMT \citep{Valkov2015} & Present approach \\ \midrule
Grid & Staggered & Collocated \\[0.1cm] 
Nature of the solver & Compressible & Incompressible \\[0.1cm] 
\begin{tabular}[c]{@{}l@{}}Reference map\\ extrapolation\end{tabular} & PDE based & Least-squares based \\[0.1cm] 
\begin{tabular}[c]{@{}l@{}}Discrete momentum conservation \\ (inertial terms)\end{tabular} & No & Yes \\[0.1cm] 
Discretization stencil & One-sided (artificial damping) & Central \\[0.1cm] 
Smoothing routines & Required (artificial damping) & No \\[0.1cm] 
Global damping & Needed (artificial damping) & Not needed \\ \hline
\end{tabular}
\caption{Comparison between the present method and the original RMT method by \citet{Valkov2015}.}
\label{tab:differences}
\end{table}

\section{Results and discussion}\label{sec:results}

In this section, we first present some basic validation test cases to assess the accuracy and cost of our fluid and coupled fluid-solid solver. This is then followed by more complex cases involving solid-solid and fluid-solid contact conditions. Since the fluid-solid coupled problem involves multiple time scales, and an explicit time integration is adopted for solving the system of equations, care must be taken to satisfy all the time step constraints involved in the problem. Time step restriction due to CFL criterion from the advection can be written (for forward Euler in one dimension) as $\Delta t\le C \Delta x/u$, where $C$ represents the Courant number. Time step restriction from the diffusion equation for fluids yields $\Delta t\le0.5\rho^f(\Delta x)^2/\mu^f$. Similarly, shear waves in the solids need to be resolved, and the speed of this shear wave is given by, $u=\sqrt[]{\mu^s/\rho^s}$. Hence a time constraint based on this shear wave speed can be defined as $\Delta t \le P \Delta x \sqrt[]{\rho^s/\mu^s}$, where $P$ represents an appropriate pre-factor that depends on the numerical method. If the ratio of $\mu^s/\rho^s$ is high, travelling shear waves in the solid typically imposes the most restrictive time constraint of all. Hence, in the stiff solid limit such as in the metals, imposed time step constraints are so strict that the simulation time close to solid length scales is virtually impossible with the explicit time stepping approach. Therefore, this formulation is best suited for the simulation of soft solids in fluids. 

\subsection{Validation of the fluid solver}
\begin{figure}
\centering
\includegraphics[width=\textwidth]{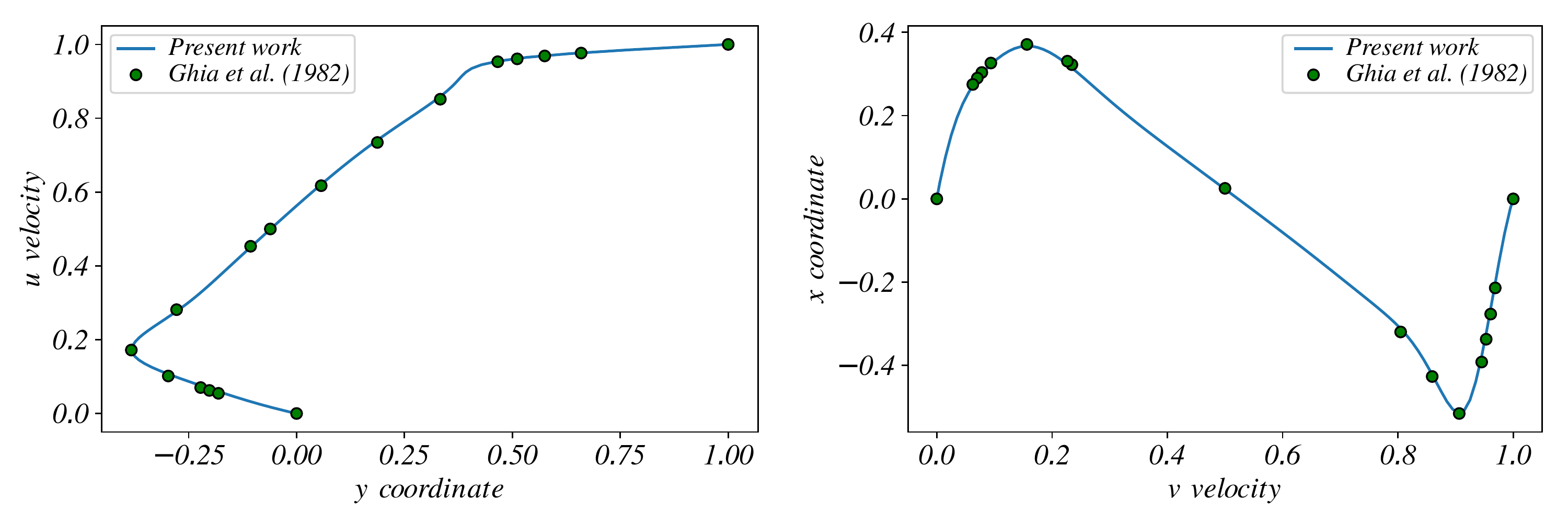}
\caption{\label{fig:lid_driven} Classical lid-driven cavity test case at $Re=1000$. (a) The x component of velocity $\vec{u}\cdot\hat{i}$ along a vertical line through the center of the domain. (b) The y component of velocity $\vec{u}\cdot\hat{j}$ along a horizontal line through the center of the domain.}
\end{figure}

The incompressible Navier-Stokes solver on a collocated grid was validated for the lid-driven cavity case against the benchmark results from \cite{Ghia1982}. A $100\times100$ grid was used for the simulation, and the results are reported for $\RR=1000$.  Figure \ref{fig:lid_driven} shows a good match of the  $u$ and $v$ velocities along the vertical and horizontal lines through the center of the domain with the results from \cite{Ghia1982}. 

\begin{figure}
\centering
\includegraphics[width=0.85\textwidth]{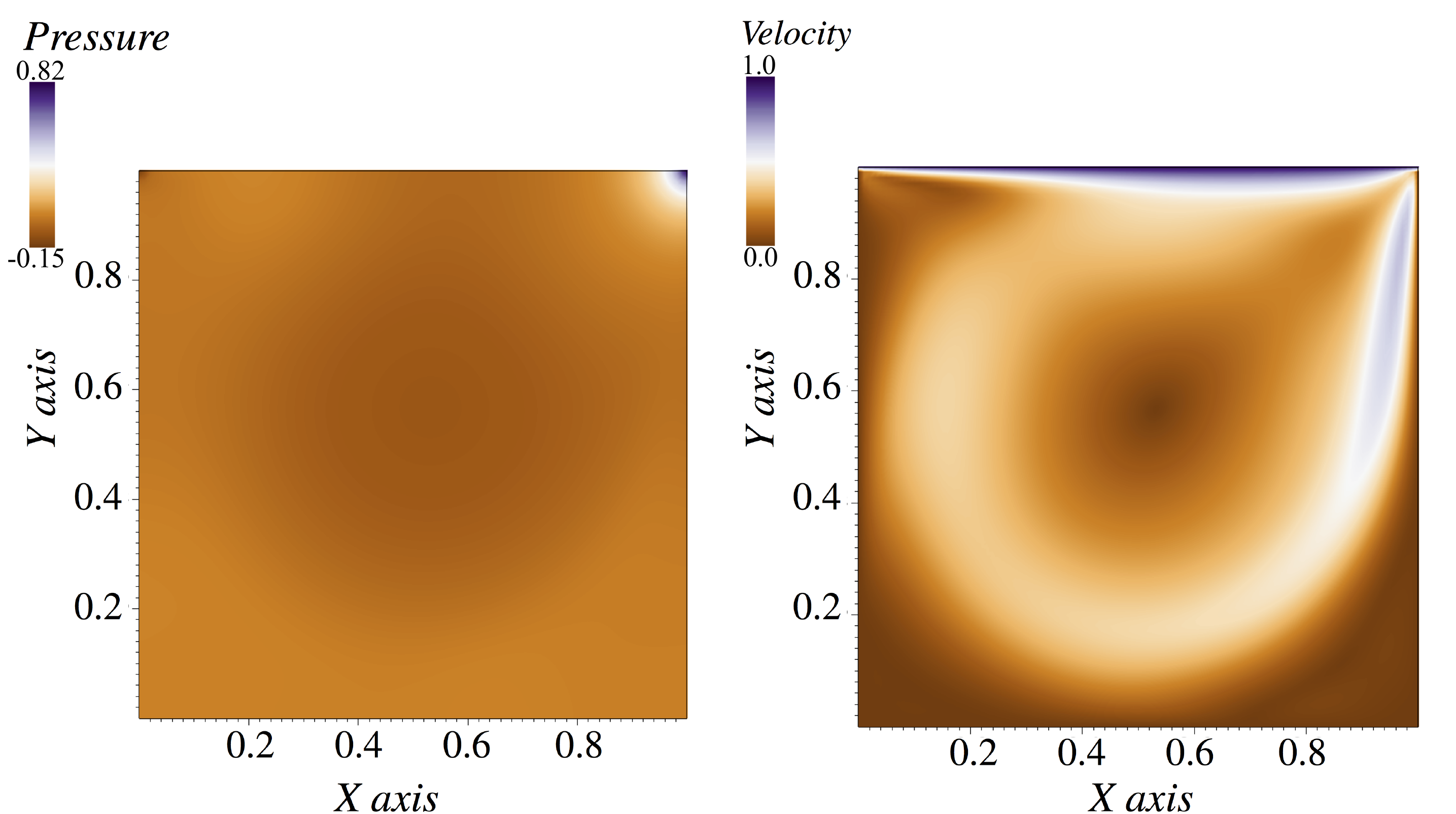}
\caption{Pseudo-color plots of velocity and pressure computed in the lid-driven cavity test case on a $100\times100$ grid at $t=100$ showing smooth fields free of checkerboard oscillations.}
\label{fig:checkerboard}
\end{figure}
Since the equations are solved on a collocated grid, to eliminate the checkerboard fields a Rhie-Chow-like interpolation was performed, as described in Section \ref{sec:conservative}. Figure \ref{fig:checkerboard} presents pseudocolor plots of the velocity and pressure fields from the lid-driven cavity case, illustrating the smoothness of the solution fields obtained.

\subsection{Cost and accuracy of the extrapolation procedure}
\begin{figure}
\centering
\includegraphics[width=\textwidth]{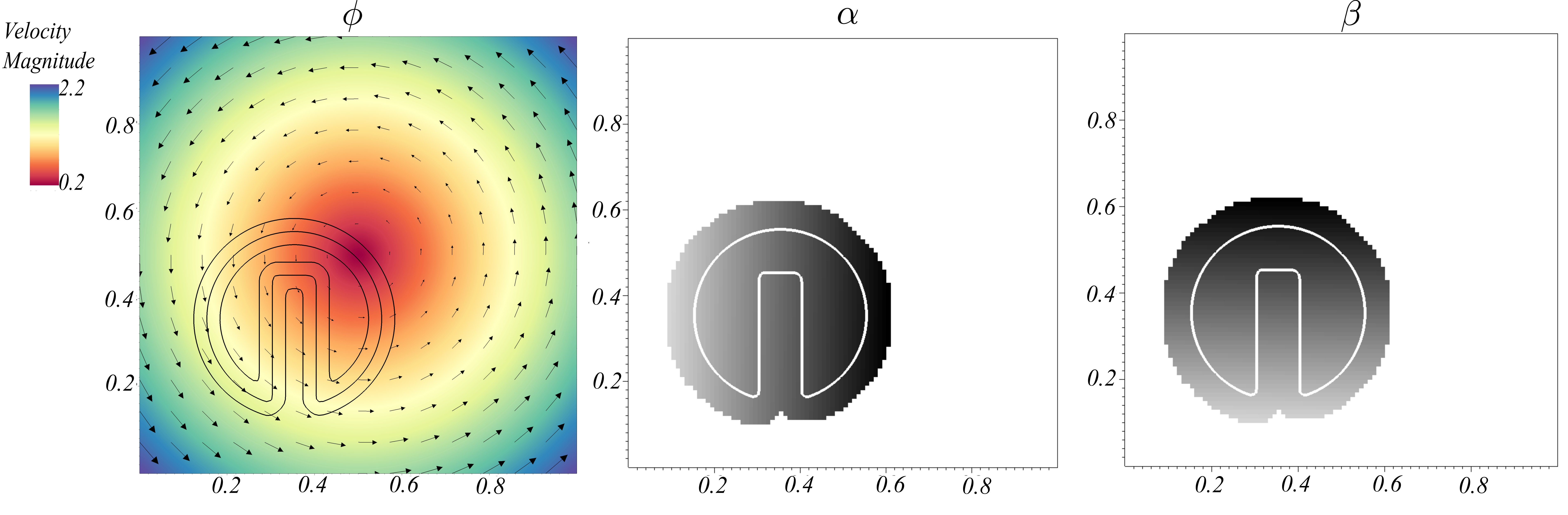}
\caption{Rotation of the Zalesak disk. Disk is initialized at an off-center location as shown. (a) Three solid lines represent the initial and final $\phi$ fields of $\partial \Omega_S$, $+\partial \Omega_E$ and $-\partial \Omega_E$. Color represents the background velocity magnitude. Vector field represents the velocity. (b) Solid line represents $\partial \Omega_S$ and the shaded region represents $\alpha=\vec{\xi}\cdot\hat{i}$ field (c) Solid line represents $\partial \Omega_S$ and the shaded region represents $\beta=\vec{\xi}\cdot\hat{j}$ field.}
\label{fig:zalesak}
\end{figure}

We compared the accuracy of our least-squares extrapolation procedure with that of the hyperbolic partial differential equation (PDE) approach used in RMT (using a RK2-minmod scheme to solve the hyperbolic PDEs). Figure \ref{fig:zalesak} shows the results of the Zalesak disk test case, wherein a slotted disk (a $\vec{\xi}$ field) that is placed off-center is advected with a given background rotational velocity field and compared against the initial conditions after one full rotation. Three solid lines in (a) represent the initial and final $\phi$ fields of $\partial \Omega_S$ (fluid-solid interface), $+\partial \Omega_E$ and $-\partial \Omega_E$ (boundaries of the extended solid region). Solid lines in (b) and (c) represents $\partial \Omega_S$ and the shaded region represents $\alpha=\vec{\xi}\cdot\hat{i}$ and $\beta=\vec{\xi}\cdot\hat{j}$ fields.

The initial and final $\phi$ fields in Figure \ref{fig:zalesak} (a) are exactly on top of each other, showing that the extrapolation procedure by itself is very accurate. One should be careful in interpreting this result, and should not relate this with the rotation of Zalesak disk usually presented in the literature that is obtained as a result of direct advection of $\phi$ field . Here $\phi$ field is reconstructed using the condition in the Eq. \ref{equ:compatibility} and the high accuracy of this $\phi$ field could only be achieved due to the advection of $\vec{\xi}$ that was linear and smooth using a second-order central scheme. This clearly shows the advantage of using the compatibility condition in the Eq. \ref{equ:compatibility} as opposed to the advection of the $\phi$ field.

Further, since the errors in the extrapolation procedure manifests as the error in the advection of the $\vec{\xi}$ field, we computed the $L_2$ norm error $E_{\xi}= ||\vec{\xi_i} - \vec{\xi_f}||_2$ for the advection, where $\vec{\xi_i}$ and $\vec{\xi_f}$ are the initial and final fields obtained after one full rotation, and report them in Table \ref{tab:extrapolate}. It is evident that our least-squares procedure is considerably more accurate when compared to the PDE approach.

\begin{table}
\centering
\begin{tabular}{ccc}
\hline
\textbf{} & Least-squares approach & PDE approach \\ \midrule
\textbf{$E_{\vec{\xi}.\hat{i}}$} & $6.72\times10^{-9}$ & $8.32\times10^{-4}$ \\ 
\textbf{$E_{\vec{\xi}.\hat{j}}$} & $7.34\times10^{-9}$ & $4.67\times10^{-4}$ \\ \hline
\end{tabular}
\caption{Comparison of the error in the cost effective least-square extrapolation procedure vs the hyperbolic PDE extrapolation procedure of \citet{Aslam2004}. $E_{\vec{\xi}\cdot\hat{i}}$, $E_{\vec{\xi}\cdot\hat{j}}$ represents the $L_2$ norm error of the $x$ and $y$ components of $\vec{\xi}$ field computed after one full rotation of the Zalesak disk on a $100\times100$ grid.}
\label{tab:extrapolate}
\end{table}

The above test case was performed on a $100\times100$ grid. Moreover, we also compared the cost of the extrapolation procedure using both the approaches and found that on an average the least-squares procedure required $\approx 100 \mathrm{ms}$ per extrapolation, whereas the PDE approach required $\approx 1550 \mathrm{ms}$ per extrapolation on this grid (close to the time taken by a Poisson solver), for an extrapolation band region of $5\Delta x$. This also proves that our least-squares procedure is extremely cost-effective when compared to the PDE approach. 

\begin{figure}
\centering
\includegraphics[width=0.5\textwidth]{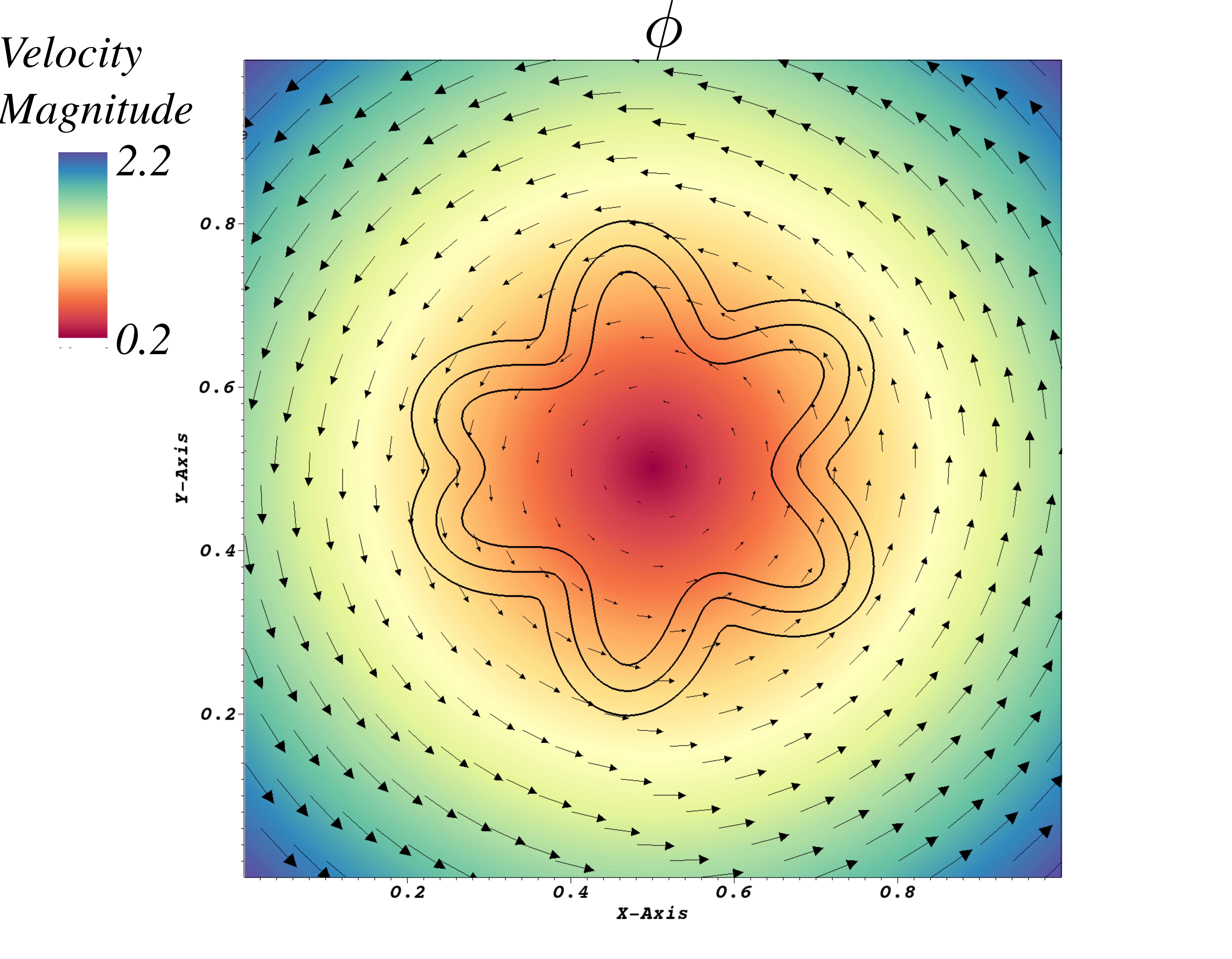}
\caption{Rotation of an asymmetric star-looking object simulated on a grid of size $100\times100$. Three solid lines represent the initial and final $\phi$ fields of $\partial \Omega_S$, $+\partial \Omega_E$ and $-\partial \Omega_E$. Color represents the background velocity magnitude. Vector field represents the velocity.}
\label{fig:star}
\end{figure}

Additionally, to demonstrate that the use of compatibility condition in the Eq. \ref{equ:compatibility} to reconstruct $\phi$ field is not limited to simple shapes that have analytical expression, we considered an asymmetric star-looking object that has sharp regions and repeated the exercise above. Figure \ref{fig:star} shows the results of the one full rotation of the star-looking object advected with a given background rotational velocity field. Three solid lines represent the initial and final $\phi$ fields of $\partial \Omega_S$ (fluid-solid interface), $+\partial \Omega_E$ and $-\partial \Omega_E$ (boundaries of the extended solid region). The initial and final $\phi$ fields in Figure \ref{fig:star} (a) are again exactly on top of each other, showing the high accuracy of the method even for objects with sharp regions. Further to quantify the error, we computed the $L_2$ norm error $E_{\xi}= ||\vec{\xi_i} - \vec{\xi_f}||_2$, where $\vec{\xi_i}$ and $\vec{\xi_f}$ are the initial and final fields obtained after one full rotation. The error values are $E_{\vec{\xi}.\hat{i}} = 6.39\times10^{-9}$ and $E_{\vec{\xi}.\hat{i}} = 6.77\times10^{-9}$, which are of the same order as the ones reported in Table \ref{tab:extrapolate} for a grid of size $100\times100$. 

\begin{figure}
\centering
\includegraphics[width=\textwidth]{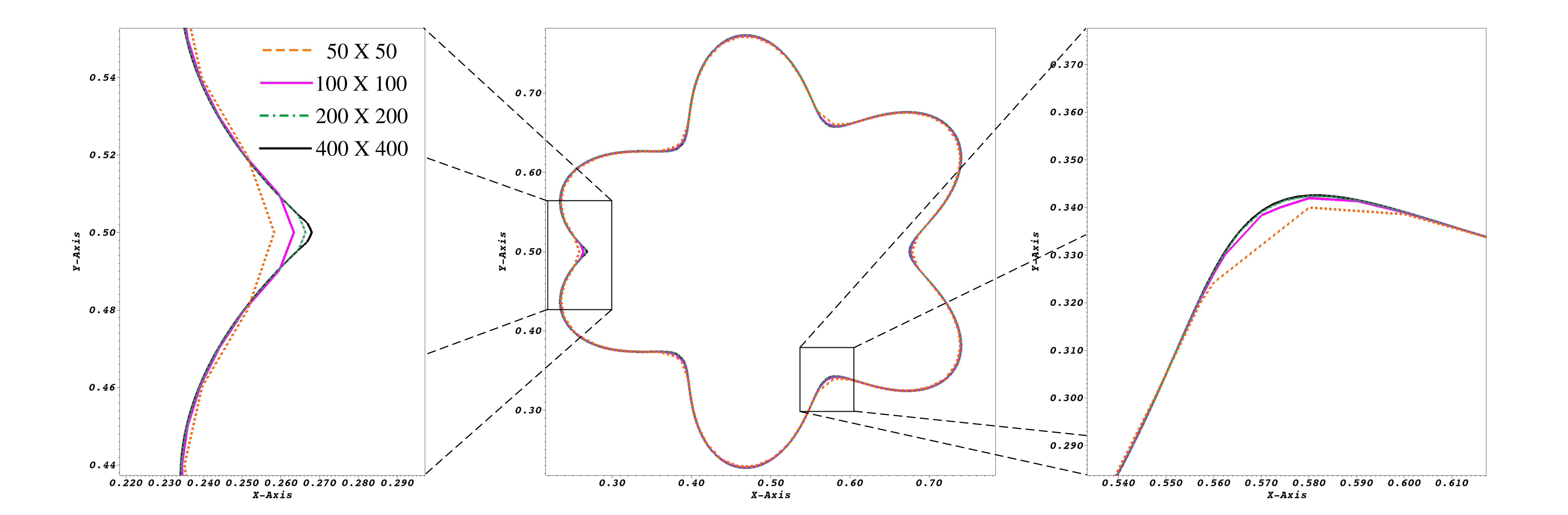}
\caption{Final shape of the asymmetric star-looking object simulated on various grids.}
\label{fig:allstar}
\end{figure}

Finally to study the effect of grid size on the sharp corners, we repeated the same test case for various grids of sizes $50\times50, 100\times100, 200\times200$ and $400\times400$. The final shape of the object for various grid sizes is shown in Figure \ref{fig:allstar}, along with close-up views around a sharp corner and a smooth corner showing the grid convergence. We also computed a normalized $L_2$ norm error $E_{\xi norm} = ||\vec{\xi_i} - \vec{\xi_f}||_2/(N_x\times N_y)$, where $N_x$ and $N_y$ are the number of grid points along $x$ and $y$ directions, and report them in Table \ref{tab:errorstar}. Normalization is done in such a way that the error quantity $E_{\xi norm}$ being compared is grid-size independent and that it represents the error incurred per grid cell in the domain. Clearly, the error per grid cell is very close to machine accuracy for all grid sizes and are roughly independent of the grid size. 

\begin{table}
\centering
\begin{tabular}{ccc}
\hline
Grid size & \textbf{$E_{\vec{\xi}.\hat{i} norm}$} & \textbf{$E_{\vec{\xi}.\hat{j} norm}$} \\ \midrule
\textbf{$50\times50$} & $1.28\times10^{-12}$ & $1.35\times10^{-12}$ \\ 
\textbf{$100\times100$} & $6.39\times10^{-13}$ & $6.77\times10^{-13}$ \\ 
\textbf{$200\times200$} & $3.53\times10^{-13}$ & $3.70\times10^{-13}$ \\ 
\textbf{$400\times400$} & $4.52\times10^{-13}$ & $4.59\times10^{-13}$ \\ \hline
\end{tabular}
\caption{Comparison of the normalized $L_2$ norm error $E_{\xi norm}$ for the case of asymmetric star-looking object for various grid sizes.}
\label{tab:errorstar}
\end{table}

\subsection{Conservative vs Non-conservative implementation \label{sec:conserve}}
Here we would like to highlight that a careful implementation of the blending of fluid and solid Cauchy stresses is crucial in obtaining a discretely conservative momentum formulation. For example, one approach is to compute fluid and solid Cauchy stresses ($\doubleunderline\sigma^s,\doubleunderline\sigma^f$), combine them to obtain a global Cauchy stress ($\doubleunderline\sigma$), and then calculate the divergence of this stress to obtain the force per unit volume ($\vec{f}$) due to stresses as
\begin{equation}
\doubleunderline\sigma \leftarrow\ Blend(\hat{H}(\hat{\phi}),\doubleunderline\sigma^s,\doubleunderline\sigma^f),
\end{equation}
\begin{equation}
\vec{f} = \vec{\nabla}\cdot\doubleunderline{\sigma}.
\end{equation}
The second approach is to compute the divergence of the solid and fluid Cauchy stresses ($\vec{\nabla}\cdot\doubleunderline\sigma^s,\vec{\nabla}\cdot\doubleunderline\sigma^f$) and combine them to obtain the force per unit volume as
\begin{equation}
\vec{f} \leftarrow Blend(\hat{H}(\hat{\phi}),\vec{\nabla}\cdot\doubleunderline\sigma^s,\vec{\nabla}\cdot\doubleunderline\sigma^f).
\end{equation}
The first approach is the one that leads to a conservative formulation, due to the presence of divergence outside the blending operation. This divergence operator, when summed up over adjacent control volumes, leads to an exact cancellation of the terms (analogous to a telescoping series). Hence, we use the conservative formulation in our solver. 

\begin{figure}
\centering
\includegraphics[width=0.3\textwidth]{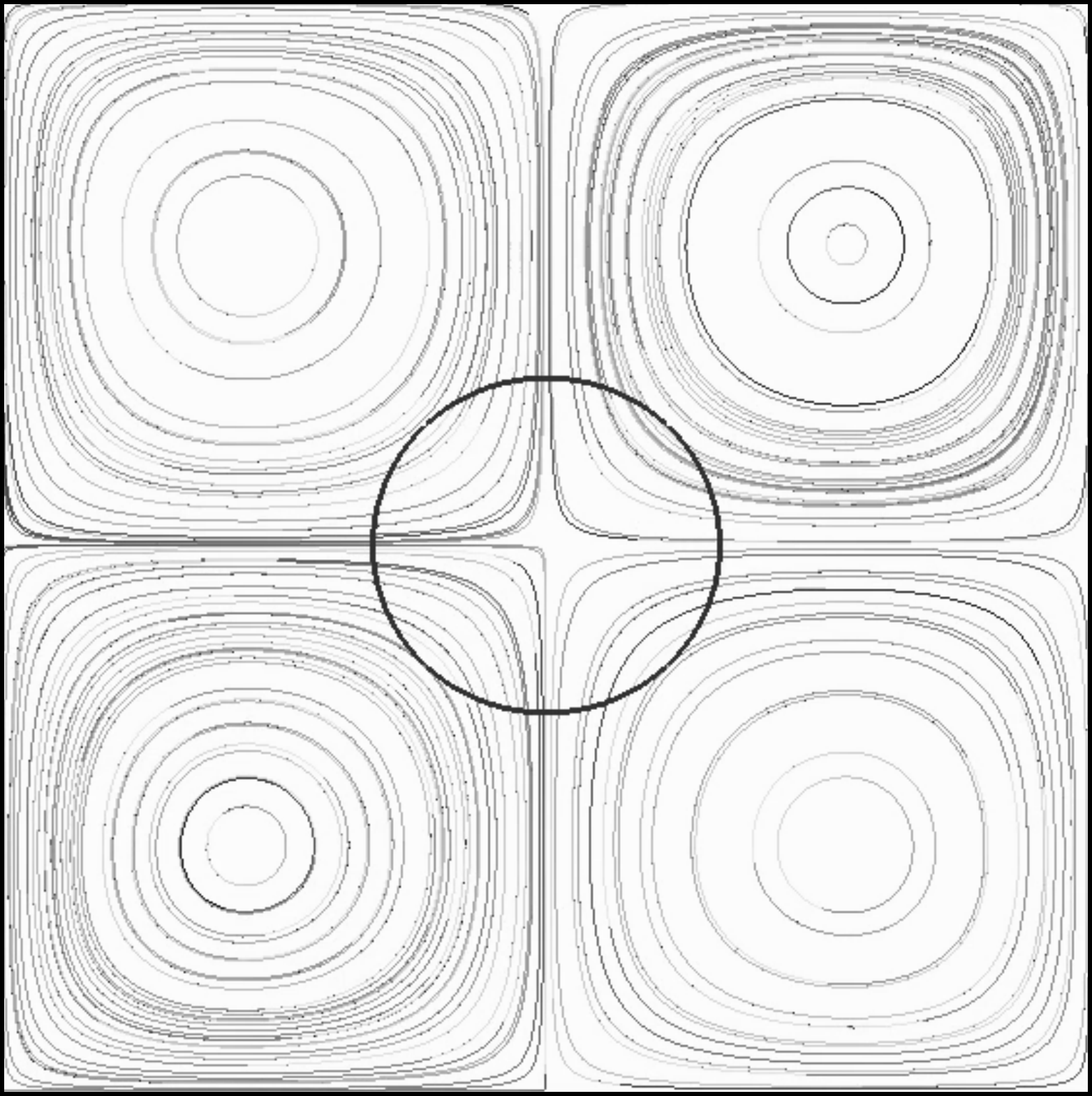}
\caption{Initial state of a circular solid placed in a Taylor-Green vortex field.}
\label{fig:initial_non_conserv}
\end{figure}

\begin{figure}
\centering
\includegraphics[width=\textwidth]{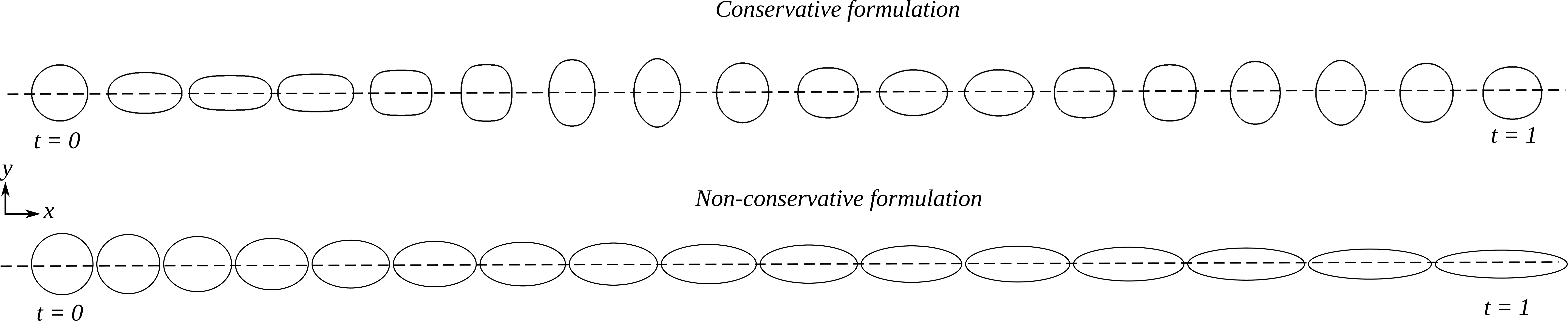}
\caption{Time evolution of the interface of the solid placed in an initially Taylor-Green vortex field, showing the comparison between the results obtained using a conservative formulation and a non-conservative formulation. Radius along the y direction $r$ is also plotted as a function of time $t$ for both the formulations.}
\label{fig:conserv_nonconserv}
\end{figure}

A simulation of a solid placed in a Taylor-Green vortex was performed to qualitatively study the differences between these two formulations. Consider Figure \ref{fig:initial_non_conserv}, which shows the initial state of a solid placed in a Taylor-Green vortex field. Initial flow field should stretch the solid to a certain extent, beyond which the internal stresses developed in the solid should retract it back resulting in an oscillating motion of the solid that stretches and retracts back and forth until all the energy is lost in the viscous dissipation of the fluid. Figure \ref{fig:conserv_nonconserv} shows the result of the simulation performed using both the non-conservative formulation and conservative formulation described above. Clearly, the results are completely unphysical for the non-conservative formulation wherein the solid extends indefinitely with no signs of retraction. By contrast, the conservative formulation for the exact same problem resulted in a more physically meaningful calculation. This simple demonstration illustrates the importance of a conservative numerical implementation, very much similar to the one in compressible flows to achieve correct shock speeds \citep{laney1998computational} and in high-density ratio two-phase flows (see Figure 7 in \citet{raessi2012consistent}).

\subsection{Convergence study}
\begin{figure}
\centering
\includegraphics[width=0.49\textwidth]{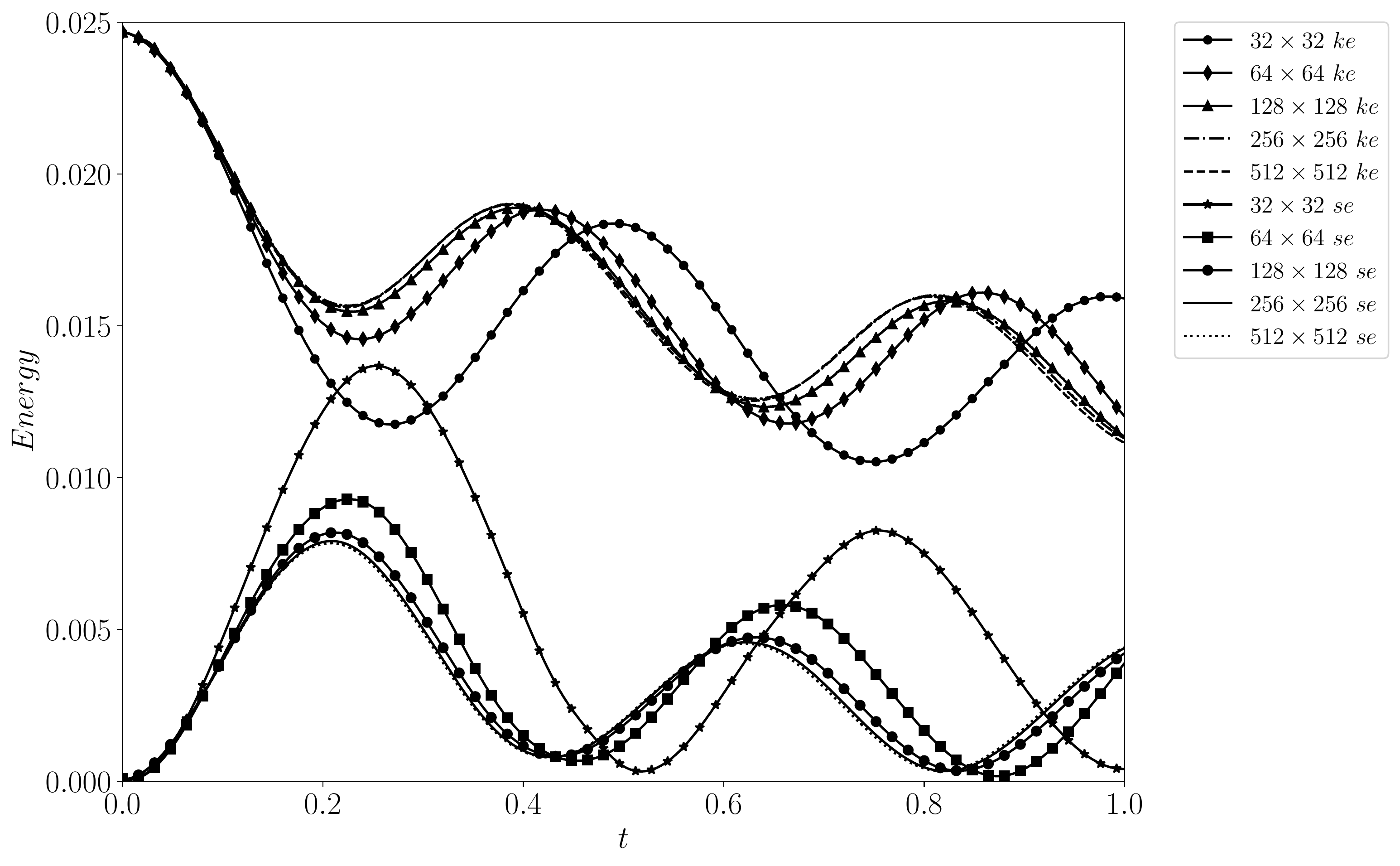}
\includegraphics[width=0.49\textwidth]{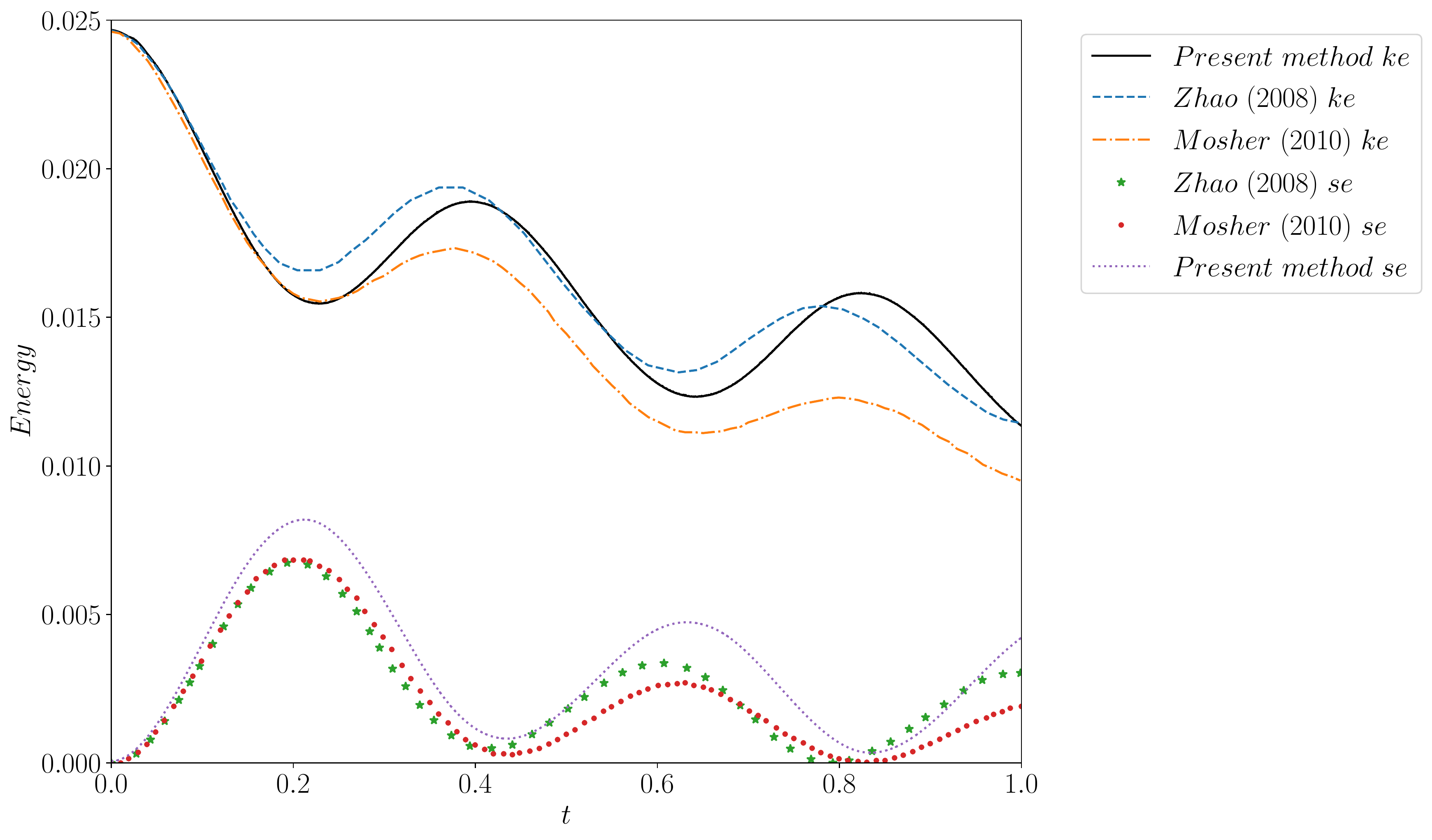}
\caption{Time evolution of the kinetic energy ($ke$) and strain energy ($se$) for the case of a solid placed in an initially Taylor-Green Vortex field. (a) Results are plotted for various grid sizes from $32\times32$ to $512\times512$. (b) Comparison with previous studies by \citet{robinson2011symmetric,zhao2008fixed} for the simulation on a grid size of $128\times128$.}
\label{fig:grid_convergence}
\end{figure}

\begin{figure}
\centering
\includegraphics[width=\textwidth]{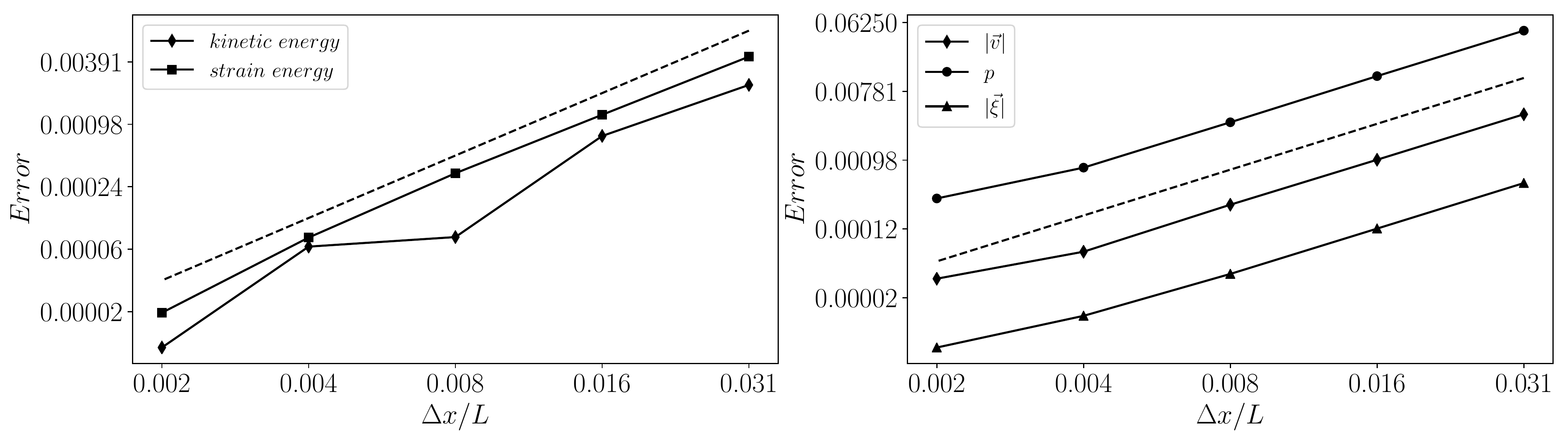}
\caption{Errors for the kinetic energy ($E_{ke}$), strain energy ($E_{se}$), pressure ($E_p$), velocity ($E_v$) and ($E_{\xi}$) fields computed at $t=0.25$. Dashed line represents an $O(\Delta x^2)$ convergence rate.}
\label{fig:order_convergence}
\end{figure}

Above demonstrated test case of a solid placed in a Taylor-Green vortex field was repeated for the values used in \citep{zhao2008fixed,robinson2011symmetric} to validate our solver against the results from a mixed Eulerian-Lagrangian based approach. A solid of radius $r=0.2$ is placed in an initially imposed Taylor-Green vortex field given by the streamfunction $\psi=\psi_0sin(k_xx)sin(k_yy)$ where $\psi_0=5\times10^{-2}$ and $k_x=k_y=2\pi$. Domain size used is $1\times1$ and is discretized into a $128\times128$ grid. Other parameters used in the simulation are fluid viscosity $\mu^f=10^{-3}$, shear modulus $\mu^s=0.5$, solid density $\rho^s=1$ and fluid density $\rho^f=1$. For the sake of consistency with the results of \cite{robinson2011symmetric,zhao2008fixed}, a small amount of viscosity equal to the fluid viscosity of $\mu^f=10^{-3}$ is added in the solid regions. But in general our solver is stable without any viscous damping in the solid regions (see section \ref{sec:fluid-solid results} for simulations without any viscosity in the solid regions). Time evolution of kinetic energy ($ke$) and strain energy ($se$) is plotted in Figure \ref{fig:grid_convergence} for various grid size and also compared against previous studies, where

\begin{equation}
ke = \int \frac{1}{2}u_i u_i\ d\Omega,
\end{equation}
and 
\begin{equation}
se = \int \mu^s (tr(\mathbb{F}^{T}\mathbb{F}) - 2)\ d\Omega.
\end{equation}
Clearly the results are independent of the grid for sizes $128\times128$ and above. Further, the viscous dissipation ($\varepsilon$) in the fluid and solid regions combined was computed using the expression
\begin{equation}
\varepsilon = \int \mu^f \frac{\partial u_i}{\partial x_j} \frac{\partial u_i}{\partial x_j}\ d\Omega,
\end{equation}
and the conservation of total energy $E$ was assessed at the final time of $t=1$ and was found to decrease less than $1\%$ of the initial time value, where $E$ is given by 

\begin{equation}
E = ke + se + \int_0^t \varepsilon(t')\ dt'.
\end{equation}
Frequency of oscillation of the solid matches well with the results of \citet{robinson2011symmetric,zhao2008fixed}. The time evolution of the kinetic energy in the present work matches well with that of \citet{robinson2011symmetric} during the first period of oscillation, but eventually the kinetic energy decays faster in the simulations by \citet{robinson2011symmetric,zhao2008fixed} compared to the current results. Similarly, the strain energy is under-predicted and decays faster in the previous works compared to the current results. This highlights the non-dissipative nature of central-difference scheme used in the current work. Further, using the same test case we also assessed the order of convergence of all the primitive variables ($\vec{u},p,\vec{\xi}$), the kinetic energy ($ke$) and the strain energy ($se$) of the solid used in our solver against a refined case on a $1024\times1024$ grid. Figure \ref{fig:order_convergence} shows that the order of convergence is roughly $O(\Delta x^2)$ for all the variables. Errors are defined as 

\begin{equation}
E_{ke} = |ke/N^2 - ke_{ref}/1024^2|,
\end{equation}

\begin{equation}
E_{se} = |se/N^2 - se_{ref}/1024^2|,
\end{equation}

\begin{equation}
E_{v} = |||\vec{v}| - |\vec{v}|_{ref}||_{\infty},
\end{equation}

\begin{equation}
E_{p} = ||p - p_{ref}||_{\infty},
\end{equation}

\begin{equation}
E_{\xi} = |||\vec{\xi}| - |\vec{\xi}|_{ref}||_{\infty},
\end{equation}
where the subscript $ref$ refers to the most refined case on a $1024\times1024$ grid. 
 

\subsection{Solid in a driven cavity}

\begin{figure}
    \centering
    \includegraphics[width=\textwidth]{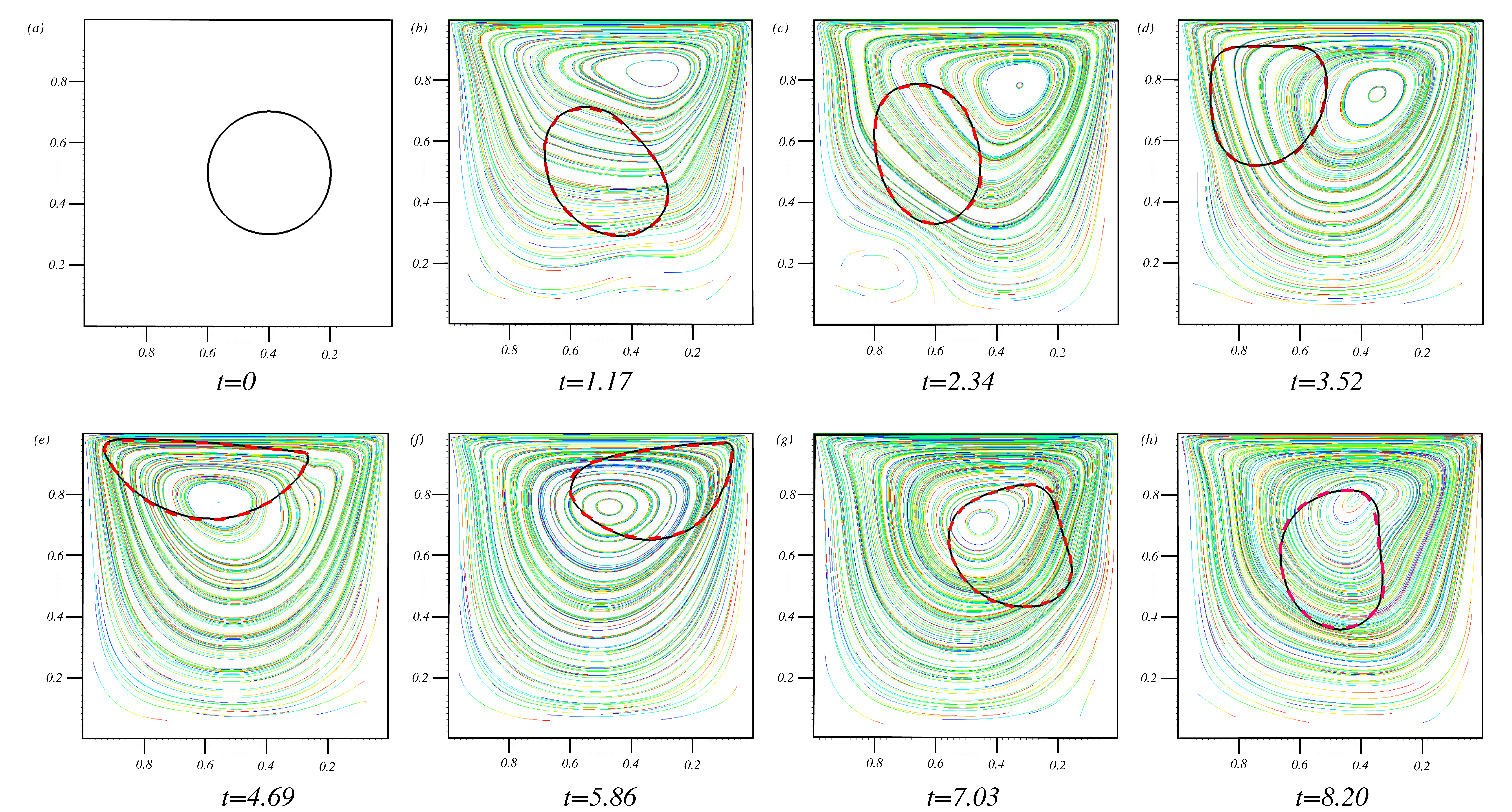}
    \caption{Comparison of the interface of a deforming solid placed in a driven cavity obtained using the present method with that of the results by \citet{sugiyama2011full} for various time instances. The black solid line represents current method and the red dashed line represents the results by \citet{sugiyama2011full}. Colored thin lines represents the flow streamlines.}
    \label{fig:solid_cavity}
\end{figure}

\begin{figure}
    \centering
    \includegraphics[width=0.6\textwidth]{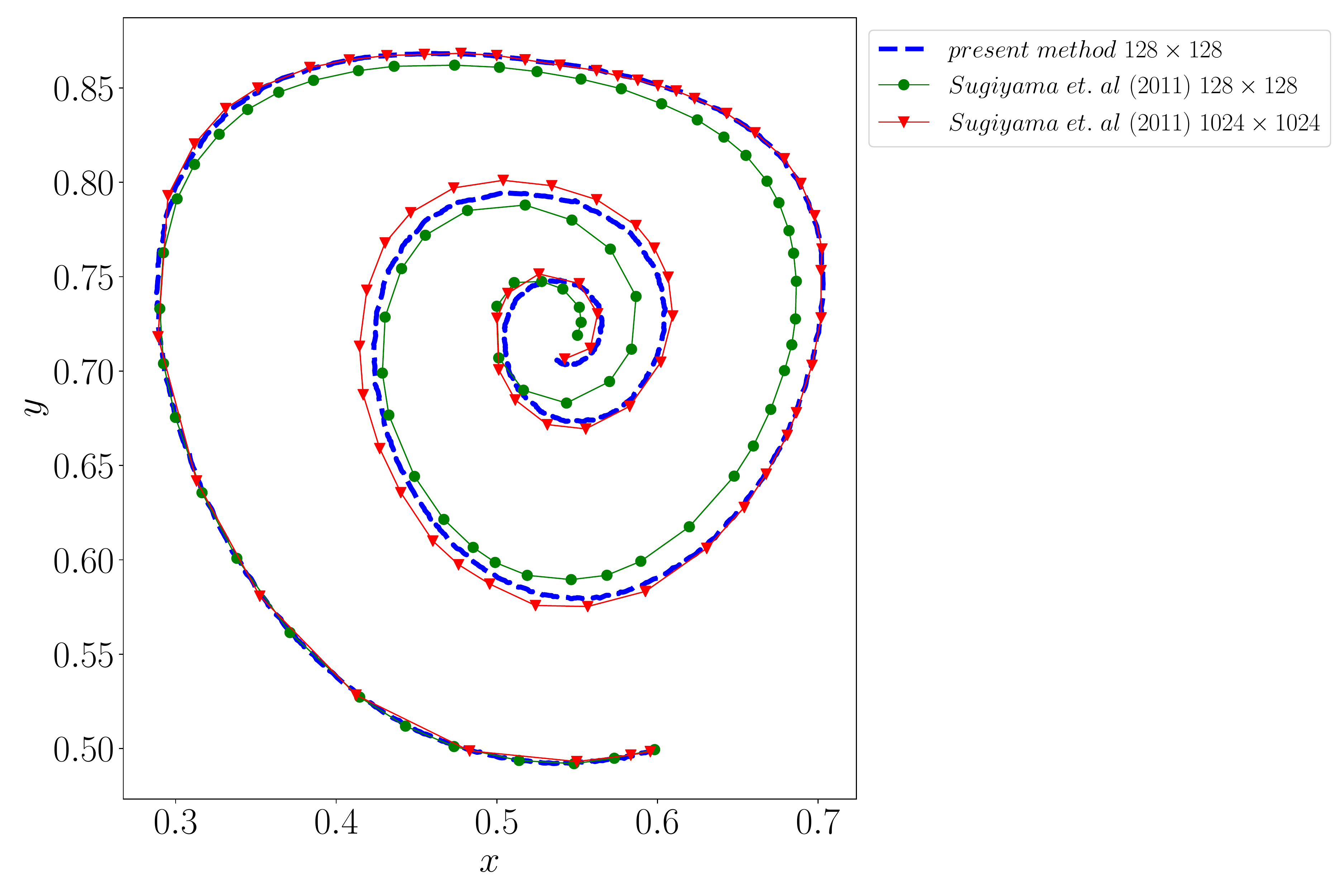}
    \caption{Comparison of the centroid of the solid placed in a driven cavity obtained using the present method on a grid of size $128\times128$ with that of the results by \citet{sugiyama2011full} on grids of size $128\times128$ and $1024\times1024$.}
    \label{fig:solid_cavity_centroid}
\end{figure}

To further validate the fluid-solid coupling of the solver, we simulated a deformable solid in a lid-driven cavity. This test case was previously simulated using mixed-Lagrangian-Eulerian based approach by \citet{zhao2008fixed} and using a VOF based Eulerian approach by \citet{sugiyama2011full}. Figure \ref{fig:solid_cavity}(a) shows the initial configuration of the solid in the domain. Domain used for this simulation is $[0,1]\times[0,1]$ and is discretized into a $128\times128$ grid. The solid is initially circular in shape with a radius of $r=0.2$ and is placed at ($0.6,0.5$) location. Other parameters used in the simulation are fluid viscosity $\mu^f=10^{-2}$, solid viscosity $10^{-2}$, shear modulus $\mu^s=0.05$, solid density $\rho^s=1$ and fluid density $\rho^f=1$. Time evolution of the interface of two solids are shown in Figure \ref{fig:solid_cavity}, where the black solid line represents the current method and the dashed red represents results by \citet{sugiyama2011full}, which shows a pretty good match. Further, we also plot the centroid of the solid in space in Figure \ref{fig:solid_cavity_centroid} against the results by \citet{sugiyama2011full}. This shows that the centroid obtained using the present conservative Reference-Map-Technique on a grid of $128\times128$ is very close to the centroid obtained using a VOF based Eulerian method of \citet{sugiyama2011full} on a grid of $1024\times1024$.  

\subsection{Simulations of solids in a fluid \label{sec:fluid-solid results}}

\begin{figure}
\centering
\includegraphics[width=\textwidth]{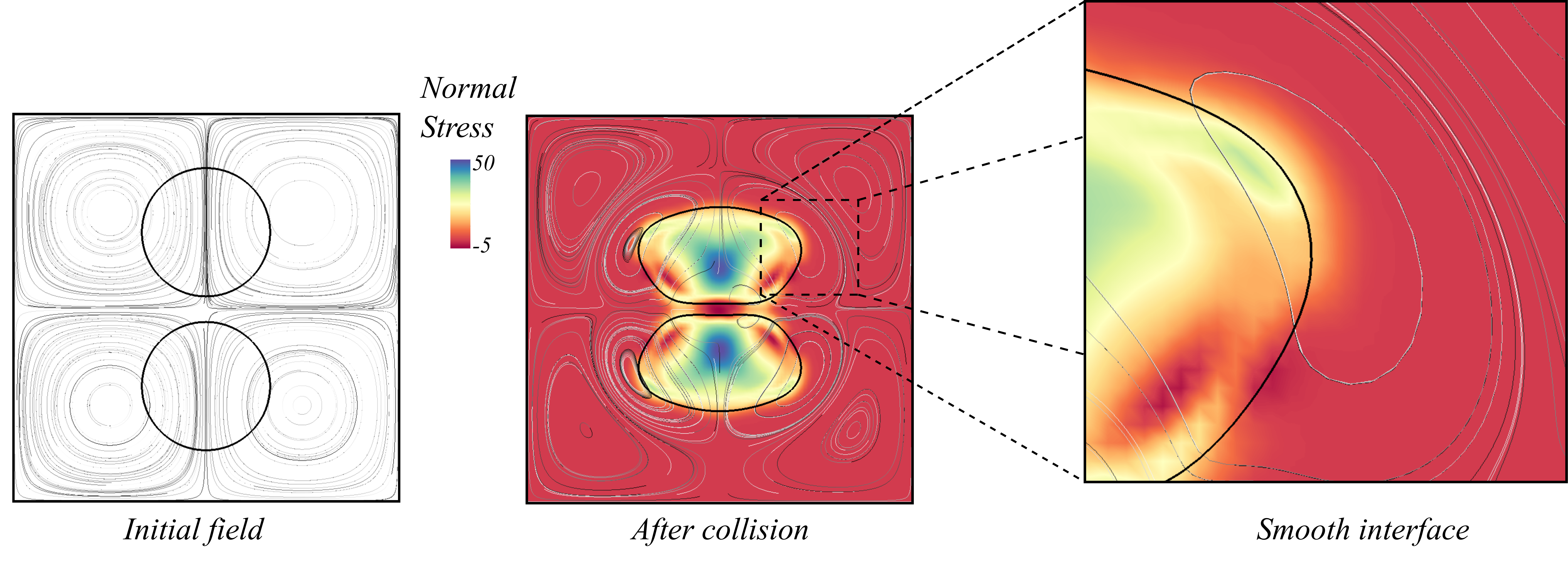}
\caption{Collision of two solids placed in a Taylor-Green vortex, showing the smoothness of the interface and the absence of any striations. Color represents the normal stress = $(\doubleunderline{\sigma}_{11}+\doubleunderline{\sigma}_{22})/2$ in the solid.}
\label{fig:no_striation}
\end{figure}

\begin{figure}
\centering
\includegraphics[width=\textwidth]{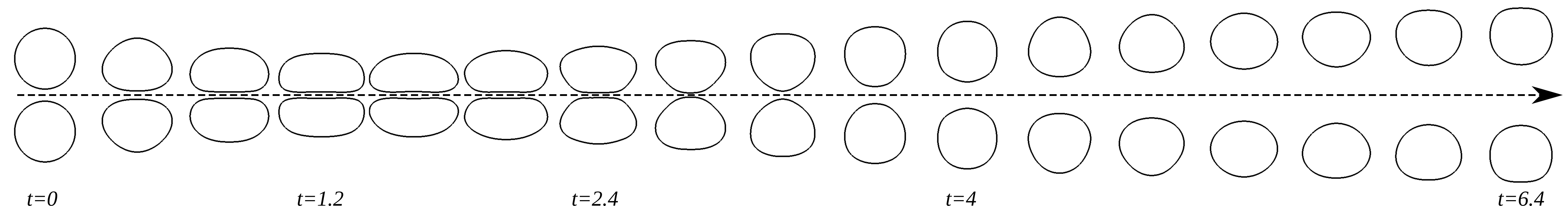}
\caption{Time evolution of the interface of two solids placed in an initially Taylor-Green vortex field, showing the collision and subsequent rebounce of both the solids obtained from the simulation on a grid of size $100\times100$. A plot of centroids of both the solids (solid lines) as a function of time is also included. Dashed line represents the axis.}
\label{fig:2solid}
\end{figure}

In this section we present the simulations of more complex configurations of incompressible solid(s) in a fluid domain such as solid-solid contact, solid-wall contact situations. First, a case of solid-solid contact is considered. Figure \ref{fig:no_striation} shows a configuration of two solids placed in an initially imposed Taylor-Green vortex field given by the streamfunction $\psi=\psi_0sin(k_xx)sin(k_yy)$ where $\psi_0=1$ and $k_x=k_y=1$. Domain used for this simulation is $[-\pi,\pi]\times[-\pi,\pi]$ and is discretized into a $100\times100$ grid. Two solids are initially circular in shape with radii $r_1=r_2=\pi/3$ and are placed at ($\pi,1.4\pi$) and ($\pi,0.6\pi$) locations respectively. Other parameters used in the simulation are fluid viscosity $\mu^f=1$, shear modulii $\mu_1^s=\mu_2^s=100$, solid densities $\rho_1^s=\rho_2^s=100$ and fluid density $\rho^f=100$. Time evolution of the interface of two solids are shown in Figure \ref{fig:2solid}. Solids collide and subsequently rebounce due to the internal stresses developed in them as a result of deformation. Centroid of both the solids are also plotted as a function of time. 

\begin{figure}
    \centering
    \includegraphics[width=0.6\textwidth]{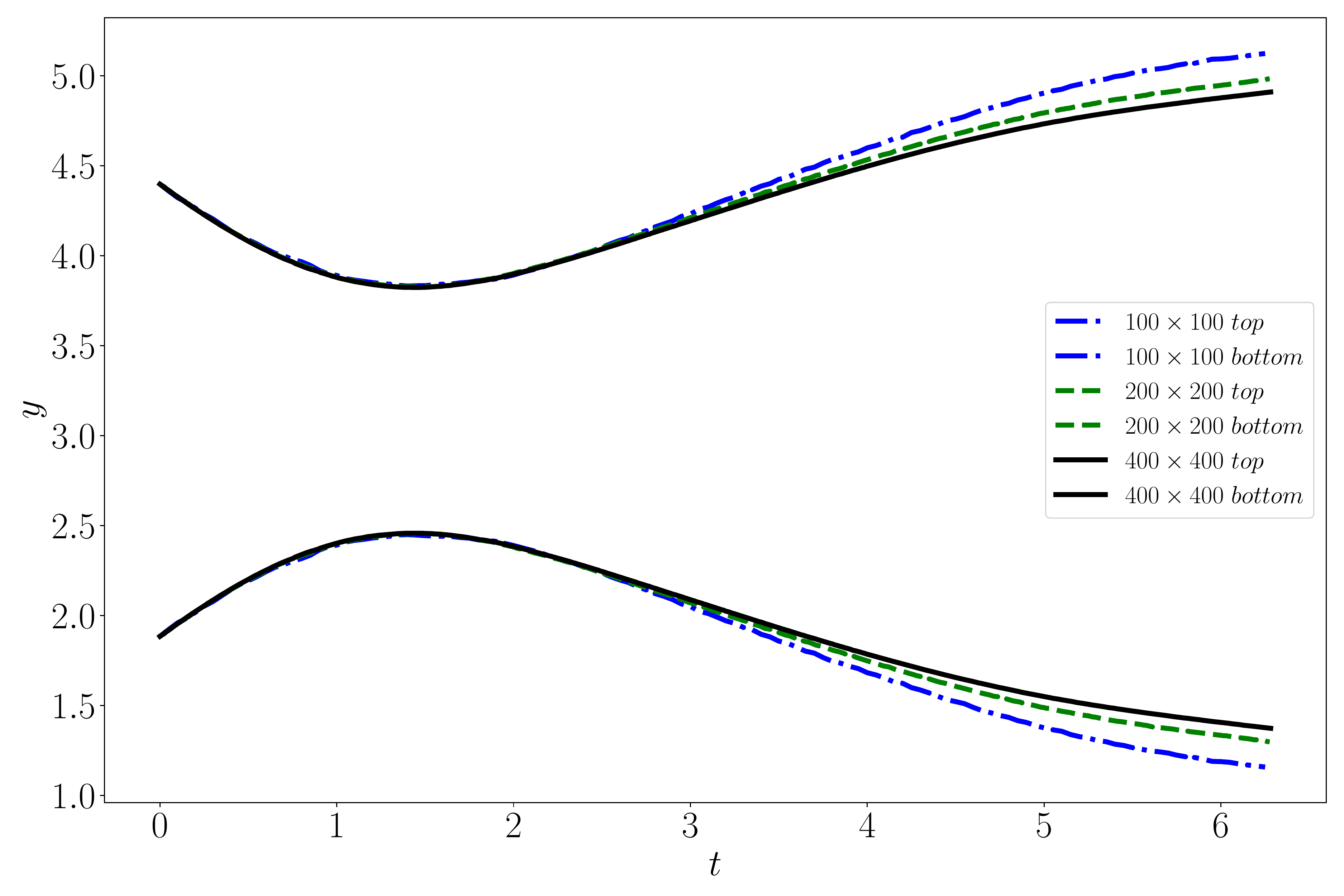}
    \caption{Time evolution of centroids of colliding solids placed in an initially Taylor-Green vortex field. Results from the simulation using grid sizes of $100\times100$, $200\times200$ and $400\times400$ are plotted.}
    \label{fig:centroid_collide}
\end{figure}

The deformed configuration of the solids along with the normal stresses $\doubleunderline\sigma_n=(\doubleunderline\sigma_{11} + \doubleunderline\sigma_{22})/2$ are shown in Figure \ref{fig:no_striation} for the time $t=0.024$. Since the solids are in the rebouncing stage, formation of the four symmetric counter-rotating vortices can be clearly seen around the solids. A zoomed-in view of the solid is also shown in this Figure to illustrate the smoothness of the interface obtained in our approach even at such coarse resolution of $100\times100$ grid points (due to the exact match between the level-set field $\phi$ and $\vec{\xi}$ fields at all times; see Section \ref{sec:level-set}). We also do not see any striations in the extrapolated $\vec{\xi}$ fields that was observed in the original RMT (see Figure 10 in \cite{Valkov2015}), thus eliminating the requirement of the artificial smoothing routines that were used to remove the striations in the extrapolated region. This test case shows the robustness of our solver in handling the solid-solid contact situations. Further, the time evolution of the centroid of the colliding solids are plotted in Figure \ref{fig:centroid_collide} for the grid sizes $100\times100$, $200\times200$ and $400\times400$, which shows that the results (including the collision model) converge with the increase in grid size.

\subsubsection{Deviation of $det(\mathbb{F})$ from $1$ (volumetric error)} \label{sec:vol-error}

In compressible flows, the conservative form of the momentum equation results in inconsistency between density advected using Eq. \ref{equ:continuity} and density computed using $\rho_{0}[det(\mathbb{F})]^{-1}$. To alleviate this, \citet{Kamrin2012} proposed an alternative approach for the computation of density in a one-dimensional setting, where the density is always defined in terms of the motion as opposed to solving the continuity equation. However, in the incompressible limit, density is constant within the solid. Therefore the Eq. \ref{equ:continuity} reduces to $\vec{\nabla}.\vec{u}=0$, which is satisfied discretely using the projection method. Hence the inconsistency is only in maintaining $\rho=\rho_{0}$ within the solid region, i.e., $det(\mathbb{F})$ equal to $1$. This condition is satisfied in the continous limit (see, Appendix A), and is generally not satisfied discretely. However, a good numerical implementation holds the value of $det(\mathbb{F})$ close to $1$. From Eq. (\ref{eq:volume_conserv}) we can write

\begin{equation}
det(\mathbb{F})=dv/dV.    
\end{equation}

\begin{figure}
\centering
\includegraphics[width=0.9\textwidth]{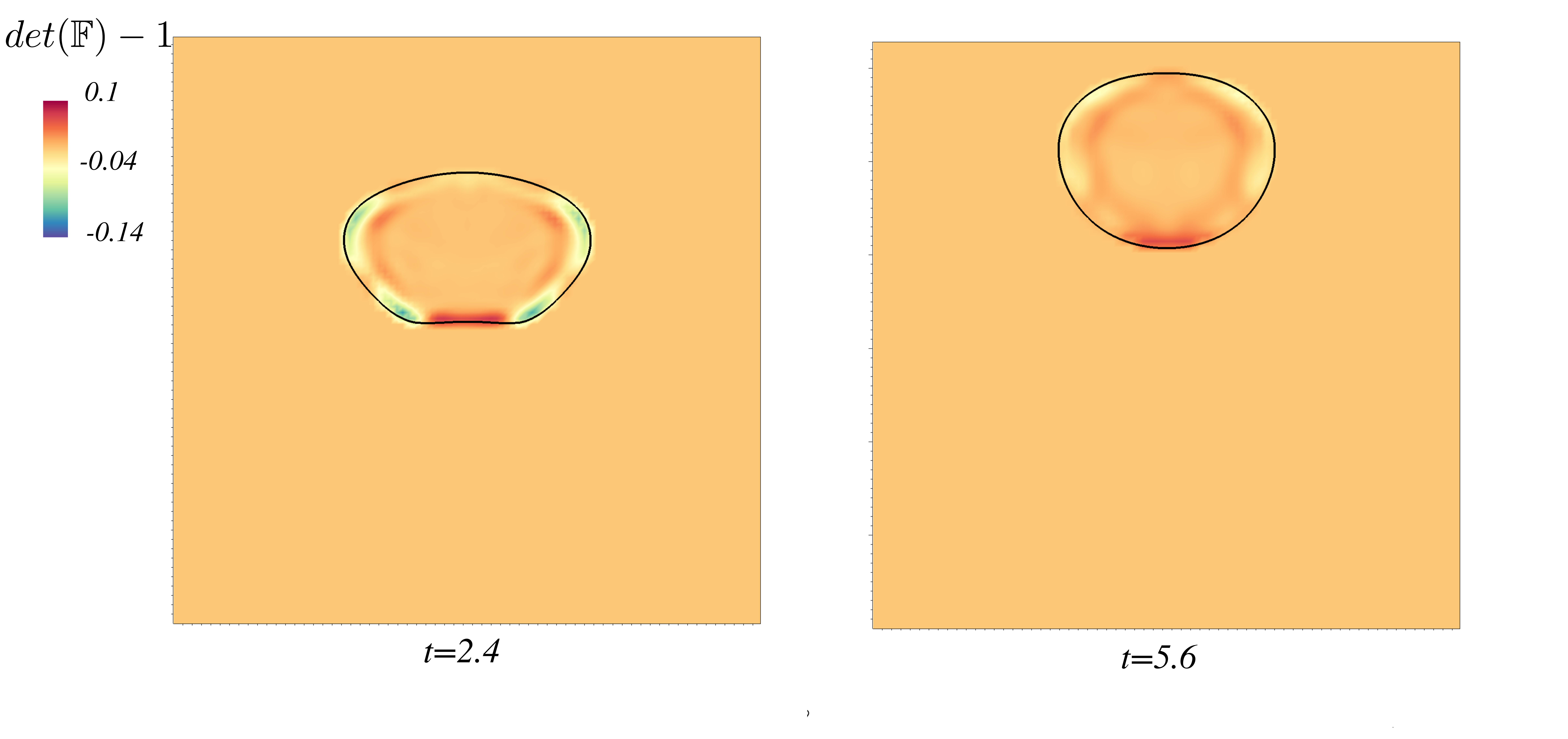}
\caption{Volumetric error at two different time instances for the case of two solids initially placed in a Taylor-Green vortex (see, Section \ref{sec:fluid-solid results}). The error is only shown for the top solid here.}
\label{fig:volume_error}
\end{figure}

Hence $det(\mathbb{F})-1=(dv-dV)/dV$ represents the local volumetric error in an incompressible solid due to the numerical discretization. Here, we present the volumetric error in the solid obtained using the present approach for the test case of two solids placed in an initially Taylor-Green vortex, described in Section \ref{sec:fluid-solid results}. Figure \ref{fig:volume_error} shows the error for the top solid at two different time instances (t=2.4, 5.6). Evidently, the volumetric error in the solid at $t=5.4$ is lower compared to that at $t=2.4$. Hence, the error ($det(\mathbb{F})-1$) does not seem to be accumulating with time, instead it is roughly proportional to the deformation of the solid. Moreover, the error is localized within the transition zone ($\Omega_T$) of the solid and the max value is around $0.14 (14\%)$ and is independent of the grid size chosen. This error occurs due to the presence of mixture region where the stress is computed as a weighted average of the fluid and solid stresses and is typical of any Eulerian approach that uses a diffuse-interface approach. However, since this error is localized to the transition zone, the total error in the mean sense is negligible. 

\begin{figure}
\centering
\includegraphics[width=0.9\textwidth]{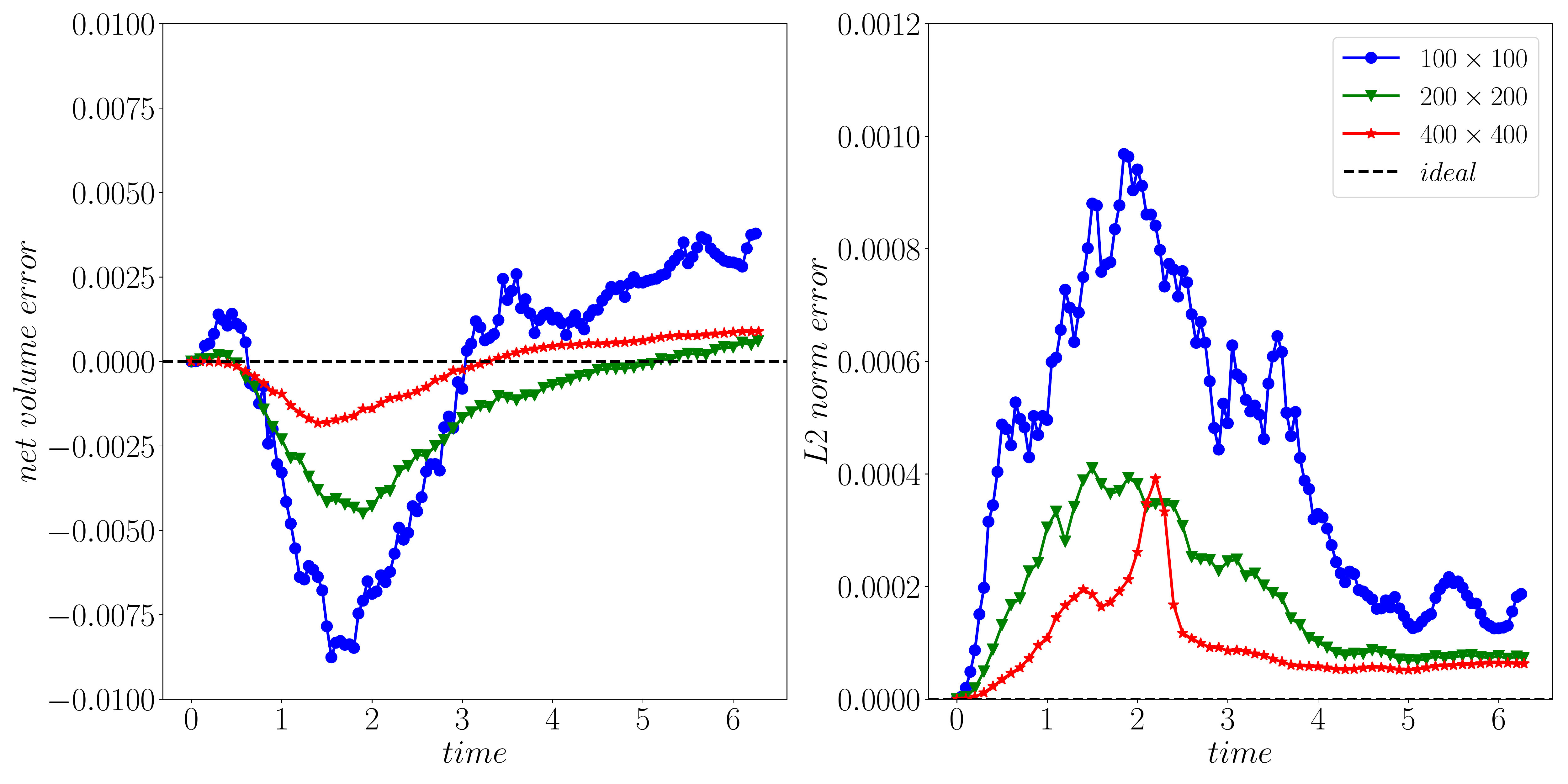}
\caption{(a) Net volumetric error and (b) $L_2$ norm error, as a function of simulation time for the top solid in the test case of two solids placed in an initially Taylor-Green vortex.}
\label{fig:net_volume_error}
\end{figure}

Additionally, to quantify the local volumetric error incurred throughout the solid the normalized $L_2$ norm of $det(\mathbb{F})$ from $1$ can be computed. This quantity is defined as $||det(\mathbb{F})_i-1||_2/n$, where $i$ is the cell index and $n$ is the number of cells inside the solid and is plotted as a function of time in Figure \ref{fig:net_volume_error} (b) for three different grid sizes. Another similar measure that could be used to evaluate the deviation of $det(\mathbb{F})$ from $1$ is the net volumetric error of the solid, which represents the total volume loss or gain of the solid during the simulation. This quantity is defined as the normalized discrete summation of the $det(\mathbb{F})-1$ quantity, i.e., $\sum_{i=0}^n (det(\mathbb{F})_i-1)/n$ and is plotted as a function of time in Figure \ref{fig:net_volume_error} (a). On a uniform grid this quantity can be expressed as 
\begin{equation}
   \sum_{i=0}^n \frac{(det(\mathbb{F})_i-1)}{n} = \sum_{i=0}^n \frac{(dv_i-dV_i)}{(n\ dV_i)} = \frac{(V_{fin} - V_{init})}{V_{init}}
\end{equation}
where $V_{fin}$ and $V_{init}$ are the final and initial volumes of the solid. Figure \ref{fig:net_volume_error} (a) shows the net volumetric error and Figure \ref{fig:net_volume_error} (b) shows the normalized $L_2$ norm error as a function of simulation time for the top solid in the test case of two solids placed in an initially Taylor-Green vortex for various grid sizes. As the solid deforms, $det(\mathbb{F})$ deviates from $1$ and the volumetric error reaches a value of roughly $1\%$ and the $L_2$ norm error reaches a value of roughly $0.1\%$ for the $100\times100$ grid case, however when the solid retracts back, $det(\mathbb{F})$ gets closer to $1$ and the volumetric error reduces down to $0.25\%$ and the $L_2$ norm error reduces to a value of roughly $0.02\%$. Therefore, there is no increase in error $det(\mathbb{F})-1$ or accumulation with time in the present approach. Furthermore, both volumetric and $L_2$ norm errors reduce significantly with increase in the grid size. Hence, the inconsistency does not pose a critical problem in the present method. 

\subsubsection{Simulations of solid-wall contact}

\begin{figure}
\centering
\includegraphics[width=\textwidth]{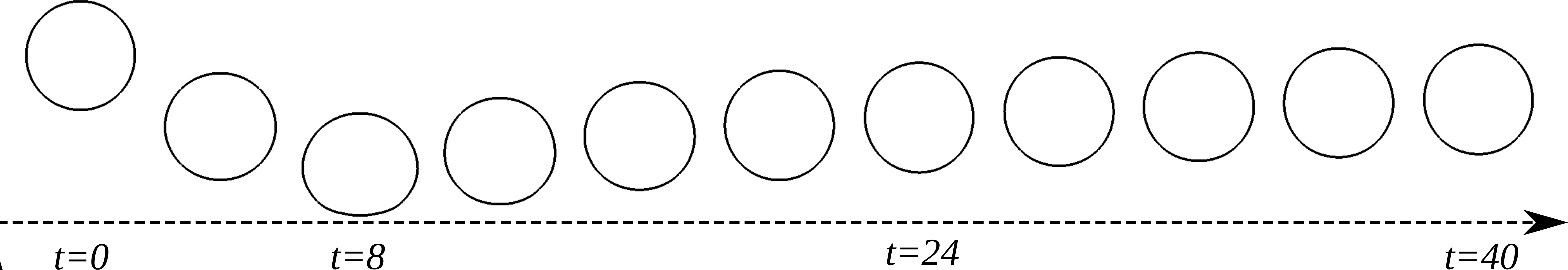}
\caption{Time evolution of the interface of a solid colliding with a rigid wall in a microgravity condition. A plot of centroid of the solid (solid line) as a function of time is also included. Dashed line represents the axis.}
\label{fig:collide}
\end{figure}

Next, a sequence of three test cases named (a) collision, (b) bounce ($\mu^s=100$) and (c) bounce ($\mu^s=1000$) that involve solid-wall contact situations are considered. These classic test cases involving the collision of elastic solids with a rigid wall can be very useful and are of practical relevance in many engineering fields of research. In all the three cases, a domain of $[0,2\pi]\times[0,2\pi]$ is used and a circular shaped solid of radius $\pi/3$ is placed at ($\pi,\pi$) in an initially quiescent surrounding fluid. Case (a) is simulated in a microgravity condition ($g=0$) and the values of other simulation parameters used in this case are $\rho^s=100,\rho^f=100,\mu^s=100,\mu^f=1$. Solid is given an initial velocity of $\vec{u}=-1\hat{j}$ and since this initial condition is fictious and doesn't satisfy incompressibility condition, the solver adjusts the velocity to achieve incompressibility in the first time step. Hence the effective velocity of the solid after one time step was $\vec{u}=0.48\hat{j}$. Time evolution of the interface of the solid is plotted as a function of time as shown in Figure \ref{fig:collide}. Solid encounters the rigid wall and bounces back and goes to a state of rest after losing all its kinetic energy to the surrounding fluid.

\begin{figure}
\centering
\includegraphics[width=\textwidth]{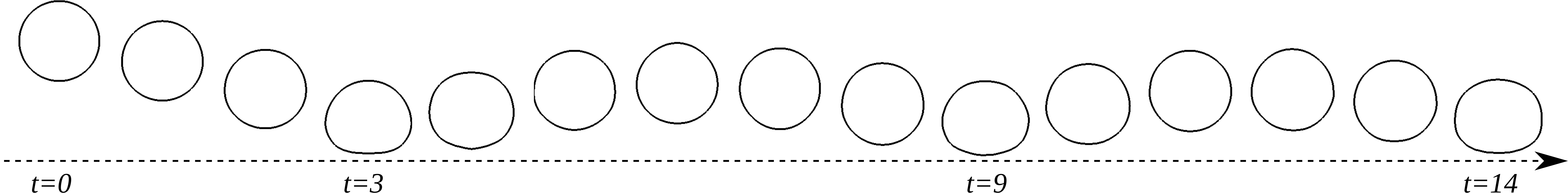}
\caption{Time evolution of the interface of a solid (more stiff $\mu^s=1000$) colliding with a rigid wall under a non-zero gravity condition. A plot of centroid of the solid (solid line) as a function of time is also included. Dashed line represents the axis.}
\label{fig:1000mu_bounce}
\end{figure}

\begin{figure}
\centering
\includegraphics[width=\textwidth]{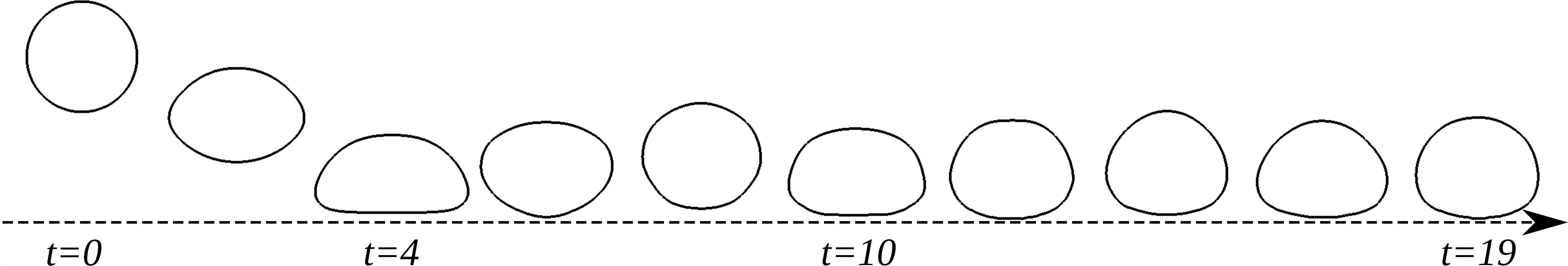}
\caption{Time evolution of the interface of a solid (less stiff $\mu^s=100$) colliding with a rigid wall under a non-zero gravity condition. . Dashed line represents the axis.}
\label{fig:100mu_bounce}
\end{figure}

\begin{figure}
    \centering
    \includegraphics[width=0.8\textwidth]{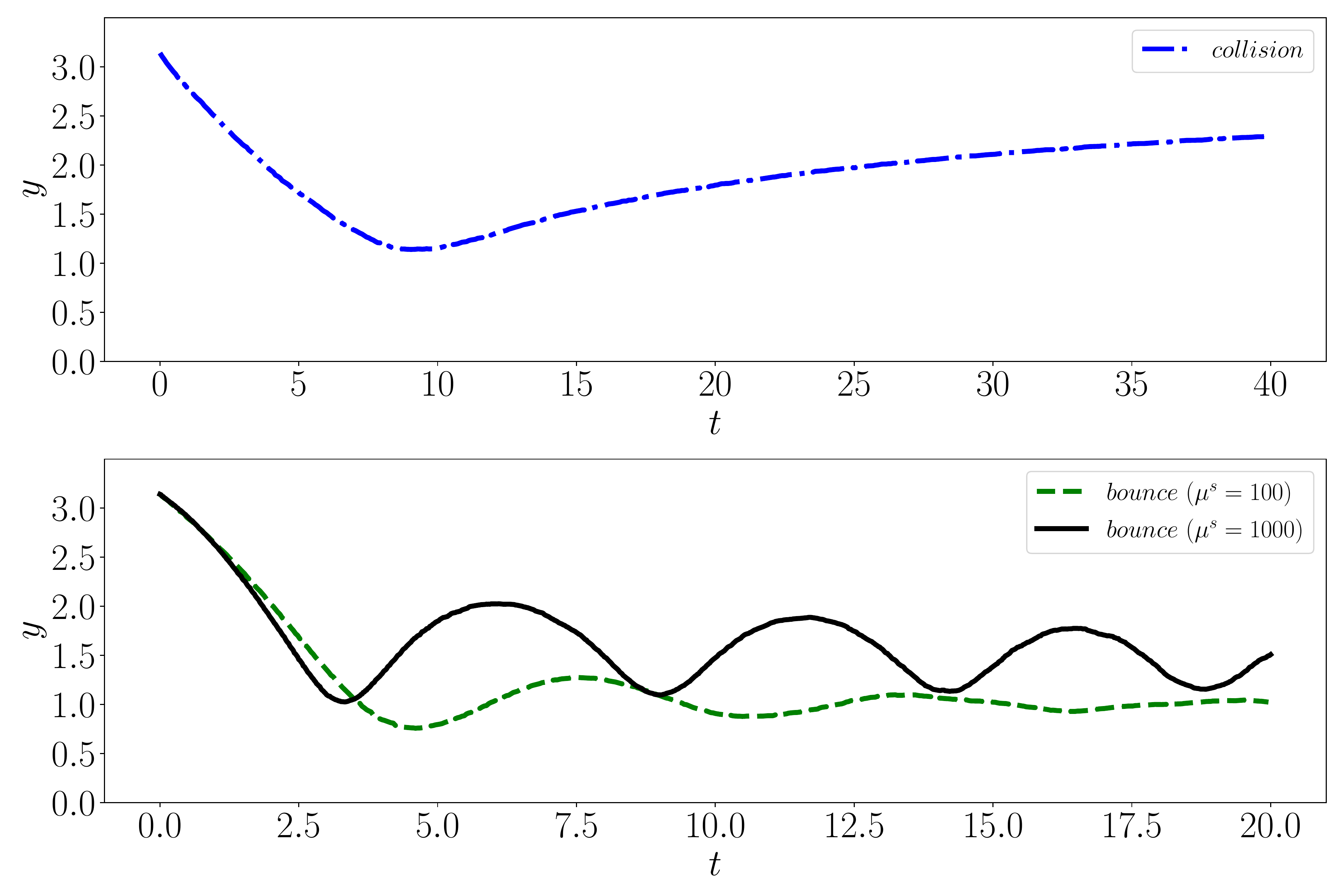}
    \caption{A plot of centroid of the solid as a function of time for all three cases (a) collision, (b) bounce ($\mu^s=100$), and (c) bounce ($\mu^s=1000$), that tests the implementation of solid-wall contact conditions.}
    \label{fig:centroid_solid_wall}
\end{figure}

Case (b) is simulated in a gravity condition with $g=0.0981$ and the values of other simulation parameters used in this case are $\rho^s=1000,\rho^f=100$ hence a density ratio of $\rho^s/\rho^f=10$, $\mu^s=1000,\mu^f=10$. Solid is driven by the gravity and is initialized with a zero velocity. Time evolution of the interface of the solid is plotted as a function of time as shown in Figure \ref{fig:1000mu_bounce}. Solid encounters the rigid wall and bounces back and forth until it goes to a state of rest after losing all its kinetic and potential energy to the surrounding fluid. Case (c) is similar to case (b), but with parameters $\mu^s=100,\mu^f=1$. Solid is initialized with zero velocity and the time evolution of the interface of the solid is plotted as a function of time in Figure \ref{fig:100mu_bounce}. Similar to case (b), here the solid bounces back and forth until it goes to a state of rest, but loses most of its energy to the fluid at its first encounter with the rigid wall due to a large deformation. The energy transferred to the fluid is eventually dissipated due to the action of viscosity. Though the strain energy stored in the solid in the event of a deformation is fully reversible/recoverable (non-viscous solid), energy spent in moving the surrounding fluid is large in the case of large deformations and hence the solid in the case (c), where the shear modulus is $\mu^s=100$, goes to rest much quicker when compared to the case (b) where the shear modulus is $\mu^s=1000$, with other parameters such as density ratio and gravity being identical. The centroid of the solid plotted as a function of the time for all three cases (a), (b) and (c) are shown in Figure \ref{fig:centroid_solid_wall}.
        
\section{Summary and Conclusions} \label{sec:summary}

We have presented an Eulerian formulation for the simulation of incompressible soft solids in a fluid. Methods that handle solids in a Lagrangian fashion are known to be too expensive for highly deforming solids due to large grid deformations and severe time step restrictions. On the other hand an Eulerian approach appears to be a more natural choice for such situations. Hence we have adopted the recently proposed ``reference map technique" (RMT) by \citet{Valkov2015} to simulate solids and fluid-solid problems on an Eulerian grid. We extended this formulation for incompressible settings with the use of an approximate Projection method by \citep{Almgren2000} to achieve divergence-free velocity condition.    

Our formulation discretely conserves momentum and is very cost-effective. Furthermore, we introduced (a) a least-squares extrapolation procedure that is more accurate and cost-effective, (b) a modified advection equation for the reference map field that improves the robustness of the method, (c) a simple, cost-effective way to reconstruct the level-set field that removes any inconsistencies between the reference map field and the level-set field at all times and thereby eliminating the need to have more subroutines to fix the issue of striations of the interface (d) use of simple central-difference schemes to compute the fluxes that improves the stability of the numerical method and to eliminate any spurious dissipation of the kinetic energy. 

We evaluated our solver on a variety of test cases involving solid-wall and solid-solid contact situations and showed that it is stable for all the cases. Furthermore, the test cases that we formulated can serve as a reference for future developers to compare and evaluate their models. Overall, this novel approach opens up a new pathway for the high fidelity numerical simulations of complex, large scale, coupled fluid-solid problems involving large deformations at lower costs compared to the Lagrangian or ALE methods.



\section*{Acknowledgments}

This investigation was funded by the Office of Naval Research, Grant \#119675. The first author was also supported by a Franklin P. and Caroline M. Johnson Fellowship. Authors would also like to thank the reviewer for her/his comments, that helped in improving this work.



\section*{Appendix A: Equivalence of $\vec{\nabla}\cdot\vec{u}=0$ and $det(\mathbb{F})=1$ relations}

\begin{figure}
\centering
\includegraphics[width=0.6\textwidth]{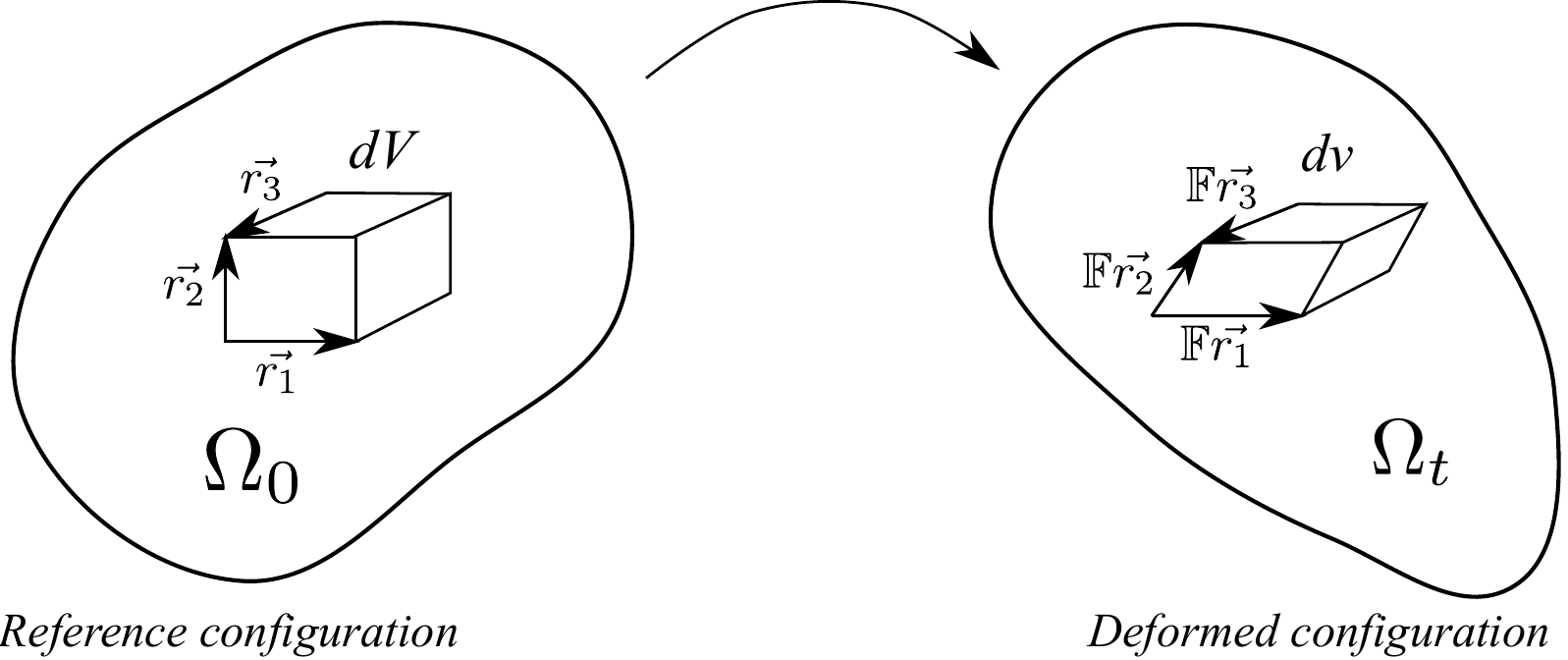}
\caption{Elementary volume before and after deformation.}
\label{fig:volume}
\end{figure}

We can relate the divergence of velocity to the normal strain rate as
\begin{equation}
\vec{\nabla}\cdot\vec{u}=\frac{\partial u_i}{\partial x_i} =\epsilon_{ii} 
\end{equation}
where $\epsilon$ represents the strain rate. Therefore, the normal strain rate / total dilatation ($\epsilon_{ii}=0$) is zero because of the divergence free condition ($\vec{\nabla}\cdot\vec{u}=0$). Now, relating the total dilation to the rate of change of elementary volume with respect to a unit volume as
\begin{equation}
\epsilon_{ii} = \frac{1}{\mathcal{V}}\frac{D(\mathcal{V})}{Dt}=0
\end{equation}
we can therefore show that the rate of change of elementary volume ($\frac{D(\mathcal{V})}{Dt}$) is zero. This implies that the elementary volume $\mathcal{V}$ remains constant. Now rewriting this condition in a convective coordinate system, we obtain
\begin{equation}
dv = dV
\end{equation}
where $dv$ is the elementary volume after deformation and $dV$ is the elementary volume before deformation as shown in Figure \ref{fig:volume}. These two elementary volumes $dv$ and $dV$ can be re-expressed in terms of vectors $\vec{r}_1$, $\vec{r}_2$ and $\vec{r}_3$ that form the undeformed elementary volume $dV$ as
\begin{equation}
dV=(\vec{r_1}\times\vec{r_2})\cdot\vec{r_3}=(\vec{N}dA)\cdot\vec{r_3}
\label{equ:vol1}
\end{equation}
\begin{equation}
dv=(\mathbb{F}\vec{r_1}\times\mathbb{F}\vec{r_2})\cdot\mathbb{F}\vec{r_3}=(\vec{n}da)\cdot\mathbb{F}\vec{r_3}
\label{equ:vol2}
\end{equation}
where $\vec{N}dA$ represents the area vector of the face of the elementary volume $dV$ formed by the vectors $\vec{r}_1$ and $\vec{r}_2$ and $\vec{n}da$ represents the area vector of the face of the elementary volume $dv$ formed by the vectors $\mathbb{F}\vec{r}_1$ and $\mathbb{F}\vec{r}_2$. Now making use of the Nanson's formula ($\vec{n}da = det(\mathbb{F})\mathbb{F}^{-T}\vec{N}dA$) that relates these two area vectors, and making use of the relations in the Eqs. (\ref{equ:vol1}-\ref{equ:vol2}), we can show that

\begin{equation}
dv=det(\mathbb{F})dV\Rightarrow det(\mathbb{F})=1
\label{eq:volume_conserv}
\end{equation}

\section*{Appendix B: Derivation of the incompressible solid Cauchy stress $\doubleunderline\sigma^s$ in terms of components of the reference map $\vec{\xi}$}

As described in Section \ref{sec:govern_eq}, we can write the Cauchy stress in terms of the \textit{left Cauchy-Green's deformation tensor} (or stretch tensor, $\mathbbm{b} = \mathbb{F}\mathbb{F}^T$) for an incompressible neo-Hookean solid as 
\begin{equation}
\doubleunderline{\sigma}^s= 2\mu^s\mathbbm{b} - P \mathds{1}
\label{equ:cauchy_stress_append}
\end{equation}
Now, expressing $\mathbbm{b}$ in terms of $\vec{\xi}$ using the relation in Eq. \ref{equ:def_grad}, we obtain
\begin{equation}
\mathbbm{b} = (\vec{\nabla}\vec{\xi})^{-1}(\vec{\nabla}\vec{\xi})^{-T} = ((\vec{\nabla}\vec{\xi})^T(\vec{\nabla}\vec{\xi}))^{-1}.
\end{equation}
Further, rewriting $\vec{\xi}$ in terms of it's components $\alpha=\vec{\xi}.\hat{i}$ and $\beta=\vec{\xi}.\hat{j}$ for a 2D system, we obtain   
\[
\mathbbm{b} = \Bigg[
\begin{bmatrix}
    \frac{\partial \alpha}{\partial x}  & \frac{\partial \alpha}{\partial y} \\
    \frac{\partial \beta}{\partial x}   &  \frac{\partial \beta}{\partial y} 
\end{bmatrix}^T 
\begin{bmatrix}
    \frac{\partial \alpha}{\partial x}  & \frac{\partial \alpha}{\partial y} \\
    \frac{\partial \beta}{\partial x}   &  \frac{\partial \beta}{\partial y}  
\end{bmatrix}
\Bigg]^{-1} = \begin{bmatrix}
   (\frac{\partial \alpha}{\partial x}) ^2 + (\frac{\partial \beta}{\partial x})^2       &  (\frac{\partial \alpha}{\partial x}) (\frac{\partial \alpha}{\partial y}) + (\frac{\partial \beta}{\partial x}) (\frac{\partial \beta}{\partial y})  \\
     (\frac{\partial \alpha}{\partial x}) (\frac{\partial \alpha}{\partial y}) + (\frac{\partial \beta}{\partial x}) (\frac{\partial \beta}{\partial y})      &  (\frac{\partial \alpha}{\partial y}) ^2 + (\frac{\partial \beta}{\partial y})^2 
\end{bmatrix}^{-1}
\]
which can be further simplified by evaluating the inverse of the matrix as
\begin{equation}
\mathbbm{b} = \underbrace{det((\vec{\nabla}\vec{\xi})^T (\vec{\nabla}\vec{\xi}))}_{C} \begin{bmatrix}
   (\frac{\partial \alpha}{\partial y}) ^2 + (\frac{\partial \beta}{\partial y})^2       & -\Big\{ (\frac{\partial \alpha}{\partial x}) (\frac{\partial \alpha}{\partial y}) + (\frac{\partial \beta}{\partial x}) (\frac{\partial \beta}{\partial y})\Big\}  \\
    -\Big\{ (\frac{\partial \alpha}{\partial x}) (\frac{\partial \alpha}{\partial y}) + (\frac{\partial \beta}{\partial x}) (\frac{\partial \beta}{\partial y})\Big\}      &  (\frac{\partial \alpha}{\partial x}) ^2 + (\frac{\partial \beta}{\partial x})^2 
\end{bmatrix}.
\label{equ:b_mat}
\end{equation}
We can further simplify the coefficient $C$ in the Eq. \ref{equ:b_mat} using standard linear algebra identities and show that it is equal to $1$ with the use of the incompressibility condition for solids ($det(\mathbb{F})=1$) as
\begin{equation}
C = det((\vec{\nabla}\vec{\xi})^T (\vec{\nabla}\vec{\xi})) =det(\vec{\nabla}\vec{\xi})^2=\Big(\frac{1}{det(\mathbb{F})}\Big)^2=1.
\end{equation}
Finally, substituting the expression for $\mathbbm{b}$ in Eq. \ref{equ:b_mat} into the Eq. \ref{equ:cauchy_stress_append}, we obtain the final expression for Cauchy stress $\doubleunderline{\sigma}^s$ in terms of the components of the reference map field $\vec{\xi}$ as
\begin{equation}
\doubleunderline{\sigma}^s= 2\mu^s\begin{bmatrix}
   (\frac{\partial \alpha}{\partial y}) ^2 + (\frac{\partial \beta}{\partial y})^2       & -\Big\{ (\frac{\partial \alpha}{\partial x}) (\frac{\partial \alpha}{\partial y}) + (\frac{\partial \beta}{\partial x}) (\frac{\partial \beta}{\partial y}) \Big\}  \\
    -\Big\{ (\frac{\partial \alpha}{\partial x}) (\frac{\partial \alpha}{\partial y}) + (\frac{\partial \beta}{\partial x}) (\frac{\partial \beta}{\partial y})\Big\}      &  (\frac{\partial \alpha}{\partial x}) ^2 + (\frac{\partial \beta}{\partial x})^2 
\end{bmatrix} - P \mathds{1}
\label{equ:solid_cauchy2app}
\end{equation}

\section*{Appendix C: Discretization of the stress terms in the momentum equation}

As shown in Appendix B, we can express the Cauchy stress in terms of the components of the reference map field $\vec{\xi}$ as
\begin{equation}
\doubleunderline{\sigma}^s= 2\mu^s\begin{bmatrix}
   (\frac{\partial \alpha}{\partial y}) ^2 + (\frac{\partial \beta}{\partial y})^2       & -\Big\{ (\frac{\partial \alpha}{\partial x}) (\frac{\partial \alpha}{\partial y}) + (\frac{\partial \beta}{\partial x}) (\frac{\partial \beta}{\partial y}) \Big\}  \\
    -\Big\{ (\frac{\partial \alpha}{\partial x}) (\frac{\partial \alpha}{\partial y}) + (\frac{\partial \beta}{\partial x}) (\frac{\partial \beta}{\partial y})\Big\}      &  (\frac{\partial \alpha}{\partial x}) ^2 + (\frac{\partial \beta}{\partial x})^2 
\end{bmatrix} - P \mathds{1}
\end{equation}
where, $\alpha=\vec{\xi}.\hat{i}$ and $\beta=\vec{\xi}.\hat{j}$ for a 2D system. We use the second-order central difference scheme for the discretization of the gradients of $\alpha$ and $\beta$, which results in a conservative and non-dissipative formulation. If $i,j$ represents the cell index along $x$ and $y$ directions, we can write the discrete form of $\doubleunderline{\sigma}^s$ as
\begin{equation}
    \sigma^s_{i,j,11} = 2\mu^s \Big[(\frac{\alpha_{i,j+1} - \alpha_{i,j-1}}{2 \Delta y}) ^2 + (\frac{\beta_{i,j+1} - \beta_{i,j-1}}{2 \Delta y})^2\Big] - P_{i,j}
\end{equation}
\begin{equation}
    \sigma^s_{i,j,12} = \sigma^s_{i,j,21} = -2\mu^s \Big[(\frac{\alpha_{i+1,j} - \alpha_{i-1,j}}{2 \Delta x})(\frac{\alpha_{i,j+1} - \alpha_{i,j-1}}{2 \Delta y}) + (\frac{\beta_{i+1,j} - \beta_{i-1,j}}{2 \Delta x}) (\frac{\beta_{i,j+1} - \beta_{i,j-1}}{2 \Delta y}) \Big]
\end{equation}
\begin{equation}
    \sigma^s_{i,j,22} = 2\mu^s \Big[(\frac{\alpha_{i+1,j} - \alpha_{i-1,j}}{2 \Delta x}) ^2 + (\frac{\beta_{i+1,j} - \beta_{i-1,j}}{2 \Delta x})^2\Big] - P_{i,j}
\end{equation}
where $\sigma^s_{i,j,11}$,  $\sigma^s_{i,j,12}$, $\sigma^s_{i,j,21}$ and $\sigma^s_{i,j,22}$ are the components of the tensor $\doubleunderline{\sigma}^s_{i,j}$. Notice that we use a wider stencil that uses $i+1$ and $i-1$ points to obtain the gradient at $i$ as opposed to a compact stencil that uses $i+1/2$ and $i-1/2$. Once $\doubleunderline{\sigma}^s_{i,j}$ is evaluated at $i,j$, we evaluate the fluid Cauchy stress $\doubleunderline{\sigma}^f_{i,j}$ in a similar fashion using the same stencil. We then obtain the total Cauchy stress at $i,j$ as
\begin{equation}
    \doubleunderline{\sigma}_{i,j} = \hat{H}[\hat{\phi}(\vec{x},t)_{i,j}]_{i,j}  \doubleunderline{\sigma}^f_{i,j} + \left\lbrace1 - \hat{H}[\hat{\phi}(\vec{x},t)_{i,j}]_{i,j}\right\rbrace\doubleunderline{\sigma}^s_{i,j}.
\end{equation}
Further, to evaluate the divergence of $\doubleunderline{\sigma}$ at $i,j$, we use the same stencil as
\begin{equation}
    \vec{\nabla}\cdot\doubleunderline{\sigma}_{i,j} = \Big\{ (\frac{\sigma_{11,i+1,j} - \sigma_{11,i-1,j}}{2 \Delta x}) + (\frac{\sigma_{12,i,j+1} - \sigma_{12,i,j-1}}{2 \Delta y}) , (\frac{\sigma_{21,i+1,j} - \sigma_{21,i-1,j}}{2 \Delta x}) + (\frac{\sigma_{22,i,j+1} - \sigma_{22,i,j-1}}{2 \Delta y})  \Big\}
\end{equation}
which results in a conservative and consistent discretization of the stress terms that results in correct physical behavior of the solid as described in Section \ref{sec:conserve}.

\section*{References}

\bibliography{fsi.bib}

\begin{thebibliography}{137}
\expandafter\ifx\csname natexlab\endcsname\relax\def\natexlab#1{#1}\fi
\providecommand{\url}[1]{\texttt{#1}}
\providecommand{\href}[2]{#2}
\providecommand{\path}[1]{#1}
\providecommand{\DOIprefix}{doi:}
\providecommand{\ArXivprefix}{arXiv:}
\providecommand{\URLprefix}{URL: }
\providecommand{\Pubmedprefix}{pmid:}
\providecommand{\doi}[1]{\href{http://dx.doi.org/#1}{\path{#1}}}
\providecommand{\Pubmed}[1]{\href{pmid:#1}{\path{#1}}}
\providecommand{\bibinfo}[2]{#2}
\ifx\xfnm\relax \def\xfnm[#1]{\unskip,\space#1}\fi
\bibitem[{Adami et~al.(2016)Adami, Kaiser, Adams and
  Bermejo-Moreno}]{Adami2016}
\bibinfo{author}{Adami, S.}, \bibinfo{author}{Kaiser, J.},
  \bibinfo{author}{Adams, N.A.}, \bibinfo{author}{Bermejo-Moreno, I.},
  \bibinfo{year}{2016}.
\newblock \bibinfo{title}{Numerical modeling of shock waves in biomedicine}.
\newblock \bibinfo{journal}{Center for Turbulence Research, Proceedings of the
  Summer Program} , \bibinfo{pages}{15--24}.
\bibitem[{Almgren et~al.(2000)Almgren, Bell and Crutchfield}]{Almgren2000}
\bibinfo{author}{Almgren, A.S.}, \bibinfo{author}{Bell, J.B.},
  \bibinfo{author}{Crutchfield, W.Y.}, \bibinfo{year}{2000}.
\newblock \bibinfo{title}{{Approximate Projection Methods: Part I. Inviscid
  Analysis}}.
\newblock \bibinfo{journal}{SIAM Journal on Scientific Computing}
  \bibinfo{volume}{22}, \bibinfo{pages}{1139--1159}.
\newblock \DOIprefix\doi{10.1137/S1064827599357024}.
\bibitem[{Andrews and Low(1999)}]{andrews1999role}
\bibinfo{author}{Andrews, D.A.}, \bibinfo{author}{Low, P.S.},
  \bibinfo{year}{1999}.
\newblock \bibinfo{title}{Role of red blood cells in thrombosis}.
\newblock \bibinfo{journal}{Current opinion in hematology} \bibinfo{volume}{6},
  \bibinfo{pages}{76}.
\bibitem[{Aslam(2004)}]{Aslam2004}
\bibinfo{author}{Aslam, T.D.}, \bibinfo{year}{2004}.
\newblock \bibinfo{title}{{A partial differential equation approach to
  multidimensional extrapolation}}.
\newblock \bibinfo{journal}{Journal of Computational Physics}
  \bibinfo{volume}{193}, \bibinfo{pages}{349--355}.
\newblock \DOIprefix\doi{10.1016/j.jcp.2003.08.001}.
\bibitem[{Barton et~al.(2010)Barton, Drikakis and
  Romenski}]{barton2010eulerian}
\bibinfo{author}{Barton, P.T.}, \bibinfo{author}{Drikakis, D.},
  \bibinfo{author}{Romenski, E.}, \bibinfo{year}{2010}.
\newblock \bibinfo{title}{An eulerian finite-volume scheme for large
  elastoplastic deformations in solids}.
\newblock \bibinfo{journal}{International journal for numerical methods in
  engineering} \bibinfo{volume}{81}, \bibinfo{pages}{453--484}.
\bibitem[{Belytschko(1980)}]{belytschko1980fluid}
\bibinfo{author}{Belytschko, T.}, \bibinfo{year}{1980}.
\newblock \bibinfo{title}{Fluid-structure interaction}.
\newblock \bibinfo{journal}{Computers \& Structures} \bibinfo{volume}{12},
  \bibinfo{pages}{459--469}.
\bibitem[{Beyer~Jr(1992)}]{beyer1992computational}
\bibinfo{author}{Beyer~Jr, R.P.}, \bibinfo{year}{1992}.
\newblock \bibinfo{title}{A computational model of the cochlea using the
  immersed boundary method}.
\newblock \bibinfo{journal}{Journal of Computational Physics}
  \bibinfo{volume}{98}, \bibinfo{pages}{145--162}.
\bibitem[{Chopp(2001)}]{Chopp2001}
\bibinfo{author}{Chopp, D.L.}, \bibinfo{year}{2001}.
\newblock \bibinfo{title}{Some improvements of the fast marching method}.
\newblock \bibinfo{journal}{SIAM Journal on Scientific Computing}
  \bibinfo{volume}{23}, \bibinfo{pages}{230--244}.
\newblock \URLprefix \url{https://doi.org/10.1137/S106482750037617X},
  \DOIprefix\doi{10.1137/S106482750037617X},
  \href{http://arxiv.org/abs/https://doi.org/10.1137/S106482750037617X}{\tt
  arXiv:https://doi.org/10.1137/S106482750037617X}.
\bibitem[{Clarke et~al.(1986)Clarke, Hassan and Salas}]{Clarke1986}
\bibinfo{author}{Clarke, D.K.}, \bibinfo{author}{Hassan, H.A.},
  \bibinfo{author}{Salas, M.D.}, \bibinfo{year}{1986}.
\newblock \bibinfo{title}{{Euler calculations for multielement airfoils using
  Cartesian grids}}.
\newblock \bibinfo{journal}{AIAA Journal} \bibinfo{volume}{24},
  \bibinfo{pages}{353--358}.
\newblock \URLprefix \url{https://doi.org/10.2514/3.9273},
  \DOIprefix\doi{10.2514/3.9273}.
\bibitem[{Cottet and Maitre(2016)}]{cottet2016semi}
\bibinfo{author}{Cottet, G.H.}, \bibinfo{author}{Maitre, E.},
  \bibinfo{year}{2016}.
\newblock \bibinfo{title}{A semi-implicit level set method for multiphase flows
  and fluid--structure interaction problems}.
\newblock \bibinfo{journal}{Journal of Computational Physics}
  \bibinfo{volume}{314}, \bibinfo{pages}{80--92}.
\bibitem[{Cottet et~al.(2008)Cottet, Maitre and Milcent}]{cottet2008eulerian}
\bibinfo{author}{Cottet, G.H.}, \bibinfo{author}{Maitre, E.},
  \bibinfo{author}{Milcent, T.}, \bibinfo{year}{2008}.
\newblock \bibinfo{title}{Eulerian formulation and level set models for
  incompressible fluid-structure interaction}.
\newblock \bibinfo{journal}{ESAIM: Mathematical Modelling and Numerical
  Analysis} \bibinfo{volume}{42}, \bibinfo{pages}{471--492}.
\bibitem[{Dillon et~al.(1995)Dillon, Fauci and
  Gaver~III}]{dillon1995microscale}
\bibinfo{author}{Dillon, R.}, \bibinfo{author}{Fauci, L.},
  \bibinfo{author}{Gaver~III, D.}, \bibinfo{year}{1995}.
\newblock \bibinfo{title}{A microscale model of bacterial swimming, chemotaxis
  and substrate transport}.
\newblock \bibinfo{journal}{Journal of theoretical biology}
  \bibinfo{volume}{177}, \bibinfo{pages}{325--340}.
\bibitem[{Dunne(2006)}]{dunne2006eulerian}
\bibinfo{author}{Dunne, T.}, \bibinfo{year}{2006}.
\newblock \bibinfo{title}{An eulerian approach to fluid--structure interaction
  and goal-oriented mesh adaptation}.
\newblock \bibinfo{journal}{International journal for numerical methods in
  fluids} \bibinfo{volume}{51}, \bibinfo{pages}{1017--1039}.
\bibitem[{Eggleton and Popel(1998)}]{eggleton1998large}
\bibinfo{author}{Eggleton, C.D.}, \bibinfo{author}{Popel, A.S.},
  \bibinfo{year}{1998}.
\newblock \bibinfo{title}{Large deformation of red blood cell ghosts in a
  simple shear flow}.
\newblock \bibinfo{journal}{Physics of fluids} \bibinfo{volume}{10},
  \bibinfo{pages}{1834--1845}.
\bibitem[{Fadlun et~al.(2000)Fadlun, Verzicco, Orlandi and
  Mohd-Yusof}]{fadlun2000combined}
\bibinfo{author}{Fadlun, E.}, \bibinfo{author}{Verzicco, R.},
  \bibinfo{author}{Orlandi, P.}, \bibinfo{author}{Mohd-Yusof, J.},
  \bibinfo{year}{2000}.
\newblock \bibinfo{title}{Combined immersed-boundary finite-difference methods
  for three-dimensional complex flow simulations}.
\newblock \bibinfo{journal}{Journal of computational physics}
  \bibinfo{volume}{161}, \bibinfo{pages}{35--60}.
\bibitem[{Fauci and McDonald(1995)}]{fauci1995sperm}
\bibinfo{author}{Fauci, L.J.}, \bibinfo{author}{McDonald, A.},
  \bibinfo{year}{1995}.
\newblock \bibinfo{title}{Sperm motility in the presence of boundaries}.
\newblock \bibinfo{journal}{Bulletin of mathematical biology}
  \bibinfo{volume}{57}, \bibinfo{pages}{679--699}.
\bibitem[{Fogelson and Guy(2004)}]{fogelson2004platelet}
\bibinfo{author}{Fogelson, A.L.}, \bibinfo{author}{Guy, R.D.},
  \bibinfo{year}{2004}.
\newblock \bibinfo{title}{Platelet--wall interactions in continuum models of
  platelet thrombosis: formulation and numerical solution}.
\newblock \bibinfo{journal}{Mathematical Medicine and Biology}
  \bibinfo{volume}{21}, \bibinfo{pages}{293--334}.
\bibitem[{Francois et~al.(2006)Francois, Cummins, Dendy, Kothe, Sicilian and
  Williams}]{Francois2006}
\bibinfo{author}{Francois, M.M.}, \bibinfo{author}{Cummins, S.J.},
  \bibinfo{author}{Dendy, E.D.}, \bibinfo{author}{Kothe, D.B.},
  \bibinfo{author}{Sicilian, J.M.}, \bibinfo{author}{Williams, M.W.},
  \bibinfo{year}{2006}.
\newblock \bibinfo{title}{{A balanced-force algorithm for continuous and sharp
  interfacial surface tension models within a volume tracking framework}}.
\newblock \bibinfo{journal}{Journal of Computational Physics}
  \bibinfo{volume}{213}, \bibinfo{pages}{141--173}.
\newblock \DOIprefix\doi{10.1016/j.jcp.2005.08.004}.
\bibitem[{Gao and Hu(2009)}]{gao2009deformation}
\bibinfo{author}{Gao, T.}, \bibinfo{author}{Hu, H.H.}, \bibinfo{year}{2009}.
\newblock \bibinfo{title}{Deformation of elastic particles in viscous shear
  flow}.
\newblock \bibinfo{journal}{Journal of Computational Physics}
  \bibinfo{volume}{228}, \bibinfo{pages}{2132--2151}.
\bibitem[{Ghaisas et~al.(2018)Ghaisas, Subramaniam and
  Lele}]{ghaisas2018unified}
\bibinfo{author}{Ghaisas, N.S.}, \bibinfo{author}{Subramaniam, A.},
  \bibinfo{author}{Lele, S.K.}, \bibinfo{year}{2018}.
\newblock \bibinfo{title}{A unified high-order eulerian method for continuum
  simulations of fluid flow and of elastic--plastic deformations in solids}.
\newblock \bibinfo{journal}{Journal of Computational Physics}
  \bibinfo{volume}{371}, \bibinfo{pages}{452--482}.
\bibitem[{Ghia et~al.(1982)Ghia, Ghia and C.T.Shin}]{Ghia1982}
\bibinfo{author}{Ghia, U.}, \bibinfo{author}{Ghia, K.N.},
  \bibinfo{author}{C.T.Shin}, \bibinfo{year}{1982}.
\newblock \bibinfo{title}{{High-Re solutions for incompressible flow using the
  Navier-Stokes equations and a multigrid method}}.
\newblock \bibinfo{journal}{Journal of Computational Physics}
  \bibinfo{volume}{48}, \bibinfo{pages}{387--411}.
\newblock \URLprefix
  \url{http://linkinghub.elsevier.com/retrieve/pii/0021999182900584},
  \DOIprefix\doi{10.1016/0021-9991(82)90058-4}.
\bibitem[{Ghias et~al.(2004)Ghias, Mittal and Lund}]{ghias2004non}
\bibinfo{author}{Ghias, R.}, \bibinfo{author}{Mittal, R.},
  \bibinfo{author}{Lund, T.}, \bibinfo{year}{2004}.
\newblock \bibinfo{title}{A non-body conformal grid method for simulation of
  compressible flows with complex immersed boundaries}, in:
  \bibinfo{booktitle}{42nd AIAA Aerospace Sciences Meeting and Exhibit},
  p.~\bibinfo{pages}{80}.
\bibitem[{Glowinski et~al.(1999)Glowinski, Pan, Hesla, Joseph and
  P{\'e}riaux}]{glowinski1999distributed}
\bibinfo{author}{Glowinski, R.}, \bibinfo{author}{Pan, T.W.},
  \bibinfo{author}{Hesla, T.I.}, \bibinfo{author}{Joseph, D.D.},
  \bibinfo{author}{P{\'e}riaux, J.}, \bibinfo{year}{1999}.
\newblock \bibinfo{title}{A distributed lagrange multiplier/fictitious domain
  method for flows around moving rigid bodies: application to particulate
  flow}.
\newblock \bibinfo{journal}{International Journal for Numerical Methods in
  Fluids} \bibinfo{volume}{30}, \bibinfo{pages}{1043--1066}.
\bibitem[{Glowinski et~al.(2001)Glowinski, Pan, Hesla, Joseph and
  Periaux}]{glowinski2001fictitious}
\bibinfo{author}{Glowinski, R.}, \bibinfo{author}{Pan, T.W.},
  \bibinfo{author}{Hesla, T.I.}, \bibinfo{author}{Joseph, D.D.},
  \bibinfo{author}{Periaux, J.}, \bibinfo{year}{2001}.
\newblock \bibinfo{title}{A fictitious domain approach to the direct numerical
  simulation of incompressible viscous flow past moving rigid bodies:
  application to particulate flow}.
\newblock \bibinfo{journal}{Journal of Computational Physics}
  \bibinfo{volume}{169}, \bibinfo{pages}{363--426}.
\bibitem[{Gong et~al.(2009)Gong, Sugiyama, Takagi and
  Matsumoto}]{gong2009deformation}
\bibinfo{author}{Gong, X.}, \bibinfo{author}{Sugiyama, K.},
  \bibinfo{author}{Takagi, S.}, \bibinfo{author}{Matsumoto, Y.},
  \bibinfo{year}{2009}.
\newblock \bibinfo{title}{The deformation behavior of multiple red blood cells
  in a capillary vessel}.
\newblock \bibinfo{journal}{Journal of biomechanical engineering}
  \bibinfo{volume}{131}, \bibinfo{pages}{074504}.
\bibitem[{Govindjee and Mihalic(1996)}]{govindjee1996computational}
\bibinfo{author}{Govindjee, S.}, \bibinfo{author}{Mihalic, P.A.},
  \bibinfo{year}{1996}.
\newblock \bibinfo{title}{Computational methods for inverse finite
  elastostatics}.
\newblock \bibinfo{journal}{Computer Methods in Applied Mechanics and
  Engineering} \bibinfo{volume}{136}, \bibinfo{pages}{47--57}.
\bibitem[{Griffith(2005)}]{griffith2005simulating}
\bibinfo{author}{Griffith, B.E.}, \bibinfo{year}{2005}.
\newblock \bibinfo{title}{Simulating the blood-muscle-valve mechanics of the
  heart by an adaptive and parallel version of the immersed boundary method}.
\newblock Ph.D. thesis. New York University, Graduate School of Arts and
  Science.
\bibitem[{Grigoriadis et~al.(2009)Grigoriadis, Kassinos and
  Votyakov}]{grigoriadis2009immersed}
\bibinfo{author}{Grigoriadis, D.}, \bibinfo{author}{Kassinos, S.C.},
  \bibinfo{author}{Votyakov, E.}, \bibinfo{year}{2009}.
\newblock \bibinfo{title}{Immersed boundary method for the mhd flows of liquid
  metals}.
\newblock \bibinfo{journal}{Journal of Computational physics}
  \bibinfo{volume}{228}, \bibinfo{pages}{903--920}.
\bibitem[{Guy and Hartenstine(2010)}]{guy2010accuracy}
\bibinfo{author}{Guy, R.D.}, \bibinfo{author}{Hartenstine, D.A.},
  \bibinfo{year}{2010}.
\newblock \bibinfo{title}{On the accuracy of direct forcing immersed boundary
  methods with projection methods}.
\newblock \bibinfo{journal}{Journal of Computational Physics}
  \bibinfo{volume}{229}, \bibinfo{pages}{2479--2496}.
\bibitem[{Hirt et~al.(1974)Hirt, Amsden and Cook}]{hirt1974arbitrary}
\bibinfo{author}{Hirt, C.}, \bibinfo{author}{Amsden, A.A.},
  \bibinfo{author}{Cook, J.}, \bibinfo{year}{1974}.
\newblock \bibinfo{title}{An arbitrary lagrangian-eulerian computing method for
  all flow speeds}.
\newblock \bibinfo{journal}{Journal of computational physics}
  \bibinfo{volume}{14}, \bibinfo{pages}{227--253}.
\bibitem[{Holzapfel(2000)}]{holzapfelnonlinear}
\bibinfo{author}{Holzapfel, G.}, \bibinfo{year}{2000}.
\newblock \bibinfo{title}{Nonlinear solid mechanics: a continuum approach for
  engineering. 2000}.
\newblock \bibinfo{publisher}{West Sussex, England: John Wiley \& Sons, Ltd}.
\bibitem[{van Hoogstraten et~al.(1994)van Hoogstraten, Slaats and
  Baaijens}]{van1994eulerian}
\bibinfo{author}{van Hoogstraten, P.A.}, \bibinfo{author}{Slaats, P.M.},
  \bibinfo{author}{Baaijens, F.P.}, \bibinfo{year}{1994}.
\newblock \bibinfo{title}{A eulerian approach to the finite element modelling
  of neo-hookean rubber material}.
\newblock \bibinfo{journal}{Applied Scientific Research} \bibinfo{volume}{48},
  \bibinfo{pages}{193--210}.
\bibitem[{Hou et~al.(2012)Hou, Wang and Layton}]{hou2012numerical}
\bibinfo{author}{Hou, G.}, \bibinfo{author}{Wang, J.}, \bibinfo{author}{Layton,
  A.}, \bibinfo{year}{2012}.
\newblock \bibinfo{title}{Numerical methods for fluid-structure interaction—a
  review}.
\newblock \bibinfo{journal}{Communications in Computational Physics}
  \bibinfo{volume}{12}, \bibinfo{pages}{337--377}.
\bibitem[{Hu(1996)}]{hu1996direct}
\bibinfo{author}{Hu, H.H.}, \bibinfo{year}{1996}.
\newblock \bibinfo{title}{Direct simulation of flows of solid-liquid mixtures}.
\newblock \bibinfo{journal}{International Journal of Multiphase Flow}
  \bibinfo{volume}{22}, \bibinfo{pages}{335--352}.
\bibitem[{Hu et~al.(2001)Hu, Patankar and Zhu}]{Hu2001}
\bibinfo{author}{Hu, H.H.}, \bibinfo{author}{Patankar, N.},
  \bibinfo{author}{Zhu, M.}, \bibinfo{year}{2001}.
\newblock \bibinfo{title}{Direct numerical simulations of fluid–solid systems
  using the arbitrary lagrangian–eulerian technique}.
\newblock \bibinfo{journal}{Journal of Computational Physics}
  \bibinfo{volume}{169}, \bibinfo{pages}{427 -- 462}.
\newblock \URLprefix
  \url{http://www.sciencedirect.com/science/article/pii/S0021999100965926},
  \DOIprefix\doi{https://doi.org/10.1006/jcph.2000.6592}.
\bibitem[{Huang and Sung(2009)}]{huang2009immersed}
\bibinfo{author}{Huang, W.X.}, \bibinfo{author}{Sung, H.J.},
  \bibinfo{year}{2009}.
\newblock \bibinfo{title}{An immersed boundary method for fluid--flexible
  structure interaction}.
\newblock \bibinfo{journal}{Computer Methods in Applied Mechanics and
  Engineering} \bibinfo{volume}{198}, \bibinfo{pages}{2650--2661}.
\bibitem[{H{\"u}bner et~al.(2004)H{\"u}bner, Walhorn and
  Dinkler}]{hubner2004monolithic}
\bibinfo{author}{H{\"u}bner, B.}, \bibinfo{author}{Walhorn, E.},
  \bibinfo{author}{Dinkler, D.}, \bibinfo{year}{2004}.
\newblock \bibinfo{title}{A monolithic approach to fluid--structure interaction
  using space--time finite elements}.
\newblock \bibinfo{journal}{Computer methods in applied mechanics and
  engineering} \bibinfo{volume}{193}, \bibinfo{pages}{2087--2104}.
\bibitem[{Hughes et~al.(1981)Hughes, Liu and Zimmermann}]{hughes1981lagrangian}
\bibinfo{author}{Hughes, T.J.}, \bibinfo{author}{Liu, W.K.},
  \bibinfo{author}{Zimmermann, T.K.}, \bibinfo{year}{1981}.
\newblock \bibinfo{title}{Lagrangian-eulerian finite element formulation for
  incompressible viscous flows}.
\newblock \bibinfo{journal}{Computer methods in applied mechanics and
  engineering} \bibinfo{volume}{29}, \bibinfo{pages}{329--349}.
\bibitem[{Hughes and Stewart(1996)}]{hughes1996space}
\bibinfo{author}{Hughes, T.J.}, \bibinfo{author}{Stewart, J.R.},
  \bibinfo{year}{1996}.
\newblock \bibinfo{title}{A space-time formulation for multiscale phenomena}.
\newblock \bibinfo{journal}{Journal of Computational and Applied Mathematics}
  \bibinfo{volume}{74}, \bibinfo{pages}{217--229}.
\bibitem[{Iaccarino et~al.(2003)Iaccarino, Kalitzin and
  Elkins}]{iaccarino2003numerical}
\bibinfo{author}{Iaccarino, G.}, \bibinfo{author}{Kalitzin, G.},
  \bibinfo{author}{Elkins, C.J.}, \bibinfo{year}{2003}.
\newblock \bibinfo{title}{Numerical and experimental investigation of the
  turbulent flow in a ribbed serpentine passage}.
\newblock \bibinfo{type}{Technical Report}. STANFORD UNIV CA DEPT OF MECHANICAL
  ENGINEERING.
\bibitem[{Ii et~al.(2012)Ii, Gong, Sugiyama, Wu, Huang and Takagi}]{ii2012full}
\bibinfo{author}{Ii, S.}, \bibinfo{author}{Gong, X.},
  \bibinfo{author}{Sugiyama, K.}, \bibinfo{author}{Wu, J.},
  \bibinfo{author}{Huang, H.}, \bibinfo{author}{Takagi, S.},
  \bibinfo{year}{2012}.
\newblock \bibinfo{title}{A full eulerian fluid-membrane coupling method with a
  smoothed volume-of-fluid approach}.
\newblock \bibinfo{journal}{Communications in Computational Physics}
  \bibinfo{volume}{12}, \bibinfo{pages}{544--576}.
\bibitem[{Ii et~al.(2011)Ii, Sugiyama, Takeuchi, Takagi and
  Matsumoto}]{ii2011implicit}
\bibinfo{author}{Ii, S.}, \bibinfo{author}{Sugiyama, K.},
  \bibinfo{author}{Takeuchi, S.}, \bibinfo{author}{Takagi, S.},
  \bibinfo{author}{Matsumoto, Y.}, \bibinfo{year}{2011}.
\newblock \bibinfo{title}{An implicit full eulerian method for the
  fluid--structure interaction problem}.
\newblock \bibinfo{journal}{International Journal for Numerical Methods in
  Fluids} \bibinfo{volume}{65}, \bibinfo{pages}{150--165}.
\bibitem[{Jain and Mani(2017)}]{Jain2017}
\bibinfo{author}{Jain, S.S.}, \bibinfo{author}{Mani, A.}, \bibinfo{year}{2017}.
\newblock \bibinfo{title}{An incompressible eulerian formulation for soft
  solids in fluids}.
\newblock \bibinfo{journal}{Center for Turbulence Research, Annual Research
  briefs} , \bibinfo{pages}{349--362}.
\bibitem[{Johnson and Tezduyar(2001)}]{johnson2001methods}
\bibinfo{author}{Johnson, A.}, \bibinfo{author}{Tezduyar, T.},
  \bibinfo{year}{2001}.
\newblock \bibinfo{title}{Methods for 3d computation of fluid--object
  interactions in spatially periodic flows}.
\newblock \bibinfo{journal}{Computer Methods in Applied Mechanics and
  Engineering} \bibinfo{volume}{190}, \bibinfo{pages}{3201--3221}.
\bibitem[{Johnson and Tezduyar(1996)}]{johnson1996simulation}
\bibinfo{author}{Johnson, A.A.}, \bibinfo{author}{Tezduyar, T.E.},
  \bibinfo{year}{1996}.
\newblock \bibinfo{title}{Simulation of multiple spheres falling in a
  liquid-filled tube}.
\newblock \bibinfo{journal}{Computer Methods in Applied Mechanics and
  Engineering} \bibinfo{volume}{134}, \bibinfo{pages}{351--373}.
\bibitem[{Johnson and Tezduyar(1997a)}]{johnson19973d}
\bibinfo{author}{Johnson, A.A.}, \bibinfo{author}{Tezduyar, T.E.},
  \bibinfo{year}{1997}a.
\newblock \bibinfo{title}{3d simulation of fluid-particle interactions with the
  number of particles reaching 100}.
\newblock \bibinfo{journal}{Computer Methods in Applied Mechanics and
  Engineering} \bibinfo{volume}{145}, \bibinfo{pages}{301--321}.
\bibitem[{Johnson and Tezduyar(1997b)}]{johnson1997parallel}
\bibinfo{author}{Johnson, A.A.}, \bibinfo{author}{Tezduyar, T.E.},
  \bibinfo{year}{1997}b.
\newblock \bibinfo{title}{Parallel computation of incompressible flows with
  complex geometries}.
\newblock \bibinfo{journal}{International Journal for Numerical Methods in
  Fluids} \bibinfo{volume}{24}, \bibinfo{pages}{1321--1340}.
\bibitem[{Johnson and Tezduyar(1999)}]{johnson1999advanced}
\bibinfo{author}{Johnson, A.A.}, \bibinfo{author}{Tezduyar, T.E.},
  \bibinfo{year}{1999}.
\newblock \bibinfo{title}{Advanced mesh generation and update methods for 3d
  flow simulations}.
\newblock \bibinfo{journal}{Computational Mechanics} \bibinfo{volume}{23},
  \bibinfo{pages}{130--143}.
\bibitem[{Kalro and Tezduyar(2000)}]{kalro2000parallel}
\bibinfo{author}{Kalro, V.}, \bibinfo{author}{Tezduyar, T.E.},
  \bibinfo{year}{2000}.
\newblock \bibinfo{title}{A parallel 3d computational method for
  fluid--structure interactions in parachute systems}.
\newblock \bibinfo{journal}{Computer Methods in Applied Mechanics and
  Engineering} \bibinfo{volume}{190}, \bibinfo{pages}{321--332}.
\bibitem[{Kamrin et~al.(2012)Kamrin, Rycroft and Nave}]{Kamrin2012}
\bibinfo{author}{Kamrin, K.}, \bibinfo{author}{Rycroft, C.H.},
  \bibinfo{author}{Nave, J.C.}, \bibinfo{year}{2012}.
\newblock \bibinfo{title}{{Reference map technique for finite-strain elasticity
  and fluid-solid interaction}}.
\newblock \bibinfo{journal}{Journal of the Mechanics and Physics of Solids}
  \bibinfo{volume}{60}, \bibinfo{pages}{1952--1969}.
\newblock \URLprefix \url{http://dx.doi.org/10.1016/j.jmps.2012.06.003},
  \DOIprefix\doi{10.1016/j.jmps.2012.06.003}.
\bibitem[{Kataoka(1986)}]{kataoka1986local}
\bibinfo{author}{Kataoka, I.}, \bibinfo{year}{1986}.
\newblock \bibinfo{title}{Local instant formulation of two-phase flow}.
\newblock \bibinfo{journal}{International Journal of Multiphase Flow}
  \bibinfo{volume}{12}, \bibinfo{pages}{745--758}.
\bibitem[{Kim and Choi(2006)}]{kim2006immersed}
\bibinfo{author}{Kim, D.}, \bibinfo{author}{Choi, H.}, \bibinfo{year}{2006}.
\newblock \bibinfo{title}{Immersed boundary method for flow around an
  arbitrarily moving body}.
\newblock \bibinfo{journal}{Journal of Computational Physics}
  \bibinfo{volume}{212}, \bibinfo{pages}{662--680}.
\bibitem[{Kim et~al.(2001)Kim, Kim and Choi}]{kim2001immersed}
\bibinfo{author}{Kim, J.}, \bibinfo{author}{Kim, D.}, \bibinfo{author}{Choi,
  H.}, \bibinfo{year}{2001}.
\newblock \bibinfo{title}{An immersed-boundary finite-volume method for
  simulations of flow in complex geometries}.
\newblock \bibinfo{journal}{Journal of Computational Physics}
  \bibinfo{volume}{171}, \bibinfo{pages}{132--150}.
\bibitem[{Kim and Peskin(2007)}]{kim2007penalty}
\bibinfo{author}{Kim, Y.}, \bibinfo{author}{Peskin, C.S.},
  \bibinfo{year}{2007}.
\newblock \bibinfo{title}{Penalty immersed boundary method for an elastic
  boundary with mass}.
\newblock \bibinfo{journal}{Physics of Fluids} \bibinfo{volume}{19},
  \bibinfo{pages}{053103}.
\bibitem[{Laney(1998)}]{laney1998computational}
\bibinfo{author}{Laney, C.B.}, \bibinfo{year}{1998}.
\newblock \bibinfo{title}{Computational gasdynamics}.
\newblock \bibinfo{publisher}{Cambridge university press}.
\bibitem[{Layton(2009)}]{layton2009using}
\bibinfo{author}{Layton, A.T.}, \bibinfo{year}{2009}.
\newblock \bibinfo{title}{Using integral equations and the immersed interface
  method to solve immersed boundary problems with stiff forces}.
\newblock \bibinfo{journal}{Computers \& Fluids} \bibinfo{volume}{38},
  \bibinfo{pages}{266--272}.
\bibitem[{Le et~al.(2008)Le, Khoo and Lim}]{le2008implicit}
\bibinfo{author}{Le, D.}, \bibinfo{author}{Khoo, B.}, \bibinfo{author}{Lim,
  K.}, \bibinfo{year}{2008}.
\newblock \bibinfo{title}{An implicit-forcing immersed boundary method for
  simulating viscous flows in irregular domains}.
\newblock \bibinfo{journal}{Computer methods in applied mechanics and
  engineering} \bibinfo{volume}{197}, \bibinfo{pages}{2119--2130}.
\bibitem[{LeVeque and Li(1994)}]{Leveque1994}
\bibinfo{author}{LeVeque, R.J.}, \bibinfo{author}{Li, Z.},
  \bibinfo{year}{1994}.
\newblock \bibinfo{title}{The immersed interface method for elliptic equations
  with discontinuous coefficients and singular sources}.
\newblock \bibinfo{journal}{SIAM Journal on Numerical Analysis}
  \bibinfo{volume}{31}, \bibinfo{pages}{1019--1044}.
\newblock \URLprefix \url{https://doi.org/10.1137/0731054},
  \DOIprefix\doi{10.1137/0731054},
  \href{http://arxiv.org/abs/https://doi.org/10.1137/0731054}{\tt
  arXiv:https://doi.org/10.1137/0731054}.
\bibitem[{Li and Ito(2006)}]{li2006immersed}
\bibinfo{author}{Li, Z.}, \bibinfo{author}{Ito, K.}, \bibinfo{year}{2006}.
\newblock \bibinfo{title}{The immersed interface method: numerical solutions of
  PDEs involving interfaces and irregular domains}.
  volume~\bibinfo{volume}{33}.
\newblock \bibinfo{publisher}{Siam}.
\bibitem[{Li and Lai(2001)}]{li2001immersed}
\bibinfo{author}{Li, Z.}, \bibinfo{author}{Lai, M.C.}, \bibinfo{year}{2001}.
\newblock \bibinfo{title}{The immersed interface method for the navier--stokes
  equations with singular forces}.
\newblock \bibinfo{journal}{Journal of Computational Physics}
  \bibinfo{volume}{171}, \bibinfo{pages}{822--842}.
\bibitem[{Li et~al.(2003)}]{li2003overview}
\bibinfo{author}{Li, Z.}, et~al., \bibinfo{year}{2003}.
\newblock \bibinfo{title}{An overview of the immersed interface method and its
  applications}.
\newblock \bibinfo{journal}{Taiwanese journal of mathematics}
  \bibinfo{volume}{7}, \bibinfo{pages}{1--49}.
\bibitem[{Liu and Walkington(2001)}]{liu2001eulerian}
\bibinfo{author}{Liu, C.}, \bibinfo{author}{Walkington, N.J.},
  \bibinfo{year}{2001}.
\newblock \bibinfo{title}{An eulerian description of fluids containing
  visco-elastic particles}.
\newblock \bibinfo{journal}{Archive for rational mechanics and analysis}
  \bibinfo{volume}{159}, \bibinfo{pages}{229--252}.
\bibitem[{Liu et~al.(2006)Liu, Liu, Farrell, Zhang, Wang, Fukui, Patankar,
  Zhang, Bajaj, Lee et~al.}]{liu2006immersed}
\bibinfo{author}{Liu, W.K.}, \bibinfo{author}{Liu, Y.},
  \bibinfo{author}{Farrell, D.}, \bibinfo{author}{Zhang, L.},
  \bibinfo{author}{Wang, X.S.}, \bibinfo{author}{Fukui, Y.},
  \bibinfo{author}{Patankar, N.}, \bibinfo{author}{Zhang, Y.},
  \bibinfo{author}{Bajaj, C.}, \bibinfo{author}{Lee, J.}, et~al.,
  \bibinfo{year}{2006}.
\newblock \bibinfo{title}{Immersed finite element method and its applications
  to biological systems}.
\newblock \bibinfo{journal}{Computer methods in applied mechanics and
  engineering} \bibinfo{volume}{195}, \bibinfo{pages}{1722--1749}.
\bibitem[{Liu et~al.(2007)Liu, Tang et~al.}]{liu2007mathematical}
\bibinfo{author}{Liu, W.K.}, \bibinfo{author}{Tang, S.}, et~al.,
  \bibinfo{year}{2007}.
\newblock \bibinfo{title}{Mathematical foundations of the immersed finite
  element method}.
\newblock \bibinfo{journal}{Computational Mechanics} \bibinfo{volume}{39},
  \bibinfo{pages}{211--222}.
\bibitem[{Luo et~al.(2007)Luo, Wang and Fan}]{luo2007modified}
\bibinfo{author}{Luo, K.}, \bibinfo{author}{Wang, Z.}, \bibinfo{author}{Fan,
  J.}, \bibinfo{year}{2007}.
\newblock \bibinfo{title}{A modified immersed boundary method for simulations
  of fluid--particle interactions}.
\newblock \bibinfo{journal}{Computer methods in applied mechanics and
  engineering} \bibinfo{volume}{197}, \bibinfo{pages}{36--46}.
\bibitem[{Maitre et~al.(2009)Maitre, Milcent, Cottet, Raoult and
  Usson}]{maitre2009applications}
\bibinfo{author}{Maitre, E.}, \bibinfo{author}{Milcent, T.},
  \bibinfo{author}{Cottet, G.H.}, \bibinfo{author}{Raoult, A.},
  \bibinfo{author}{Usson, Y.}, \bibinfo{year}{2009}.
\newblock \bibinfo{title}{Applications of level set methods in computational
  biophysics}.
\newblock \bibinfo{journal}{Mathematical and Computer Modelling}
  \bibinfo{volume}{49}, \bibinfo{pages}{2161--2169}.
\bibitem[{Mark and van Wachem(2008)}]{mark2008derivation}
\bibinfo{author}{Mark, A.}, \bibinfo{author}{van Wachem, B.G.},
  \bibinfo{year}{2008}.
\newblock \bibinfo{title}{Derivation and validation of a novel implicit
  second-order accurate immersed boundary method}.
\newblock \bibinfo{journal}{Journal of Computational Physics}
  \bibinfo{volume}{227}, \bibinfo{pages}{6660--6680}.
\bibitem[{Michler et~al.(2004)Michler, Hulshoff, Van~Brummelen and
  De~Borst}]{michler2004monolithic}
\bibinfo{author}{Michler, C.}, \bibinfo{author}{Hulshoff, S.},
  \bibinfo{author}{Van~Brummelen, E.}, \bibinfo{author}{De~Borst, R.},
  \bibinfo{year}{2004}.
\newblock \bibinfo{title}{A monolithic approach to fluid--structure
  interaction}.
\newblock \bibinfo{journal}{Computers \& fluids} \bibinfo{volume}{33},
  \bibinfo{pages}{839--848}.
\bibitem[{Miller and Colella(2001)}]{Miller2001}
\bibinfo{author}{Miller, G.H.}, \bibinfo{author}{Colella, P.},
  \bibinfo{year}{2001}.
\newblock \bibinfo{title}{A high-order eulerian godunov method for
  elastic-plastic flow in solids}.
\newblock \bibinfo{journal}{J. Comput. Phys.} \bibinfo{volume}{167},
  \bibinfo{pages}{131--176}.
\newblock \URLprefix \url{http://dx.doi.org/10.1006/jcph.2000.6665},
  \DOIprefix\doi{10.1006/jcph.2000.6665}.
\bibitem[{Mirjalili et~al.(2017)Mirjalili, Jain and Dodd}]{Mirjalili2017}
\bibinfo{author}{Mirjalili, S.}, \bibinfo{author}{Jain, S.S.},
  \bibinfo{author}{Dodd, M.}, \bibinfo{year}{2017}.
\newblock \bibinfo{title}{Interface-capturing methods for two-phase flows: An
  overview and recent developments}.
\newblock \bibinfo{journal}{Center for Turbulence Research, Annual Research
  Briefs} , \bibinfo{pages}{117--135}.
\bibitem[{Mittal et~al.(2003)Mittal, Bonilla and
  Udaykumar}]{mittal2003cartesian}
\bibinfo{author}{Mittal, R.}, \bibinfo{author}{Bonilla, C.},
  \bibinfo{author}{Udaykumar, H.}, \bibinfo{year}{2003}.
\newblock \bibinfo{title}{Cartesian grid methods for simulating flows with
  moving boundaries}.
\newblock \bibinfo{journal}{Computational methods and experimental
  measurements-XI} , \bibinfo{pages}{557--566}.
\bibitem[{Mittal and Iaccarino(2005)}]{mittal2005immersed}
\bibinfo{author}{Mittal, R.}, \bibinfo{author}{Iaccarino, G.},
  \bibinfo{year}{2005}.
\newblock \bibinfo{title}{Immersed boundary methods}.
\newblock \bibinfo{journal}{Annu. Rev. Fluid Mech.} \bibinfo{volume}{37},
  \bibinfo{pages}{239--261}.
\bibitem[{Mittal et~al.(2004)Mittal, Seshadri and
  Udaykumar}]{mittal2004flutter}
\bibinfo{author}{Mittal, R.}, \bibinfo{author}{Seshadri, V.},
  \bibinfo{author}{Udaykumar, H.S.}, \bibinfo{year}{2004}.
\newblock \bibinfo{title}{Flutter, tumble and vortex induced autorotation}.
\newblock \bibinfo{journal}{Theoretical and Computational Fluid Dynamics}
  \bibinfo{volume}{17}, \bibinfo{pages}{165--170}.
\bibitem[{Mittal et~al.(2002)Mittal, Utturkar and
  Udaykumar}]{mittal2002computational}
\bibinfo{author}{Mittal, R.}, \bibinfo{author}{Utturkar, Y.},
  \bibinfo{author}{Udaykumar, H.}, \bibinfo{year}{2002}.
\newblock \bibinfo{title}{Computational modeling and analysis of biomimetic
  flight mechanisms}, in: \bibinfo{booktitle}{40th AIAA Aerospace Sciences
  Meeting \& Exhibit}, p. \bibinfo{pages}{865}.
\bibitem[{Mittal and Tezduyar(1994)}]{mittal1994massively}
\bibinfo{author}{Mittal, S.}, \bibinfo{author}{Tezduyar, T.E.},
  \bibinfo{year}{1994}.
\newblock \bibinfo{title}{Massively parallel finite element computation of
  incompressible flows involving fluid-body interactions}.
\newblock \bibinfo{journal}{Computer Methods in Applied Mechanics and
  Engineering} \bibinfo{volume}{112}, \bibinfo{pages}{253--282}.
\bibitem[{Mittal and Tezduyar(1995)}]{mittal1995parallel}
\bibinfo{author}{Mittal, S.}, \bibinfo{author}{Tezduyar, T.E.},
  \bibinfo{year}{1995}.
\newblock \bibinfo{title}{Parallel finite element simulation of 3d
  incompressible flows: Fluid-structure interactions}.
\newblock \bibinfo{journal}{International Journal for Numerical Methods in
  Fluids} \bibinfo{volume}{21}, \bibinfo{pages}{933--953}.
\bibitem[{Mohd-Yusof(1997)}]{mohd1997simulations}
\bibinfo{author}{Mohd-Yusof, J.}, \bibinfo{year}{1997}.
\newblock \bibinfo{title}{For simulations of flow in complex geometries}.
\newblock \bibinfo{journal}{Annual Research Briefs} \bibinfo{volume}{317}.
\bibitem[{Mori and Peskin(2008)}]{mori2008implicit}
\bibinfo{author}{Mori, Y.}, \bibinfo{author}{Peskin, C.S.},
  \bibinfo{year}{2008}.
\newblock \bibinfo{title}{Implicit second-order immersed boundary methods with
  boundary mass}.
\newblock \bibinfo{journal}{Computer methods in applied mechanics and
  engineering} \bibinfo{volume}{197}, \bibinfo{pages}{2049--2067}.
\bibitem[{Morinishi et~al.(1998)Morinishi, Lund, Vasilyev and
  Moin}]{morinishi1998fully}
\bibinfo{author}{Morinishi, Y.}, \bibinfo{author}{Lund, T.S.},
  \bibinfo{author}{Vasilyev, O.V.}, \bibinfo{author}{Moin, P.},
  \bibinfo{year}{1998}.
\newblock \bibinfo{title}{Fully conservative higher order finite difference
  schemes for incompressible flow}.
\newblock \bibinfo{journal}{Journal of computational physics}
  \bibinfo{volume}{143}, \bibinfo{pages}{90--124}.
\bibitem[{Nagano et~al.(2010)Nagano, Sugiyama, Takeuchi, II, Takagi and
  Matsumoto}]{nagano2010full}
\bibinfo{author}{Nagano, N.}, \bibinfo{author}{Sugiyama, K.},
  \bibinfo{author}{Takeuchi, S.}, \bibinfo{author}{II, S.},
  \bibinfo{author}{Takagi, S.}, \bibinfo{author}{Matsumoto, Y.},
  \bibinfo{year}{2010}.
\newblock \bibinfo{title}{Full-eulerian finite-difference simulation of fluid
  flow in hyperelastic wavy channel}.
\newblock \bibinfo{journal}{Journal of Fluid Science and Technology}
  \bibinfo{volume}{5}, \bibinfo{pages}{475--490}.
\bibitem[{Neumann et~al.(1982)Neumann, Schaefer-Ridder, Wang and
  Hofschneider}]{neumann1982gene}
\bibinfo{author}{Neumann, E.}, \bibinfo{author}{Schaefer-Ridder, M.},
  \bibinfo{author}{Wang, Y.}, \bibinfo{author}{Hofschneider, P.},
  \bibinfo{year}{1982}.
\newblock \bibinfo{title}{Gene transfer into mouse lyoma cells by
  electroporation in high electric fields.}
\newblock \bibinfo{journal}{The EMBO journal} \bibinfo{volume}{1},
  \bibinfo{pages}{841--845}.
\bibitem[{Nitikitpaiboon and Bathe(1993)}]{nitikitpaiboon1993arbitrary}
\bibinfo{author}{Nitikitpaiboon, C.}, \bibinfo{author}{Bathe, K.},
  \bibinfo{year}{1993}.
\newblock \bibinfo{title}{An arbitrary lagrangian-eulerian velocity potential
  formulation for fluid-structure interaction}.
\newblock \bibinfo{journal}{Computers \& structures} \bibinfo{volume}{47},
  \bibinfo{pages}{871--891}.
\bibitem[{Okazawa et~al.(2007)Okazawa, Kashiyama and
  Kaneko}]{okazawa2007eulerian}
\bibinfo{author}{Okazawa, S.}, \bibinfo{author}{Kashiyama, K.},
  \bibinfo{author}{Kaneko, Y.}, \bibinfo{year}{2007}.
\newblock \bibinfo{title}{Eulerian formulation using stabilized finite element
  method for large deformation solid dynamics}.
\newblock \bibinfo{journal}{International Journal for Numerical Methods in
  Engineering} \bibinfo{volume}{72}, \bibinfo{pages}{1544--1559}.
\bibitem[{Onate et~al.(2008)Onate, Idelsohn, Celigueta and
  Rossi}]{onate2008advances}
\bibinfo{author}{Onate, E.}, \bibinfo{author}{Idelsohn, S.R.},
  \bibinfo{author}{Celigueta, M.A.}, \bibinfo{author}{Rossi, R.},
  \bibinfo{year}{2008}.
\newblock \bibinfo{title}{Advances in the particle finite element method for
  the analysis of fluid--multibody interaction and bed erosion in free surface
  flows}.
\newblock \bibinfo{journal}{Computer methods in applied mechanics and
  engineering} \bibinfo{volume}{197}, \bibinfo{pages}{1777--1800}.
\bibitem[{Patankar(2001)}]{patankar2001formulation}
\bibinfo{author}{Patankar, N.}, \bibinfo{year}{2001}.
\newblock \bibinfo{title}{A formulation for fast computations of rigid
  particulate flows}.
\newblock \bibinfo{journal}{Center for Turbulence Research Annual Research
  Briefs} \bibinfo{volume}{2001}, \bibinfo{pages}{185--196}.
\bibitem[{Peskin(1972)}]{peskin1972flow}
\bibinfo{author}{Peskin, C.S.}, \bibinfo{year}{1972}.
\newblock \bibinfo{title}{Flow patterns around heart valves: a numerical
  method}.
\newblock \bibinfo{journal}{Journal of computational physics}
  \bibinfo{volume}{10}, \bibinfo{pages}{252--271}.
\bibitem[{Peskin(1982a)}]{Peskin1982}
\bibinfo{author}{Peskin, C.S.}, \bibinfo{year}{1982}a.
\newblock \bibinfo{title}{The fluid dynamics of heart valves: Experimental,
  theoretical, and computational methods}.
\newblock \bibinfo{journal}{Annual Review of Fluid Mechanics}
  \bibinfo{volume}{14}, \bibinfo{pages}{235--259}.
\newblock \URLprefix \url{https://doi.org/10.1146/annurev.fl.14.010182.001315},
  \DOIprefix\doi{10.1146/annurev.fl.14.010182.001315},
  \href{http://arxiv.org/abs/https://doi.org/10.1146/annurev.fl.14.010182.001315}{\tt
  arXiv:https://doi.org/10.1146/annurev.fl.14.010182.001315}.
\bibitem[{Peskin(1982b)}]{peskin1982fluid}
\bibinfo{author}{Peskin, C.S.}, \bibinfo{year}{1982}b.
\newblock \bibinfo{title}{The fluid dynamics of heart valves: experimental,
  theoretical, and computational methods}.
\newblock \bibinfo{journal}{Annual review of fluid mechanics}
  \bibinfo{volume}{14}, \bibinfo{pages}{235--259}.
\bibitem[{Peskin(2002)}]{peskin2002immersed}
\bibinfo{author}{Peskin, C.S.}, \bibinfo{year}{2002}.
\newblock \bibinfo{title}{The immersed boundary method}.
\newblock \bibinfo{journal}{Acta numerica} \bibinfo{volume}{11},
  \bibinfo{pages}{479--517}.
\bibitem[{Pozrikidis(2003)}]{pozrikidis2003modeling}
\bibinfo{author}{Pozrikidis, C.}, \bibinfo{year}{2003}.
\newblock \bibinfo{title}{Modeling and simulation of capsules and biological
  cells}.
\newblock \bibinfo{publisher}{CRC Press}.
\bibitem[{Pozrikidis(2010)}]{pozrikidis2010computational}
\bibinfo{author}{Pozrikidis, C.}, \bibinfo{year}{2010}.
\newblock \bibinfo{title}{Computational hydrodynamics of capsules and
  biological cells}.
\newblock \bibinfo{publisher}{CRC press}.
\bibitem[{Raessi and Pitsch(2012)}]{raessi2012consistent}
\bibinfo{author}{Raessi, M.}, \bibinfo{author}{Pitsch, H.},
  \bibinfo{year}{2012}.
\newblock \bibinfo{title}{Consistent mass and momentum transport for simulating
  incompressible interfacial flows with large density ratios using the level
  set method}.
\newblock \bibinfo{journal}{Computers \& Fluids} \bibinfo{volume}{63},
  \bibinfo{pages}{70--81}.
\bibitem[{Robinson-Mosher et~al.(2011)Robinson-Mosher, Schroeder and
  Fedkiw}]{robinson2011symmetric}
\bibinfo{author}{Robinson-Mosher, A.}, \bibinfo{author}{Schroeder, C.},
  \bibinfo{author}{Fedkiw, R.}, \bibinfo{year}{2011}.
\newblock \bibinfo{title}{A symmetric positive definite formulation for
  monolithic fluid structure interaction}.
\newblock \bibinfo{journal}{Journal of Computational Physics}
  \bibinfo{volume}{230}, \bibinfo{pages}{1547--1566}.
\bibitem[{Rosti and Brandt(2017)}]{rosti2017numerical}
\bibinfo{author}{Rosti, M.E.}, \bibinfo{author}{Brandt, L.},
  \bibinfo{year}{2017}.
\newblock \bibinfo{title}{Numerical simulation of turbulent channel flow over a
  viscous hyper-elastic wall}.
\newblock \bibinfo{journal}{Journal of Fluid Mechanics} \bibinfo{volume}{830},
  \bibinfo{pages}{708--735}.
\bibitem[{Rycroft et~al.(2018)Rycroft, Wu, Yu and
  Kamrin}]{rycroft2018reference}
\bibinfo{author}{Rycroft, C.H.}, \bibinfo{author}{Wu, C.H.},
  \bibinfo{author}{Yu, Y.}, \bibinfo{author}{Kamrin, K.}, \bibinfo{year}{2018}.
\newblock \bibinfo{title}{Reference map technique for incompressible
  fluid-structure interaction}.
\newblock \bibinfo{journal}{arXiv preprint arXiv:1810.03015} .
\bibitem[{Ryzhakov et~al.(2010)Ryzhakov, Rossi, Idelsohn and
  O{\~n}ate}]{ryzhakov2010monolithic}
\bibinfo{author}{Ryzhakov, P.}, \bibinfo{author}{Rossi, R.},
  \bibinfo{author}{Idelsohn, S.}, \bibinfo{author}{O{\~n}ate, E.},
  \bibinfo{year}{2010}.
\newblock \bibinfo{title}{A monolithic lagrangian approach for fluid--structure
  interaction problems}.
\newblock \bibinfo{journal}{Computational mechanics} \bibinfo{volume}{46},
  \bibinfo{pages}{883--899}.
\bibitem[{Stein et~al.(2000)Stein, Benney, Kalro, Tezduyar, Leonard and
  Accorsi}]{stein2000parachute}
\bibinfo{author}{Stein, K.}, \bibinfo{author}{Benney, R.},
  \bibinfo{author}{Kalro, V.}, \bibinfo{author}{Tezduyar, T.E.},
  \bibinfo{author}{Leonard, J.}, \bibinfo{author}{Accorsi, M.},
  \bibinfo{year}{2000}.
\newblock \bibinfo{title}{Parachute fluid--structure interactions: 3-d
  computation}.
\newblock \bibinfo{journal}{Computer Methods in Applied Mechanics and
  Engineering} \bibinfo{volume}{190}, \bibinfo{pages}{373--386}.
\bibitem[{Stein et~al.(2001)Stein, Benney, Tezduyar and
  Potvin}]{stein2001fluid}
\bibinfo{author}{Stein, K.}, \bibinfo{author}{Benney, R.},
  \bibinfo{author}{Tezduyar, T.}, \bibinfo{author}{Potvin, J.},
  \bibinfo{year}{2001}.
\newblock \bibinfo{title}{Fluid--structure interactions of a cross parachute:
  numerical simulation}.
\newblock \bibinfo{journal}{Computer Methods in Applied Mechanics and
  Engineering} \bibinfo{volume}{191}, \bibinfo{pages}{673--687}.
\bibitem[{Stockie and Green(1998)}]{stockie1998simulating}
\bibinfo{author}{Stockie, J.M.}, \bibinfo{author}{Green, S.I.},
  \bibinfo{year}{1998}.
\newblock \bibinfo{title}{Simulating the motion of flexible pulp fibres using
  the immersed boundary method}.
\newblock \bibinfo{journal}{Journal of Computational Physics}
  \bibinfo{volume}{147}, \bibinfo{pages}{147--165}.
\bibitem[{Sugiyama et~al.(2017)Sugiyama, Ii, Shimizu, Noda and
  Takagi}]{sugiyama2017full}
\bibinfo{author}{Sugiyama, K.}, \bibinfo{author}{Ii, S.},
  \bibinfo{author}{Shimizu, K.}, \bibinfo{author}{Noda, S.},
  \bibinfo{author}{Takagi, S.}, \bibinfo{year}{2017}.
\newblock \bibinfo{title}{A full eulerian method for fluid-structure
  interaction problems}.
\newblock \bibinfo{journal}{Procedia Iutam} \bibinfo{volume}{20},
  \bibinfo{pages}{159--166}.
\bibitem[{Sugiyama et~al.(2010)Sugiyama, Ii, Takeuchi, Takagi and
  Matsumoto}]{sugiyama2010full}
\bibinfo{author}{Sugiyama, K.}, \bibinfo{author}{Ii, S.},
  \bibinfo{author}{Takeuchi, S.}, \bibinfo{author}{Takagi, S.},
  \bibinfo{author}{Matsumoto, Y.}, \bibinfo{year}{2010}.
\newblock \bibinfo{title}{Full eulerian simulations of biconcave neo-hookean
  particles in a poiseuille flow}.
\newblock \bibinfo{journal}{Computational Mechanics} \bibinfo{volume}{46},
  \bibinfo{pages}{147--157}.
\bibitem[{Sugiyama et~al.(2011)Sugiyama, Ii, Takeuchi, Takagi and
  Matsumoto}]{sugiyama2011full}
\bibinfo{author}{Sugiyama, K.}, \bibinfo{author}{Ii, S.},
  \bibinfo{author}{Takeuchi, S.}, \bibinfo{author}{Takagi, S.},
  \bibinfo{author}{Matsumoto, Y.}, \bibinfo{year}{2011}.
\newblock \bibinfo{title}{A full eulerian finite difference approach for
  solving fluid--structure coupling problems}.
\newblock \bibinfo{journal}{Journal of Computational Physics}
  \bibinfo{volume}{230}, \bibinfo{pages}{596--627}.
\bibitem[{Takagi et~al.(2012)Takagi, Sugiyama, Ii and
  Matsumoto}]{takagi2012review}
\bibinfo{author}{Takagi, S.}, \bibinfo{author}{Sugiyama, K.},
  \bibinfo{author}{Ii, S.}, \bibinfo{author}{Matsumoto, Y.},
  \bibinfo{year}{2012}.
\newblock \bibinfo{title}{A review of full eulerian methods for fluid structure
  interaction problems}.
\newblock \bibinfo{journal}{Journal of Applied Mechanics} \bibinfo{volume}{79},
  \bibinfo{pages}{010911}.
\bibitem[{Takizawa et~al.(2011a)Takizawa, Henicke, Montes, Tezduyar, Hsu and
  Bazilevs}]{takizawa2011numerical}
\bibinfo{author}{Takizawa, K.}, \bibinfo{author}{Henicke, B.},
  \bibinfo{author}{Montes, D.}, \bibinfo{author}{Tezduyar, T.E.},
  \bibinfo{author}{Hsu, M.C.}, \bibinfo{author}{Bazilevs, Y.},
  \bibinfo{year}{2011}a.
\newblock \bibinfo{title}{Numerical-performance studies for the stabilized
  space--time computation of wind-turbine rotor aerodynamics}.
\newblock \bibinfo{journal}{Computational Mechanics} \bibinfo{volume}{48},
  \bibinfo{pages}{647--657}.
\bibitem[{Takizawa et~al.(2012)Takizawa, Henicke, Puntel, Spielman and
  Tezduyar}]{takizawa2012space}
\bibinfo{author}{Takizawa, K.}, \bibinfo{author}{Henicke, B.},
  \bibinfo{author}{Puntel, A.}, \bibinfo{author}{Spielman, T.},
  \bibinfo{author}{Tezduyar, T.E.}, \bibinfo{year}{2012}.
\newblock \bibinfo{title}{Space-time computational techniques for the
  aerodynamics of flapping wings}.
\newblock \bibinfo{journal}{Journal of Applied Mechanics} \bibinfo{volume}{79},
  \bibinfo{pages}{010903}.
\bibitem[{Takizawa et~al.(2011b)Takizawa, Henicke, Tezduyar, Hsu and
  Bazilevs}]{takizawa2011stabilized}
\bibinfo{author}{Takizawa, K.}, \bibinfo{author}{Henicke, B.},
  \bibinfo{author}{Tezduyar, T.E.}, \bibinfo{author}{Hsu, M.C.},
  \bibinfo{author}{Bazilevs, Y.}, \bibinfo{year}{2011}b.
\newblock \bibinfo{title}{Stabilized space---time computation of wind-turbine
  rotor aerodynamics}.
\newblock \bibinfo{journal}{Computational Mechanics} \bibinfo{volume}{48},
  \bibinfo{pages}{333--344}.
\bibitem[{Tezduyar et~al.(1992)Tezduyar, Behr, Mittal and
  Liou}]{tezduyar1992new}
\bibinfo{author}{Tezduyar, T.E.}, \bibinfo{author}{Behr, M.},
  \bibinfo{author}{Mittal, S.}, \bibinfo{author}{Liou, J.},
  \bibinfo{year}{1992}.
\newblock \bibinfo{title}{A new strategy for finite element computations
  involving moving boundaries and interfaces—the
  deforming-spatial-domain/space-time procedure: Ii. computation of
  free-surface flows, two-liquid flows, and flows with drifting cylinders}.
\newblock \bibinfo{journal}{Computer methods in applied mechanics and
  engineering} \bibinfo{volume}{94}, \bibinfo{pages}{353--371}.
\bibitem[{Torii et~al.(2004)Torii, Oshima, Kobayashi, Takagi and
  Tezduyar}]{torii2004influence}
\bibinfo{author}{Torii, R.}, \bibinfo{author}{Oshima, M.},
  \bibinfo{author}{Kobayashi, T.}, \bibinfo{author}{Takagi, K.},
  \bibinfo{author}{Tezduyar, T.E.}, \bibinfo{year}{2004}.
\newblock \bibinfo{title}{Influence of wall elasticity on image-based blood
  flow simulations}.
\newblock \bibinfo{journal}{Nippon Kikai Gakkai Ronbunshu, A Hen/Transactions
  of the Japan Society of Mechanical Engineers, Part A} \bibinfo{volume}{70},
  \bibinfo{pages}{1224--1231}.
\bibitem[{Torii et~al.(2006a)Torii, Oshima, Kobayashi, Takagi and
  Tezduyar}]{torii2006computer}
\bibinfo{author}{Torii, R.}, \bibinfo{author}{Oshima, M.},
  \bibinfo{author}{Kobayashi, T.}, \bibinfo{author}{Takagi, K.},
  \bibinfo{author}{Tezduyar, T.E.}, \bibinfo{year}{2006}a.
\newblock \bibinfo{title}{Computer modeling of cardiovascular fluid--structure
  interactions with the deforming-spatial-domain/stabilized space--time
  formulation}.
\newblock \bibinfo{journal}{Computer Methods in Applied Mechanics and
  Engineering} \bibinfo{volume}{195}, \bibinfo{pages}{1885--1895}.
\bibitem[{Torii et~al.(2006b)Torii, Oshima, Kobayashi, Takagi and
  Tezduyar}]{torii2006fluid}
\bibinfo{author}{Torii, R.}, \bibinfo{author}{Oshima, M.},
  \bibinfo{author}{Kobayashi, T.}, \bibinfo{author}{Takagi, K.},
  \bibinfo{author}{Tezduyar, T.E.}, \bibinfo{year}{2006}b.
\newblock \bibinfo{title}{Fluid--structure interaction modeling of aneurysmal
  conditions with high and normal blood pressures}.
\newblock \bibinfo{journal}{Computational Mechanics} \bibinfo{volume}{38},
  \bibinfo{pages}{482--490}.
\bibitem[{Torii et~al.(2007a)Torii, Oshima, Kobayashi, Takagi and
  Tezduyar}]{torii2007influence}
\bibinfo{author}{Torii, R.}, \bibinfo{author}{Oshima, M.},
  \bibinfo{author}{Kobayashi, T.}, \bibinfo{author}{Takagi, K.},
  \bibinfo{author}{Tezduyar, T.E.}, \bibinfo{year}{2007}a.
\newblock \bibinfo{title}{Influence of wall elasticity in patient-specific
  hemodynamic simulations}.
\newblock \bibinfo{journal}{Computers \& Fluids} \bibinfo{volume}{36},
  \bibinfo{pages}{160--168}.
\bibitem[{Torii et~al.(2007b)Torii, Oshima, Kobayashi, Takagi and
  Tezduyar}]{torii2007numerical}
\bibinfo{author}{Torii, R.}, \bibinfo{author}{Oshima, M.},
  \bibinfo{author}{Kobayashi, T.}, \bibinfo{author}{Takagi, K.},
  \bibinfo{author}{Tezduyar, T.E.}, \bibinfo{year}{2007}b.
\newblock \bibinfo{title}{Numerical investigation of the effect of hypertensive
  blood pressure on cerebral aneurysm—dependence of the effect on the
  aneurysm shape}.
\newblock \bibinfo{journal}{International Journal for Numerical Methods in
  Fluids} \bibinfo{volume}{54}, \bibinfo{pages}{995--1009}.
\bibitem[{Torii et~al.(2008)Torii, Oshima, Kobayashi, Takagi and
  Tezduyar}]{torii2008fluid}
\bibinfo{author}{Torii, R.}, \bibinfo{author}{Oshima, M.},
  \bibinfo{author}{Kobayashi, T.}, \bibinfo{author}{Takagi, K.},
  \bibinfo{author}{Tezduyar, T.E.}, \bibinfo{year}{2008}.
\newblock \bibinfo{title}{Fluid--structure interaction modeling of a
  patient-specific cerebral aneurysm: influence of structural modeling}.
\newblock \bibinfo{journal}{Computational Mechanics} \bibinfo{volume}{43},
  \bibinfo{pages}{151}.
\bibitem[{Torii et~al.(2009)Torii, Oshima, Kobayashi, Takagi and
  Tezduyar}]{torii2009fluid}
\bibinfo{author}{Torii, R.}, \bibinfo{author}{Oshima, M.},
  \bibinfo{author}{Kobayashi, T.}, \bibinfo{author}{Takagi, K.},
  \bibinfo{author}{Tezduyar, T.E.}, \bibinfo{year}{2009}.
\newblock \bibinfo{title}{Fluid--structure interaction modeling of blood flow
  and cerebral aneurysm: significance of artery and aneurysm shapes}.
\newblock \bibinfo{journal}{Computer Methods in Applied Mechanics and
  Engineering} \bibinfo{volume}{198}, \bibinfo{pages}{3613--3621}.
\bibitem[{Torii et~al.(2010a)Torii, Oshima, Kobayashi, Takagi and
  Tezduyar}]{torii2010influence}
\bibinfo{author}{Torii, R.}, \bibinfo{author}{Oshima, M.},
  \bibinfo{author}{Kobayashi, T.}, \bibinfo{author}{Takagi, K.},
  \bibinfo{author}{Tezduyar, T.E.}, \bibinfo{year}{2010}a.
\newblock \bibinfo{title}{Influence of wall thickness on fluid--structure
  interaction computations of cerebral aneurysms}.
\newblock \bibinfo{journal}{International Journal for Numerical Methods in
  Biomedical Engineering} \bibinfo{volume}{26}, \bibinfo{pages}{336--347}.
\bibitem[{Torii et~al.(2010b)Torii, Oshima, Kobayashi, Takagi and
  Tezduyar}]{torii2010role}
\bibinfo{author}{Torii, R.}, \bibinfo{author}{Oshima, M.},
  \bibinfo{author}{Kobayashi, T.}, \bibinfo{author}{Takagi, K.},
  \bibinfo{author}{Tezduyar, T.E.}, \bibinfo{year}{2010}b.
\newblock \bibinfo{title}{Role of 0d peripheral vasculature model in
  fluid--structure interaction modeling of aneurysms}.
\newblock \bibinfo{journal}{Computational Mechanics} \bibinfo{volume}{46},
  \bibinfo{pages}{43--52}.
\bibitem[{Torii et~al.(2011)Torii, Oshima, Kobayashi, Takagi and
  Tezduyar}]{torii2011influencing}
\bibinfo{author}{Torii, R.}, \bibinfo{author}{Oshima, M.},
  \bibinfo{author}{Kobayashi, T.}, \bibinfo{author}{Takagi, K.},
  \bibinfo{author}{Tezduyar, T.E.}, \bibinfo{year}{2011}.
\newblock \bibinfo{title}{Influencing factors in image-based fluid--structure
  interaction computation of cerebral aneurysms}.
\newblock \bibinfo{journal}{International Journal for Numerical Methods in
  Fluids} \bibinfo{volume}{65}, \bibinfo{pages}{324--340}.
\bibitem[{Tryggvason et~al.(2011)Tryggvason, Scardovelli and
  Zaleski}]{tryggvason2011direct}
\bibinfo{author}{Tryggvason, G.}, \bibinfo{author}{Scardovelli, R.},
  \bibinfo{author}{Zaleski, S.}, \bibinfo{year}{2011}.
\newblock \bibinfo{title}{Direct numerical simulations of gas--liquid
  multiphase flows}.
\newblock \bibinfo{publisher}{Cambridge University Press}.
\bibitem[{Turitto et~al.(1972)Turitto, Benis and Leonard}]{turitto1972platelet}
\bibinfo{author}{Turitto, V.T.}, \bibinfo{author}{Benis, A.M.},
  \bibinfo{author}{Leonard, E.F.}, \bibinfo{year}{1972}.
\newblock \bibinfo{title}{Platelet diffusion in flowing blood}.
\newblock \bibinfo{journal}{Industrial \& engineering chemistry fundamentals}
  \bibinfo{volume}{11}, \bibinfo{pages}{216--223}.
\bibitem[{Udaykumar et~al.(1996)Udaykumar, Shyy and Rao}]{udaykumar1996elafint}
\bibinfo{author}{Udaykumar, H.}, \bibinfo{author}{Shyy, W.},
  \bibinfo{author}{Rao, M.}, \bibinfo{year}{1996}.
\newblock \bibinfo{title}{Elafint: a mixed eulerian--lagrangian method for
  fluid flows with complex and moving boundaries}.
\newblock \bibinfo{journal}{International journal for numerical methods in
  fluids} \bibinfo{volume}{22}, \bibinfo{pages}{691--712}.
\bibitem[{Udaykumar et~al.(2003)Udaykumar, Tran, Belk and
  Vanden}]{udaykumar2003eulerian}
\bibinfo{author}{Udaykumar, H.}, \bibinfo{author}{Tran, L.},
  \bibinfo{author}{Belk, D.}, \bibinfo{author}{Vanden, K.},
  \bibinfo{year}{2003}.
\newblock \bibinfo{title}{An eulerian method for computation of multimaterial
  impact with eno shock-capturing and sharp interfaces}.
\newblock \bibinfo{journal}{Journal of Computational Physics}
  \bibinfo{volume}{186}, \bibinfo{pages}{136--177}.
\bibitem[{Valkov et~al.(2015)Valkov, Rycroft and Kamrin}]{Valkov2015}
\bibinfo{author}{Valkov, B.}, \bibinfo{author}{Rycroft, C.H.},
  \bibinfo{author}{Kamrin, K.}, \bibinfo{year}{2015}.
\newblock \bibinfo{title}{{Eulerian Method for Multiphase Interactions of Soft
  Solid Bodies in Fluids}}.
\newblock \bibinfo{journal}{Journal of Applied Mechanics} \bibinfo{volume}{82},
  \bibinfo{pages}{041011}.
\newblock \URLprefix
  \url{http://appliedmechanics.asmedigitalcollection.asme.org/article.aspx?doi=10.1115/1.4029765},
  \DOIprefix\doi{10.1115/1.4029765}, \href{http://arxiv.org/abs/1409.6183}{\tt
  arXiv:1409.6183}.
\bibitem[{Wang and Layton(2009)}]{wang2009numerical}
\bibinfo{author}{Wang, J.}, \bibinfo{author}{Layton, A.}, \bibinfo{year}{2009}.
\newblock \bibinfo{title}{Numerical simulations of fiber sedimentation in
  navier-stokes flows}.
\newblock \bibinfo{journal}{Communications in Computational Physics}
  \bibinfo{volume}{5}, \bibinfo{pages}{61}.
\bibitem[{Wang and Zhang(2013)}]{wang2013modified}
\bibinfo{author}{Wang, X.}, \bibinfo{author}{Zhang, L.T.},
  \bibinfo{year}{2013}.
\newblock \bibinfo{title}{Modified immersed finite element method for
  fully-coupled fluid--structure interactions}.
\newblock \bibinfo{journal}{Computer methods in applied mechanics and
  engineering} \bibinfo{volume}{267}, \bibinfo{pages}{150--169}.
\bibitem[{Wang(2006)}]{wang2006immersed}
\bibinfo{author}{Wang, X.S.}, \bibinfo{year}{2006}.
\newblock \bibinfo{title}{From immersed boundary method to immersed continuum
  methods}.
\newblock \bibinfo{journal}{International Journal for Multiscale Computational
  Engineering} \bibinfo{volume}{4}.
\bibitem[{Wang(2007)}]{wang2007iterative}
\bibinfo{author}{Wang, X.S.}, \bibinfo{year}{2007}.
\newblock \bibinfo{title}{An iterative matrix-free method in implicit immersed
  boundary/continuum methods}.
\newblock \bibinfo{journal}{Computers \& structures} \bibinfo{volume}{85},
  \bibinfo{pages}{739--748}.
\bibitem[{Wang(2010)}]{wang2010immersed}
\bibinfo{author}{Wang, X.S.}, \bibinfo{year}{2010}.
\newblock \bibinfo{title}{Immersed boundary/continuum methods}, in:
  \bibinfo{booktitle}{Computational Modeling in Biomechanics}.
  \bibinfo{publisher}{Springer}, pp. \bibinfo{pages}{3--48}.
\bibitem[{Watanabe et~al.(2004)Watanabe, Sugiura, Kafuku and
  Hisada}]{watanabe2004multiphysics}
\bibinfo{author}{Watanabe, H.}, \bibinfo{author}{Sugiura, S.},
  \bibinfo{author}{Kafuku, H.}, \bibinfo{author}{Hisada, T.},
  \bibinfo{year}{2004}.
\newblock \bibinfo{title}{Multiphysics simulation of left ventricular filling
  dynamics using fluid-structure interaction finite element method}.
\newblock \bibinfo{journal}{Biophysical journal} \bibinfo{volume}{87},
  \bibinfo{pages}{2074--2085}.
\bibitem[{Weymouth(2008)}]{weymouth2008physics}
\bibinfo{author}{Weymouth, G.D.}, \bibinfo{year}{2008}.
\newblock \bibinfo{title}{Physics and learning based computational models for
  breaking bow waves based on new boundary immersion approaches}.
\newblock Ph.D. thesis. Massachusetts Institute of Technology.
\bibitem[{Weymouth et~al.(2006)Weymouth, Dommermuth, Hendrickson and
  Yue}]{weymouth2006advancements}
\bibinfo{author}{Weymouth, G.D.}, \bibinfo{author}{Dommermuth, D.G.},
  \bibinfo{author}{Hendrickson, K.}, \bibinfo{author}{Yue, D.K.P.},
  \bibinfo{year}{2006}.
\newblock \bibinfo{title}{Advancements in cartesian-grid methods for
  computational ship hydrodynamics} .
\bibitem[{Wootton and Ku(1999)}]{wootton1999fluid}
\bibinfo{author}{Wootton, D.M.}, \bibinfo{author}{Ku, D.N.},
  \bibinfo{year}{1999}.
\newblock \bibinfo{title}{Fluid mechanics of vascular systems, diseases, and
  thrombosis}.
\newblock \bibinfo{journal}{Annual review of biomedical engineering}
  \bibinfo{volume}{1}, \bibinfo{pages}{299--329}.
\bibitem[{Xiao(1999)}]{xiao1999computation}
\bibinfo{author}{Xiao, F.}, \bibinfo{year}{1999}.
\newblock \bibinfo{title}{Computation of complex flow containing rheological
  bodies}.
\newblock \bibinfo{journal}{Computation Fluid Dynamics Journal}
  \bibinfo{volume}{8}, \bibinfo{pages}{43--49}.
\bibitem[{Yang and Balaras(2006)}]{yang2006embedded}
\bibinfo{author}{Yang, J.}, \bibinfo{author}{Balaras, E.},
  \bibinfo{year}{2006}.
\newblock \bibinfo{title}{An embedded-boundary formulation for large-eddy
  simulation of turbulent flows interacting with moving boundaries}.
\newblock \bibinfo{journal}{Journal of Computational Physics}
  \bibinfo{volume}{215}, \bibinfo{pages}{12--40}.
\bibitem[{Yuki et~al.(2007)Yuki, Takeuchi and Kajishima}]{yuki2007efficient}
\bibinfo{author}{Yuki, Y.}, \bibinfo{author}{Takeuchi, S.},
  \bibinfo{author}{Kajishima, T.}, \bibinfo{year}{2007}.
\newblock \bibinfo{title}{Efficient immersed boundary method for strong
  interaction problem of arbitrary shape object with the self-induced flow}.
\newblock \bibinfo{journal}{Journal of Fluid Science and Technology}
  \bibinfo{volume}{2}, \bibinfo{pages}{1--11}.
\bibitem[{Zhang and Gay(2007)}]{zhang2007immersed}
\bibinfo{author}{Zhang, L.}, \bibinfo{author}{Gay, M.}, \bibinfo{year}{2007}.
\newblock \bibinfo{title}{Immersed finite element method for fluid-structure
  interactions}.
\newblock \bibinfo{journal}{Journal of Fluids and Structures}
  \bibinfo{volume}{23}, \bibinfo{pages}{839--857}.
\bibitem[{Zhao et~al.(2008)Zhao, Freund and Moser}]{zhao2008fixed}
\bibinfo{author}{Zhao, H.}, \bibinfo{author}{Freund, J.B.},
  \bibinfo{author}{Moser, R.D.}, \bibinfo{year}{2008}.
\newblock \bibinfo{title}{A fixed-mesh method for incompressible
  flow--structure systems with finite solid deformations}.
\newblock \bibinfo{journal}{Journal of Computational Physics}
  \bibinfo{volume}{227}, \bibinfo{pages}{3114--3140}.
\bibitem[{Zhu and Peskin(2003)}]{zhu2003interaction}
\bibinfo{author}{Zhu, L.}, \bibinfo{author}{Peskin, C.S.},
  \bibinfo{year}{2003}.
\newblock \bibinfo{title}{Interaction of two flapping filaments in a flowing
  soap film}.
\newblock \bibinfo{journal}{Physics of fluids} \bibinfo{volume}{15},
  \bibinfo{pages}{1954--1960}.

\end{thebibliography}

\end{document}